\documentclass[twocolumn]{aastex63}
\newcommand{\lya}{Ly$\alpha$}
\newcommand{\oiii}{[O\,{\sc iii}]}
\newcommand{\ha}{H$\alpha$}

\newcommand{\w}{$W_r$~}
\newcommand{\kms}{km s$^{-1}$}
\newcommand{\z}{$z$}

\newcommand{\cii}{[C~{\sc ii}]~158~$\mu$m}

\received{}
\revised{}
\accepted{}
\submitjournal{AAS}

\usepackage{booktabs}
\usepackage{array}
\usepackage{tabularx}
\usepackage{xcolor}
 
\shorttitle{C3D DSFG ENV}
\shortauthors{Zou et al.}

\begin{document}

\title{COSMOS-3D: Diverse Environments and Hot-dust Signatures among Dusty Star-forming Galaxies at $z$ = 4.9--7.2}

\author[0000-0002-3983-6484]{Siwei Zou}
\affiliation{Chinese Academy of Sciences South America Center for Astronomy, National Astronomical Observatories, CAS, Beijing 100101, China}
\affiliation{Departamento de Astronom\'ia, Universidad de Chile, Casilla 36-D, Santiago, Chile}

\author[0000-0002-6290-3198]{Manuel Aravena}
\affiliation{Instituto de Estudios Astrof\'isicos, Facultad de Ingenier\'ia y Ciencias, Universidad Diego Portales, Av. Ej\'ercito 441, Santiago, Chile}
\affiliation{Millennium Nucleus for Galaxies (MINGAL)}

\author[0009-0009-4187-2095]{Shaoze Geng}
\affiliation{Chinese Academy of Sciences South America Center for Astronomy, National Astronomical Observatories, CAS, Beijing 100101, China}
\affiliation{School of Astronomy and Space Sciences, University of Chinese Academy of Sciences, Beijing 100049, China} 


\author[0000-0001-5492-4522]{Romain A. Meyer}
\affiliation{Department of Astronomy, University of Geneva, Chemin Pegasi 51,
1290 Versoix, Switzerland}

\author[0000-0002-6221-1829]{Jianwei Lyu}
\affiliation{Steward Observatory, University of Arizona,933 N Cherry Ave, Tucson, AZ, 85721, USA}

\author[0000-0002-6184-9097]{Jaclyn B. Champagne}
\affiliation{Space Telescope Science Institute, 3700 San Martin Drive, Baltimore, MD 21218, USA}

\author[0000-0001-6511-8745]{Jia-Sheng Huang}
\affiliation{Chinese Academy of Sciences South America Center for Astronomy, National Astronomical Observatories, CAS, Beijing 100101, China}
\affiliation{Harvard-Smithsonian Center for Astrophysics, 60 Garden Street, Cambridge, MA 02138, USA}

\author[0000-0002-9382-9832]{Andreas L. Faisst}
\affiliation{IPAC, California Institute of Technology, 1200 E. California Blvd. Pasadena, CA 91125, USA}

\author[0000-0003-0964-7188]{Shuqi Fu}
\affiliation{Kavli Institute for Astronomy and Astrophysics, Peking University, Beijing 100871, China}
\affiliation{Department of Astronomy, School of Physics, Peking University, Beijing 100871, People’s Republic of China}

\author[0000-0003-4569-2285]{Andrew J. Battisti}
\affiliation{International Centre for Radio Astronomy Research, University of Western Australia, 7 Fairway, Crawley, WA 6009, Australia}
\affiliation{Research School of Astronomy and Astrophysics, Australian National University, Cotter Road, Weston Creek, ACT 2611, Australia}

\author[0000-0002-4205-9567]{Hiddo Algera}
\affiliation{Institute of Astronomy and Astrophysics, Academia Sinica, 11F of Astronomy-Mathematics Building, No. 1, Sec. 4, Roosevelt Rd, Taipei 106216, Taiwan}

\author[0000-0002-0930-6466]{Caitlin M. Casey}
\affiliation{Department of Physics, University of California, Santa Barbara, Santa Barbara, CA 93106, USA}
\affiliation{Cosmic Dawn Center (DAWN), Denmark}

\author[0000-0003-3310-0131]{Xiaohui Fan}
\affiliation{Steward Observatory, University of Arizona, 933 N Cherry Ave, Tucson, AZ, 85721, USA}

\author[0000-0002-3560-8599]{Maximilien Franco}
\affiliation{Université Paris-Saclay, Université Paris Cité, CEA,
CNRS, AIM, 91191 Gif-sur-Yvette, France}

\author[0000-0002-0236-919X]{Ghassem Gozaliasl}
\affiliation{Department of Computer Science, Aalto University, PO Box 15400, Espoo, FI-00076, Finland}
\affiliation{Department of Physics, University of Helsinki, P. O. Box 64, FI-00014 Helsinki, Finland}

\author[0000-0003-0964-7188]{Linhua Jiang}
\affiliation{Kavli Institute for Astronomy and Astrophysics, Peking University, Beijing 100871, China}
\affiliation{Department of Astronomy, School of Physics, Peking University, Beijing 100871, People’s Republic of China}


\author[0000-0001-6874-1321]{Koki Kakiichi}
\affiliation{Cosmic Dawn Center (DAWN), Denmark}
\affiliation{Niels Bohr Institute, University of Copenhagen, Jagtvej 128, DK-2200, Copenhagen N, Denmark}

\author[0000-0002-2603-2639]{Darshan Kakkad}
\affiliation{Centre for Astrophysics Research, Department of Physics, Astronomy and Mathematics, University of Hertfordshire, Hatfield AL10 9AB, UK}

\author[0000-0001-5951-459X]{Zihao Li}
\affiliation{Cosmic Dawn Center (DAWN), Denmark}
\affiliation{Niels Bohr Institute, University of Copenhagen, Jagtvej 128, DK-2200, Copenhagen N, Denmark}

\author[0009-0004-1270-2373]{Lun-Jun Liu}
\affiliation{California Institute of Technology, 1200 E. California Blvd., Pasadena, CA, 91125 USA}

\author[0000-0002-9883-1413]{Felix Martinez III}
\affiliation{Laboratory for Multiwavelength Astrophysics, School of Physics and Astronomy, Rochester Institute of Technology, 84 Lomb Memorial Drive, Rochester, NY 14623, USA}

\author[0000-0002-7051-1100]{Jorge A. Zavala}
\affiliation{University of Massachusetts Amherst, 710 North Pleasant Street, Amherst, MA 01003-9305, USA}

\author[0000-0003-2716-8332]{Rasha M. Samir}
\affiliation{Department of Astronomy, National Research Institute of Astronomy and Geophysics (NRIAG), Cairo, 11421, Egypt}




\correspondingauthor{Siwei Zou}
\email{zousw@bao.ac.cn}


\begin{abstract}
Dusty star-forming galaxies (DSFGs) are expected to trace early massive-halo assembly, but the connection between dust-obscured star formation, morphology, hot dust, and environment remains unclear at $z>4$. We combine JWST/NIRCam F444W grism spectroscopy from COSMOS-3D with MIRI F1000W/F2100W imaging and a mixed ALMA-selected and ALMA-followed dusty-galaxy sample from CRISTAL, CHAMPS, REBELS, and A3COSMOS to study 18 DSFGs or DSFG candidates at $z=4.9$--7.2. We compare their environments with the parent spectroscopically confirmed emission-line sample and the COSMOS-Web photo-$z$ galaxy sample using separate and combined overdensity estimators. The parent narrow-line \ha~sample has a median observed, dust-uncorrected ${\rm SFR}_{\rm H\alpha}=11.7~M_\odot~{\rm yr^{-1}}$. Within $R=0.5$ pMpc, DSFGs have a median $\delta_{\rm spec+phot}=0.92^{+0.06}_{-0.16}$, compared with $0.80^{+0.05}_{-0.04}$ for HAEs, with a stronger contrast on smaller scales. The strongest compact overdensities are associated with merger/interacting morphology: merging DSFGs reach $\delta_{\rm spec+phot}=4.70^{+1.61}_{-1.51}$ within $R=0.5$ pMpc. In contrast, MIRI-bright DSFGs do not show an enhanced number of nearby visible \ha-emitting companions, and their F2100W fluxes are difficult to explain with the 3.3~$\mu$m PAH feature alone, suggesting an additional hot-dust component. These results are consistent with a possible phase-dependent picture in which the compact line-emitter core, merger-driven dusty phase, and MIRI-bright hot-dust phase need not be spatially or temporally identical during early structure growth.
\end{abstract}

\keywords{High-redshift galaxies (734); Submillimeter astronomy (1647); Galaxy environments (2029); Emission line galaxies (459); Galaxy mergers (608)}

\section{Introduction} \label{sec_intro}
 

Dusty star-forming galaxies (DSFGs), often selected as bright submillimeter galaxies (SMGs), have long been proposed as signposts of massive dark-matter halos and forming cluster cores \citep*{casey14,hodge20}. Yet the interplay between dust-obscured star formation, possible active galactic nuclei (AGN) activity, and large-scale environment at the earliest epochs remains poorly understood, especially at $z>4$, where photometric-redshift uncertainties and projection effects make spectroscopic confirmation essential. 

Early searches for dusty galaxies in dense environments often relied on
far-infrared or submillimeter number-count excesses around known
high-redshift structures or massive radio galaxies
\citep[e.g.,][]{ivison00,stevens03,blain04,clements14,planck15,planck16}. Herschel and Planck-based studies identified candidate overdensities of dusty star-forming sources in known or candidate protocluster fields, showing
that far-infrared selection can efficiently locate regions of intense obscured activity, but also highlighting the severe impact of projection effects and the need for spectroscopic follow-up \citep{kato16,greenslade18,calvi23,mckinney25}. More statistical approaches have reached a similar conclusion: bright SMGs are frequently associated with photometric-redshift overdensities, suggesting that they are useful tracers of large-scale structure \citep{calvi23}. These studies motivate the use of DSFGs as environmental signposts, but also show that a robust measurement of their connection to overdensity requires well-defined comparison samples and redshift information.

Several benchmark systems now demonstrate that DSFGs can indeed reside in spectroscopically confirmed overdense environments at $z>4$. In GOODS-N, the GN20 structure contains multiple CO-confirmed SMGs at $z$=4.05, showing that a large fraction of star formation in early overdensities can be heavily obscured \citep{daddi09,tan14}. More extreme examples include the Distant Red Core at $z_{\rm spec}\sim4$, which contains a compact core of spectroscopically confirmed DSFGs, and SPT2349--56 at $z\sim4.3$, where ALMA spectroscopy resolved an
exceptionally dense, gas-rich protocluster core and its surrounding
Mpc-scale structure \citep{oteo18,miller18,hill20}. At even earlier epochs, AzTEC-3 lies in the COSMOS $z\sim5.3$ protocluster and has been studied through CO, \cii, and \lya~tracers \citep{riechers10,capak11,guaita22}. The MAMBO-9 at $z=5.85$ is a merging pair of optically dark DSFGs in an overdense environment, with JWST and ALMA observations revealing \cii~velocity gradients and tidal interaction \citep{akins26}. The DSFG HDF850.1 is embedded in the GOODS-N $z\sim5.2$ overdensity, which hosts multiple dusty starbursts and JWST-confirmed \ha~emitters \citep{walter12,calvi21,helton24,sun24,herard25,lagache26}. During the epoch of reionization, SPT0311--58 and HFLS3 show that very massive dusty systems can already be embedded in compact, spectroscopically confirmed galaxy structures \citep{marrone18,spilker22,arribas24,jones24}.

The physical link between gas supply, dusty star formation, black-hole growth, and overdensity remains unclear. Observations of extended [C~{\sc ii}] emission on circumgalactic scales around $z\sim4$--7 quasars and galaxies indicate that cold, metal-enriched gas can extend beyond the rest-frame UV and dust-continuum emission \citep{cicone15,fujimoto19,fujimoto20,meyer22,bis24,meyer25b}. Similar extended cold-gas reservoirs have also been detected at lower redshift through CO and [C~{\sc i}] emission, including the molecular gas halo around the Spiderweb Galaxy at $z=2.16$ and a $\sim100$ kpc atomic-carbon stream connected to the radio galaxy 4C~41.17 at $z=3.8$ \citep{emonts16,emonts23}. In ALPINE and CRISTAL, extended [C~{\sc ii}] structures are often interpreted as diffuse neutral gas associated with interactions, satellites, outflows, or gas redistribution in the CGM \citep{ginolfi20cgm,ikeda25,liu26}. Absorption-line studies provide a complementary view, with disturbed cold gas traced by strong Mg~{\sc ii} absorbers around $z\sim2.5$ dusty disk galaxies \citep{zou25}. Together, these results suggest that overdense environments may influence DSFG evolution not only by increasing the number of companions, but also by supplying and redistributing cold gas that can fuel dusty star formation and later hot-dust or obscured-AGN activity. This picture is supported by $z\sim2$--3 protoclusters such as Spiderweb, Hyperion, and SSA22, which show enhanced dusty star formation, complex large-scale structure, and in some cases elevated AGN activity \citep{dannerbauer14,casey15,umehata15,umehata17,zavala19,jackie26,zhang26}. Recent studies further suggest enhanced AGN activity in dense environments at $2<z<4$, and possibly an elevated AGN fraction among protocluster DSFGs relative to matched field DSFGs \citep{shah25,islallave26}.

\begin{figure*}
\centering
 \includegraphics[width=0.75\textwidth]{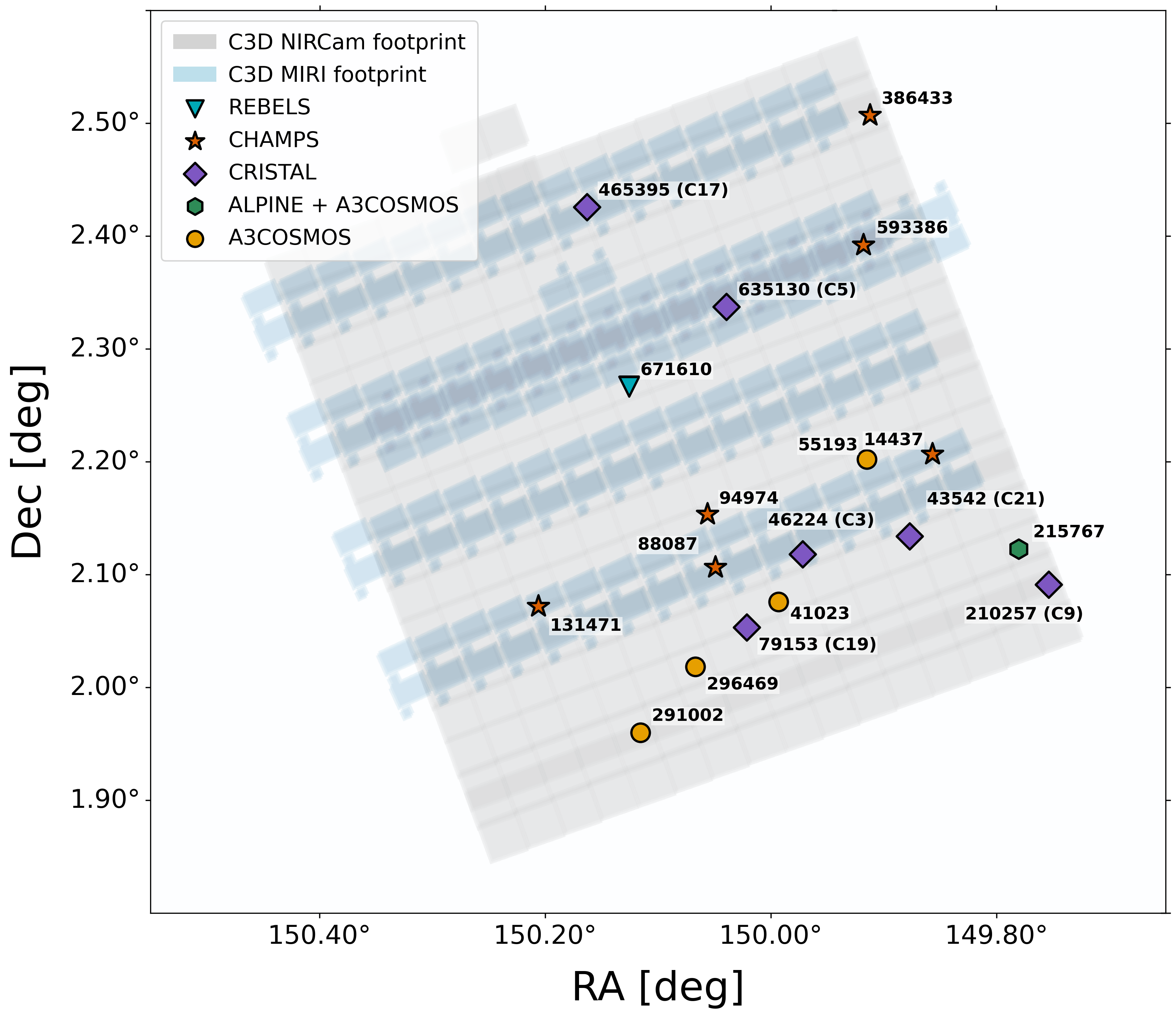}
  \caption{\small{The C3D footprints of the NIRCam F444W grism observations (grey) and the MIRI (blue) footprints, together with all DSFGs used in this work.}
}\label{fig:footprint}
\end{figure*}

However, most previous studies have searched for DSFGs and possible AGN activity within known protoclusters or large-scale overdensities, or have associated dusty galaxies with structures identified from broad photo-$z$ selections. A more direct test requires a redshift map that is constructed independently of the DSFG positions. JWST/NIRCam wide-field slitless spectroscopy (WFSS) now provides an efficient way to build such blind spectroscopic redshift maps over large extragalactic fields, allowing dusty-galaxy environments to be compared with a uniformly selected galaxy population rather than only with pre-selected protocluster fields. NIRSpec adds rest-frame optical line diagnostics of dust attenuation, ionization conditions, and gas-phase metallicity \citep[e.g.,][]{cristal25,faisst25,zavala26}, while MIRI F1000W/F2100W probes rest-frame near-infrared emission at $z\sim5$--7, including hot dust, the 3.3 $\mu$m PAH feature, and possible obscured-AGN activity \citep{lyu24}. Combined with ALMA measurements of dust continuum and the cold ISM, these data allow us to test whether the environments of DSFGs are linked primarily to dust-obscured star formation, stellar mass, hot-dust or AGN activity, or merger signatures.

In this work, we use ``DSFG'' in a broad observational sense to describe spectroscopically confirmed gas-rich star-forming galaxies with ALMA dust-continuum detections. The sample includes both ALMA-selected dust-continuum sources and UV-selected galaxies with ALMA follow-up. The sample includes dust-continuum-selected CHAMPS sources (ALMA \#2023.1.00180.L; PI: A.~Faisst; Faisst et al., in prep.), A3COSMOS continuum sources \citep{liu19a3cosmos}, and ALPINE--CRISTAL galaxies that were originally UV-selected but have ALMA \cii~and/or dust-continuum detections \citep{fevre20,faisst20,bet20,cristal25}. We also include one REBELS-P8/A3COSMOS source with an ALMA continuum detection \citep{bouwens22,inami22}. We combine these sources with JWST/NIRCam grism spectroscopy and MIRI F1000W/F2100W imaging from COSMOS-3D to measure their small-scale environments at $z=4.9$--7.2 and to connect those environments to star formation, mergers, and hot-dust or obscured-AGN candidates.

COSMOS-3D (hereafter C3D) is a JWST Cycle~3 Large Program (GO\#5893; PI: K.~Kakiichi), consisting of 124.8 hours of primary observations and 120.8 hours of parallel observations. The full program overview will be presented in Kakiichi et al. (in prep.). The NIRCam/F444W grism spectra provide an efficient redshift survey for confirming DSFGs and mapping their galaxy environments, while the two MIRI bands provide constraints on rest-frame near-infrared excess emission from hot dust, PAH emission, or obscured-AGN activity. We construct our 18-source DSFG sample from \textsc{CRISTAL}, \textsc{CHAMPS}, \textsc{REBELS}, and \textsc{A3COSMOS}, and use the surrounding NIRCam-selected emission-line galaxy population to quantify their local overdensities. We describe the NIRCam and MIRI observations in Section~\ref{sec:observation}. The construction and properties of the 18 DSFGs analyzed in this work are presented in Section~\ref{sec:analysis}. The spectroscopically selected and photometrically selected galaxy overdensities are analyzed in Sections~\ref{sec:spec_delta} and~\ref{sec:phot_delta}, respectively. We discuss the implications of our results in Section~\ref{sec:discussion}.

Throughout this work, we assume a flat $\Lambda$CDM cosmology with $H_0=70~{\rm km~s^{-1}~Mpc^{-1}}$, $\Omega_{\rm m}=0.3$, and $\Omega_{\Lambda}=0.7$. Distances are quoted in proper or comoving units as specified. For quantities expressed in $h^{-1}$ units, we use $h=0.70$.

\section{Data reduction and analysis}\label{sec:observation}

The early C3D NIRCam and MIRI data releases and reduction are presented in Wang et al. (in prep.) and Lyu et al. (in prep.), respectively. The C3D imaging is tied to the COSMOS-Web imaging products and astrometric reference frame. COSMOS-Web provides wide-area NIRCam imaging over the COSMOS field \citep{casey23,franco26}, and part of the F115W imaging used here comes from COSMOS-Web. The full C3D photometric catalog, incorporating both NIRCam and MIRI data, will be published in Champagne et al. (in prep.). The C3D data were obtained from December 2024 to January 2026. The footprints of the C3D NIRCam and MIRI coverage, together with the DSFGs studied in this work, are presented in Figure~\ref{fig:footprint}.

\subsection{COSMOS-3D NIRCam imaging and grism spectra}


We performed the NIRCam grism spectroscopy using the F444W filter and employed NIRCam F200W in the short-wavelength (SW) channel during WFSS observations, along with F115W and F356W for the accompanying direct imaging. The total NIRCam coverage is $\sim~1151~{\rm arcmin^2}$. The $5\sigma$ line flux limit is $4\times10^{-18}{\rm erg~s^{-1}cm^{-2}}$ at $\sim 4.5\mu$m. The NIRCam F115W, F200W, and F356W imaging data presented in this work were reduced using version 1.16.1 of the JWST calibration pipeline, with the CRDS calibration reference-file context \texttt{jwst\_1303.pmap}. The grism spectra were extracted using the \texttt{unfold} and \texttt{grizli} procedures. The \texttt{unfold} program uses the updated trace model from \citet{sun23}\footnote{\url{https://github.com/fengwusun/nircam_grism}}, while \texttt{grizli} is described by \citet{brammer19}. Details of the extraction are described in \citet{wang23} and \citet{meyer25}.

\subsection{COSMOS-3D MIRI imaging}

C3D obtains MIRI parallel imaging in F1000W and F2100W, which we use to study massive galaxies and identify hot-dust or obscured-AGN candidates. The total survey area for the MIRI parallel imaging is ${\sim484~{\rm arcmin^2}}$, with ${\sim346~{\rm arcmin^2}}$ overlapping the COSMOS-Web MIRI F770W coverage \citep{harish25}.

The MIRI data were reduced following the method adopted in the Systematic Mid-infrared Instrument Legacy Extragalactic Survey (SMILES; \citealt{lyu24}). We briefly summarize the procedure here. The JWST standard pipeline is used for Stage 1 and Stage 2 processing. In particular, during Stage 2, we apply a custom external background subtraction developed as part of the Rainbow Database JWST pipeline, which constructs a super-background model to mitigate cosmic ray showers (see details in \citealt{alvarez23}; Lyu et al. in prep).

\subsection{Photometry}\label{sec:photometry}
JWST/NIRCam and MIRI photometry was measured following the SEP and SE++ procedure adopted for the COSMOS2025 catalog \citep{shuntov25} and implemented for the C3D photometric catalog in \citet{zj26}. Sources were detected with SEP on square-root positive $\chi^2$ images constructed from the available broadband data in each tile, using combined ``hot'' and ``cold'' runs to recover both compact faint sources and extended bright galaxies. Kron fluxes were measured with compact and standard apertures, $(k,R)=(1.2,1.6)$ and $(2.5,3.5)$, respectively, using SEP and \texttt{photutils}, and their ratio was used for aperture correction. Although the images were not PSF-matched, Kron shapes were consistently defined from the common $\chi^2$ detection image. Fixed-aperture photometry was then obtained with SE++ in ASSOC mode at the SEP source positions \citep{bertin20}. The full C3D photometric catalog will be presented in Champagne et al. (in prep.).

\subsection{ALMA programs and catalogs used in this work}

\vspace{1mm}
\noindent{\bf ALPINE and CRISTAL}
\vspace{1mm}

The ALMA Large Program to INvestigate [C~{\sc ii}] at Early times (ALPINE) targeted \cii~emission in 118 star-forming galaxies at $z=4.4$--5.9 \citep{fevre20,faisst20,bet20}, sampling $\log(M_\star/M_\odot)\sim8.4$--11 and total ${\rm SFR}\sim3$--270~$M_\odot~{\rm yr^{-1}}$. \citet{bet20} reported 23 continuum detections among the targeted ALPINE galaxies and 57 additional serendipitous dust-continuum sources in the ALMA pointings. The survey was designed to characterize the far-infrared properties of normal star-forming galaxies at $z\sim4$--6, linking \cii~emission, dust-obscured star formation, and galaxy metallicity.

The \cii~Resolved ISM in STar-forming galaxies with ALMA (CRISTAL) survey is an ALMA Cycle~8 Large Program designed to map gas, dust, and stars in star-forming galaxies at $4 \leq z \leq 6$ on kiloparsec scales \citep{cristal25}. Based on deep ALMA observations of 19 ALPINE galaxies, CRISTAL provides spatially resolved \cii~and dust-continuum measurements of main-sequence galaxies in the early Universe. Because the targeted CRISTAL sample is drawn from ALPINE, it occupies a similar main-sequence parameter space, with $\log(M_\star/M_\odot)\simeq9.5$--11 and ${\rm SFR}\gtrsim10~M_\odot~{\rm yr^{-1}}$. These data are complemented by ancillary HST, JWST/NIRCam, and JWST/NIRSpec observations, whose high angular resolution allows different gas and stellar components to be resolved in and around these systems \citep{her21,par25}. Recent JWST/NIRSpec follow-up of the ALPINE--CRISTAL galaxies is presented in \citet{faisst25,fujimoto25}. Within the C3D footprint, 13 CRISTAL galaxies are covered, seven of which have H$\alpha$ accessible in the F444W NIRCam grism spectra over $z=4.8$--6.9.

\vspace{1mm}
\noindent{\bf CHAMPS}
\vspace{1mm}

The COSMOS High-$z$ ALMA--MIRI Population Survey (CHAMPS; ALMA \#2023.1.00180.L; PI: A.~Faisst) is a large ALMA 1.1~mm survey covering a total of 0.18~deg$^2$ in the COSMOS field. CHAMPS was designed to cover the MIRI pointings of COSMOS-Web (GO\#1727; PIs: J.~Kartaltepe and C.~Casey; \citealp{casey23}) and PRIMER (GO\#1837; PI: J.~Dunlop; \citealp{dunlop21}), and is therefore not a single contiguous ALMA mosaic. It is not selected by a fixed $M_\star$ or SFR cut; instead, its parameter space is set by the JWST/MIRI parent coverage and the ALMA Band~6 continuum depth. CHAMPS provides ALMA coverage over the JWST/MIRI-observed regions used to identify and characterize massive dusty galaxies and hot-dust or obscured-AGN candidates. In combination with COSMOS-Web and PRIMER imaging, CHAMPS enables UV-to-FIR SED constraints for $\sim$30,000 MIRI-detected sources in COSMOS, including measurements of stellar masses, dust-obscured SFRs, AGN contributions, and ALMA continuum number counts. The CHAMPS survey overview and catalog will be presented in Faisst et al. (in prep.) and Martinez et al. (in prep.).

\vspace{1mm}
\noindent{\bf REBELS}
\vspace{1mm}

The Reionization Era Bright Emission Line Survey (REBELS) is an ALMA Large Program targeting UV-luminous galaxies at $z>6.5$ with spectral scans for \cii~and dust-continuum emission, providing one of the first statistical samples of cold interstellar-medium reservoirs during the epoch of reionization \citep{bouwens22,inami22}. REBELS and its pilot programs have shown that dust and cold gas are already present in massive galaxies at $z\sim7$, with dust-continuum detections in a substantial fraction of the UV-bright sample \citep{fudamoto21,inami22}. Published analyses of dust-detected REBELS galaxies find total UV+IR SFRs of $\sim30$--$130~M_\odot~{\rm yr^{-1}}$ \citep{ferrara22}, while stacking analyses probe $\log(M_\star/M_\odot)\simeq9.4$--10.4 and find obscured star-formation fractions of $f_{\rm obs}\simeq0.3$--0.6 \citep{algera23}. These results suggest that dust-obscured star formation can contribute non-negligibly to the cosmic star-formation rate density at $z\sim7$, while \cii-based measurements constrain the molecular gas content and depletion times of early galaxies \citep{algera23,aravena24}.

\vspace{1mm}
\noindent{\bf A3COSMOS}
\vspace{1mm}

We also make use of the A3COSMOS catalog, an automated mining of public ALMA archival observations in the COSMOS field designed to provide uniformly reduced continuum images and source catalogs for studying dust-obscured star formation and interstellar-medium properties across cosmic time \citep{liu19a3cosmos}. Because A3COSMOS is an archival compilation rather than a single survey with uniform target selection, it does not correspond to one fixed $M_\star$ or SFR range. The released catalog provides redshifts, stellar masses, and SFRs for $\sim700$ ALMA-detected galaxies over $0.5<z<6$, but the effective selection function is heterogeneous. Existing A3COSMOS-based studies show that robust ALMA continuum detections are biased toward massive galaxies, typically $\log(M_\star/M_\odot)\gtrsim10$ at high redshift. The catalog includes both blind ALMA source extractions and prior-based photometry at the positions of known multi-wavelength COSMOS sources, with updated data products released through the A3COSMOS data page.\footnote{\url{https://sites.google.com/view/a3cosmos/data/dataset_v20250312}} In this work, we define reliable A3COSMOS detections using the peak signal-to-noise ratio ${\rm S/N}_{\rm peak}={\tt Peak\_flux}/{\tt RMS\_noise}$, and require ${\rm S/N}_{\rm peak}\geq4.35$ for both the real prior-selected and blind A3COSMOS catalogs.

\begin{table*}
\centering
\caption{\small The 18-source DSFG sample used in the environmental analysis. For H$\alpha$ sources, fluxes and FWHMs are measured from the 1D fits shown in Figure~\ref{fig:five_miri_dsfg_cutouts} for the five MIRI-detected sources and Appendix Figure~\ref{fig:dsfg_cutouts} for the remaining sources. For the two \oiii-selected sources, \oiii$\lambda5008$ fluxes are taken from \citet{meyer25}, while FWHMs are measured from the 1D spectra. Redshift errors are formal $1\sigma$ uncertainties from the fitted line centroids. The SFRs are observed, dust-uncorrected line-based estimates. H$\alpha$ SFRs use the Chabrier-scaled calibration of \citet{kennicutt98}. For the \oiii-selected sources, SFRs are estimated assuming \oiii$\lambda5008$/H$\beta=5$ and Case B H$\alpha$/H$\beta=2.86$ \citep{shapley23,nakajima23}. These SFRs should be regarded as lower limits. The ALMA columns give the observed-frame continuum wavelength $\lambda_{\rm obs}$ and flux density $S_\nu$. Stellar masses and infrared luminosities are Bayesian means from energy-balanced \textsc{CIGALE} SED fitting. $L_{\rm IR}$ includes only the host-galaxy dust emission integrated over rest-frame 8--1000~$\mu$m and excludes the separately modelled Fritz AGN component described in Section~\ref{sec:dsfg_miri}.}
\label{tab:dsfg18}
\scriptsize
\setlength{\tabcolsep}{1.5pt}
\renewcommand{\arraystretch}{1.08}
\begin{tabular*}{\textwidth}{@{\extracolsep{\fill}}l c c c c c c c c c c c l@{}}
\hline
ID & RA & Dec & Line & $z_{\rm line}$ & Flux & FWHM\textsuperscript{c} &
$\log M_\star$ & SFR$_{\rm obs}$ & $\lambda_{\rm obs}$ &
$S_{\nu}^{\rm obs}$ &
$\log L_{\rm IR}$ & Sample \\
 & (deg) & (deg) & & & ($10^{-18}$ cgs) & (km s$^{-1}$) &
($M_\odot$) & ($M_\odot$ yr$^{-1}$) & ($\mu$m) & (mJy) &
($L_\odot$) & \\
\hline
386433 & 149.9119 & 2.5069 & O3 & $7.197 \pm 0.0004$ & $25.24 \pm 1.52$ & $406 \pm 31$ & $9.56 \pm 0.23$ & $50 \pm 3$ & 1234 & $0.55 \pm 0.18$ & $11.25 \pm 0.27$ & CH \\
671610 & 150.1257 & 2.2666 & O3 & $6.852 \pm 0.0003$ & $65.33 \pm 1.97$ & $665 \pm 25$ & $10.02 \pm 0.05$ & $116 \pm 4$ & 705 & $0.36 \pm 0.09$ & $11.42 \pm 0.35$ & P8/A3 \\
88087\textsuperscript{a} & 150.0492 & 2.1066 & \ha & $5.968 \pm 0.0004$ & $8.03 \pm 1.02$ & $257 \pm 36$ & $11.16 \pm 0.02$ & $18 \pm 2$ & 1234 & $0.87 \pm 0.13$ & $12.41 \pm 0.02$ & CH \\
14437\textsuperscript{a} & 149.8566 & 2.2068 & \ha & $5.757 \pm 0.0001$ & $10.35 \pm 4.71^{\rm br}$ & $591 \pm 142^{\rm br}$ & $10.53 \pm 0.02$ & $68 \pm 14$ & 1234 & $0.64 \pm 0.14$ & $11.74 \pm 0.33$ & CH \\
 & & & & & $22.37 \pm 4.90^{\rm nr}$ & $287 \pm 29^{\rm nr}$ & & & & & & \\
46224\textsuperscript{a} & 149.9719 & 2.1182 & \ha & $5.689 \pm 0.0002$ & $40.99 \pm 7.20^{\rm br}$ & $1151 \pm 101^{\rm br}$ & $10.18 \pm 0.10$ & $143 \pm 21$ & 1055 & $0.11 \pm 0.02$ & $11.72 \pm 0.18$ & CRISTAL-03 \\
 & & & & & $29.70 \pm 7.35^{\rm nr}$ & $409 \pm 55^{\rm nr}$ & & & & & & \\
465395 & 150.1630 & 2.4256 & \ha & $5.636 \pm 0.0009$ & $3.67 \pm 0.81$ & $408 \pm 98$ & $9.10 \pm 0.03$ & $7 \pm 2$ & 1047 & $<0.10$ & $10.65 \pm 0.03$ & CRISTAL-17 \\
131471\textsuperscript{a} & 150.2062 & 2.0718 & \ha & $5.613 \pm 0.0004$ & $11.72 \pm 1.07$ & $461 \pm 47$ & $11.28 \pm 0.05$ & $23 \pm 2$ & 1234 & $0.99 \pm 0.16$ & $12.79 \pm 0.02$ & CH \\
41023 & 149.9932 & 2.0759 & \ha & $5.579 \pm 0.0002$ & $25.81 \pm 2.49^{\rm br}$ & $857 \pm 154^{\rm br}$ & $10.46 \pm 0.02$ & $64 \pm 6$ & 1249 & $1.12 \pm 0.29$ & $11.62 \pm 0.03$ & A3 \\
 & & & & & $7.38 \pm 1.86^{\rm nr}$ & $192 \pm 39^{\rm nr}$ & & & & & & \\
210257 & 149.7537 & 2.0910 & \ha & $5.576 \pm 0.0002$ & $35.96 \pm 1.09$ & $487 \pm 17$ & $9.72 \pm 0.06$ & $70 \pm 2$ & 1038 & $0.20 \pm 0.04$ & $11.24 \pm 0.04$ & CRISTAL-09 \\
635130 & 150.0393 & 2.3372 & \ha & $5.541 \pm 0.0008$ & $5.08 \pm 0.95$ & $400 \pm 82$ & $9.70 \pm 0.26$ & $10 \pm 2$ & 1032 & $0.36 \pm 0.08$ & $10.97 \pm 0.28$ & CRISTAL-05 \\
291002\textsuperscript{b} & 150.1156 & 1.9601 & \ha & $5.310 \pm 0.0003$ & $12.52 \pm 0.77$ & $415 \pm 36$ & $9.43 \pm 0.17$ & $21 \pm 1$ & 3325 & $0.31 \pm 0.11$ & $10.78 \pm 0.33$ & A3 \\
43542\textsuperscript{b} & 149.8769 & 2.1341 & \ha & $5.254 \pm 0.0005$ & $7.09 \pm 0.91$ & $383 \pm 54$ & $10.03 \pm 0.20$ & $12 \pm 2$ & 987 & $0.60 \pm 0.14$ & $11.61 \pm 0.25$ & CRISTAL-21 \\
79153\textsuperscript{b} & 150.0213 & 2.0534 & \ha & $5.234 \pm 0.0001$ & $17.16 \pm 0.55$ & $399 \pm 14$ & $9.71 \pm 0.11$ & $29 \pm 1$ & 984 & $0.22 \pm 0.03$ & $10.93 \pm 0.07$ & CRISTAL-19 \\
296469 & 150.0671 & 2.0184 & \ha & $5.231 \pm 0.0007$ & $22.03 \pm 3.70^{\rm br}$ & $1793 \pm 321^{\rm br}$ & $10.37 \pm 0.09$ & $50 \pm 8$ & 1035 & $0.35 \pm 0.13$ & $11.65 \pm 0.03$ & A3 \\
 & & & & & $8.05 \pm 3.08^{\rm nr}$ & $492 \pm 122^{\rm nr}$ & & & & & & \\
593386\textsuperscript{a} & 149.9180 & 2.3921 & \ha & $5.204 \pm 0.0004$ & $17.99 \pm 2.58^{\rm br}$ & $870 \pm 128^{\rm br}$ & $10.48 \pm 0.10$ & $34 \pm 6$ & 1234 & $0.74 \pm 0.13$ & $11.78 \pm 0.07$ & CH \\
 & & & & & $2.70 \pm 2.25^{\rm nr}$ & $237 \pm 116^{\rm nr}$ & & & & & & \\
215767 & 149.7803 & 2.1226 & \ha & $5.182 \pm 0.0005$ & $18.48 \pm 8.62^{\rm br}$ & $1499 \pm 515^{\rm br}$ & $9.73 \pm 0.02$ & $57 \pm 19$ & 953 & $0.19 \pm 0.05$ & $11.28 \pm 0.02$ & ALP/A3 \\
 & & & & & $16.64 \pm 7.94^{\rm nr}$ & $638 \pm 338^{\rm nr}$ & & & & & & \\
94974\textsuperscript{b} & 150.0563 & 2.1537 & \ha & $5.076 \pm 0.0003$ & $6.86 \pm 0.75$ & $256 \pm 32$ & $10.54 \pm 0.12$ & $11 \pm 1$ & 1234 & $0.46 \pm 0.09$ & $12.21 \pm 0.16$ & CH \\
55193 & 149.9148 & 2.2023 & \ha & $4.933 \pm 0.0004$ & $17.06 \pm 1.64$ & $394 \pm 44$ & $10.54 \pm 0.06$ & $25 \pm 2$ & 873 & $0.30 \pm 0.08$ & $12.06 \pm 0.09$ & A3 \\
\hline
\multicolumn{13}{l}{\footnotesize CH and A3 denote \textsc{CHAMPS} and \textsc{A3COSMOS}. P8/A3 denotes \textsc{REBELS}-P8+\textsc{A3COSMOS}.}\\
\multicolumn{13}{l}{\footnotesize ALP/A3 denotes \textsc{ALPINE}+\textsc{A3COSMOS}. CRISTAL sources are identified by their survey IDs. O3 denotes \oiii$\lambda5008$.}\\
\multicolumn{13}{l}{\footnotesize \textsuperscript{a}Robust MIRI F1000W and/or F2100W detection ($>3\sigma$).}\\
\multicolumn{13}{l}{\footnotesize \textsuperscript{b}Merger or multicomponent morphology flag used in the environment analysis.}\\
\multicolumn{13}{l}{\footnotesize \textsuperscript{c}The labels ``br'' and ``nr'' denote the broader and narrower Gaussian components in two-component fits, respectively.}\\
\end{tabular*}
\end{table*}

\subsection{\oiii~and \ha~emitter search}
The \oiii~line search and the grizli-based catalog are presented in
\citet{meyer25b}. Here we describe the search for \ha~emitters in the
JWST/NIRCam F444W grism data. We first constructed a photometric parent
sample from the COSMOS-Web catalog\citep{shuntov25}, selecting sources with
$m_{\rm F444W}<28$ whose photometric-redshift solutions overlap the
redshift range over which \ha~falls within the F444W grism coverage,
$z\simeq4.8$--6.9. Specifically, we required the 16--84 percentile
photo-$z$ interval to overlap the F444W \ha~redshift window. This loose,
uncertainty-aware preselection was designed to retain sources whose
best-fit photo-$z$ may be offset from the grism redshift, while rejecting
obvious low-redshift contaminants before the spectral search.

For each source in the parent sample, we searched the extracted
\texttt{unfold} 1D spectrum for emission-line candidates after
subtracting a local running-median continuum. Candidate peaks were
initially identified from the continuum-subtracted spectrum using the
local flux uncertainty, and were then refit with Gaussian line templates
over a narrow wavelength window centered on the candidate feature. The
templates included single \ha~and the common neighboring line complexes
\ha+[N\,\textsc{ii}], \ha+[S\,\textsc{ii}], and
\ha+[N\,\textsc{ii}]+[S\,\textsc{ii}]. The \ha~component was treated as the primary line and was required to satisfy ${\rm S/N}_{\rm 1D}>5$. We define
${\rm S/N}_{\rm 1D}=f_{\rm peak}/\sigma_{\rm local}$, where
$f_{\rm peak}$ is the maximum continuum-subtracted flux density within a
seven-pixel window centered on the candidate peak, and
$\sigma_{\rm local}$ is the median $1\sigma$ per-pixel flux-density
uncertainty over the same window. For multi-line templates, this S/N is
defined for the \ha~line only. We also required a corresponding 2D line detection with ${\rm S/N}_{\rm 2D}>3$, measured at the expected wavelength position in
the continuum-subtracted 2D grism spectrum. Here the signal is the summed continuum-subtracted flux within the spatial-spectral aperture enclosing the line trace in the grism 2D spectra, and
$N_{\rm 2D}=(\sum_i \sigma_i^2)^{1/2}$ is computed over the same pixels
from the inverse-variance map, with $\sigma_i={\rm IVAR}_i^{-1/2}$. The ${\rm IVAR}_i$ is the inverse variance. The [N\,\textsc{ii}] and
[S\,\textsc{ii}] lines were used only as supporting evidence when they
fell in the wavelength coverage and were not required to be individually
significant. Thus, many accepted sources are effectively
single-line \ha~detections, while robust multi-line systems such as
\ha+[S\,\textsc{ii}] or \ha+[N\,\textsc{ii}] provide additional support
for the redshift assignment. Candidates were ranked using the \ha~line
significance, the 2D line significance, the reduced $\chi^2$ of the
Gaussian fit, the fitted line width, the local dominance of the candidate
peak over nearby noise spikes, and the consistency of any in-band
companion lines. Known contaminants were removed before visual inspection, including sources already identified as [O\,\textsc{iii}] emitters in the final
C3D \oiii~catalog of \citet{meyer25b}, low-redshift sources with spectroscopic
redshifts \citep{cweb_specz}, and broad Pa$\alpha$ and Pa$\beta$ emitters
from \citet{jiang26}.

All remaining candidates were visually inspected using the 1D and 2D
spectra. Each candidate was inspected independently by three
assessors. In the visual-inspection stage, we assigned a score of 1 to
marginal or uncertain candidates, a score of 2 to likely \ha~emitters,
and a score of 3 to secure \ha~emitters. The default final score
was taken to be the median of the three scores. Candidates with
discrepant classifications, especially cases spanning both rejection and
secure acceptance, were re-examined jointly and assigned a consensus
score; when no clear consensus was reached, the more conservative lower
score was adopted. The final catalog was constructed by retaining only
candidates with final visual scores of 2 or 3. We further
removed sources with failed or nonphysical 1D Gaussian fits, including
cases with unstable continuum estimates, nonpositive fitted fluxes or
flux uncertainties, or fitted line centers inconsistent with the observed
emission peak. The resulting visually accepted \ha~catalog contains 1210 entries spanning $z=4.91$--$6.61$. After removing broad-line sources with ${\rm FWHM}_{\rm H\alpha}>800~{\rm km~s^{-1}}$, the parent narrow-line HAE sample contains 1158 galaxies. This sample has a median observed, dust-uncorrected ${\rm SFR}_{\rm H\alpha}=11.7~M_\odot~{\rm yr^{-1}}$, a median dust-corrected ${\rm SFR}_{\rm H\alpha}=17.7~M_\odot~{\rm yr^{-1}}$, and a median stellar mass of $\log(M_\star/M_\odot)=9.06$ from the $\log(M_\star/M_\odot)$ values in COSMOS2025 catalog.


\section{Analysis}\label{sec:analysis}

\subsection{\ha~luminosity function}\label{sec:ha_lf}

\begin{figure}
\centering
    \includegraphics[width=0.49\textwidth]{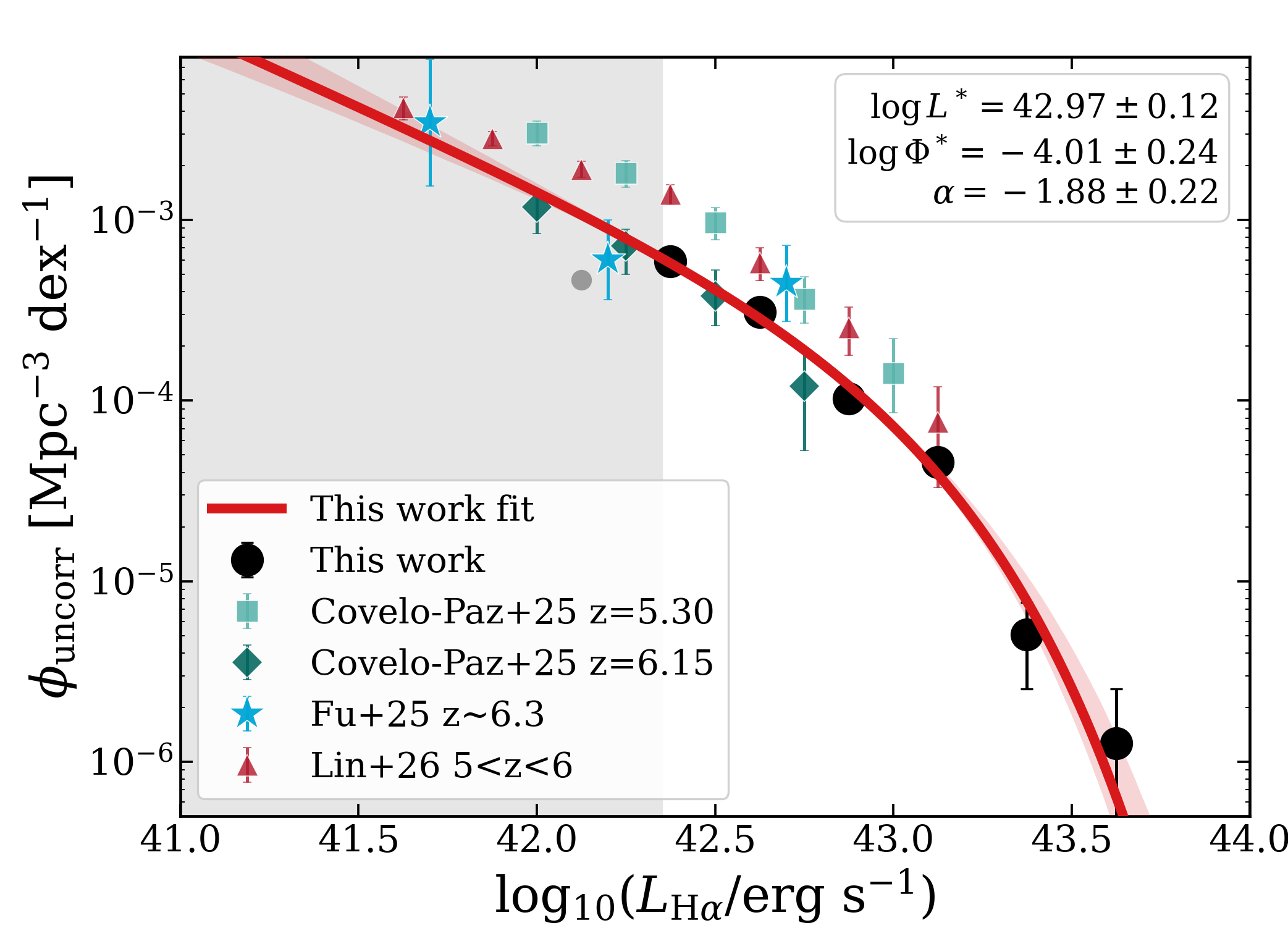}
    \caption{\small Observed H$\alpha$ luminosity function from the broad-line \ha-removed catalog, without completeness correction. Black points show the LF bins used in the Schechter-function fit. Gray points mark bins below the adopted detection limit. This observed LF is used to estimate the expected blank-field number of H$\alpha$ emitters in the DSFG environment measurements. We overplot observed H$\alpha$ LFs at similar redshifts from \citet{paz25} at $z\simeq5.3$ and $z\simeq6.2$, \citet{fu25} at $z\simeq6.3$, and \citet{lin26} at $5<z<6$.}
\label{fig:ha_lf}
\end{figure}

We derive the H$\alpha$ luminosity function (LF) to estimate the blank-field expectation for the spectroscopic overdensity calculation in Section~\ref{sec:spec_delta}. Since the observed neighbor counts used for the overdensity measurement are not completeness-corrected, we use an LF without completeness correction for the field expectation. This keeps $N_{\rm obs}$ and $N_{\rm exp}$ on the same selection scale. The LF is derived from the 1210 H$\alpha$ emitters with valid fitted fluxes, FWHM values, and redshifts. Following \citet{paz25}, we retain narrow-line emitters by requiring ${\rm FWHM}_{\rm H\alpha}<800~{\rm km~s^{-1}}$, leaving 1158 galaxies. This narrow-line sample has a median observed, dust-uncorrected ${\rm SFR}_{\rm H\alpha}=11.7~M_\odot~{\rm yr^{-1}}$ (16th--84th percentile: 6.4--24.4), a median dust-corrected ${\rm SFR}_{\rm H\alpha}=17.7~M_\odot~{\rm yr^{-1}}$ (9.3--40.3), and a median CWEB stellar mass of $\log(M_\star/M_\odot)=9.06$ (8.59--9.60). H$\alpha$ luminosities were computed from the fitted integrated line fluxes as
$L_{\rm H\alpha}=4\pi D_L^2 f_{\rm H\alpha}$,
using the measured line redshift $z_{\rm H\alpha}$. We adopted the median $5\sigma$ line-flux limit of the parent H$\alpha$ catalog,
$f_{\rm lim}=4.42\times10^{-18}~{\rm erg~s^{-1}~cm^{-2}}$,
derived from the fitted line-flux uncertainties. The luminosity function was then calculated over $4.79<z_{\rm H\alpha}<6.62$ using a $1/V_{\max}$ estimator, with $V_{\max}$ determined from this median flux limit. The Schechter function was fit to bins with $\log_{10}(L_{\rm H\alpha}/{\rm erg~s^{-1}})\geq42.35$, giving $\log_{10}(L^\star/{\rm erg~s^{-1}})=42.97\pm0.12$, $\log_{10}(\Phi^\star/{\rm Mpc^{-3}})=-4.01\pm0.24$, and $\alpha=-1.88\pm0.22$. We plot our LF in Figure~\ref{fig:ha_lf}, together with observed, completeness-uncorrected LF measurements from other JWST surveys at similar redshifts \citep{paz25,fu25,lin26}.

\subsection{DSFGs with \ha~or \oiii~emission line}
We cross-matched the DSFG parent sample with the C3D \ha~and \oiii~emitter catalogs using both positional and redshift information. We first searched for C3D line emitters within 0\farcs6 of the ALMA or parent-catalog position. For sources with an ALMA \cii~redshift or a published spectroscopic redshift, we required the C3D grism redshift to be consistent within $|\Delta v|<500~{\rm km~s^{-1}}$. For continuum-selected sources without an ALMA line redshift, the C3D grism redshift was used as the systemic redshift after visual confirmation of the counterpart. After cross-matching the \oiii~and \ha~emitter catalogs with our DSFG parent sample, we analyze a final sample of 18 DSFGs or DSFG candidates at $z=4.93$--7.20, including nine sources with newly determined C3D grism redshifts. Details of all sources, including the line fluxes and FWHM measured from the grism spectra, are presented in Table~\ref{tab:dsfg18}. Note that the listed dust-continuum flux densities are observed-frame ALMA measurements taken in different bands: CHAMPS is observed in Band~6 ($\sim1.1$~mm), ALPINE/CRISTAL is mainly in Band~7, A3COSMOS uses source-dependent archival ALMA bands, and REBELS is observed in Bands~5/6. These comprise six ALPINE--CRISTAL sources (\texttt{ID 79153, 43542, 635130, 210257, 465395, 46224}; \citealp{cristal25}), five C3D \ha+CHAMPS sources (\texttt{ID 14437, 88087, 94974, 131471, 593386}; CHAMPS, Faisst et al., in prep.), four A3COSMOS-only sources (\texttt{ID 41023, 291002, 296469, 55193}), one ALPINE+A3COSMOS source (\texttt{ID 215767}; \citealp{fevre20,faisst20,bet20}), one REBELS-P8/A3COSMOS source (\texttt{ID 671610}; \citealp{bouwens22,inami22}), and one C3D \oiii+CHAMPS source (\texttt{ID 386433}; \citealp{zavala26}).  

\subsection{Stellar masses, star formation, and dust obscuration}
\label{sec:dsfg_properties}

\begin{figure*}
\centering
    \includegraphics[width=0.49\textwidth]{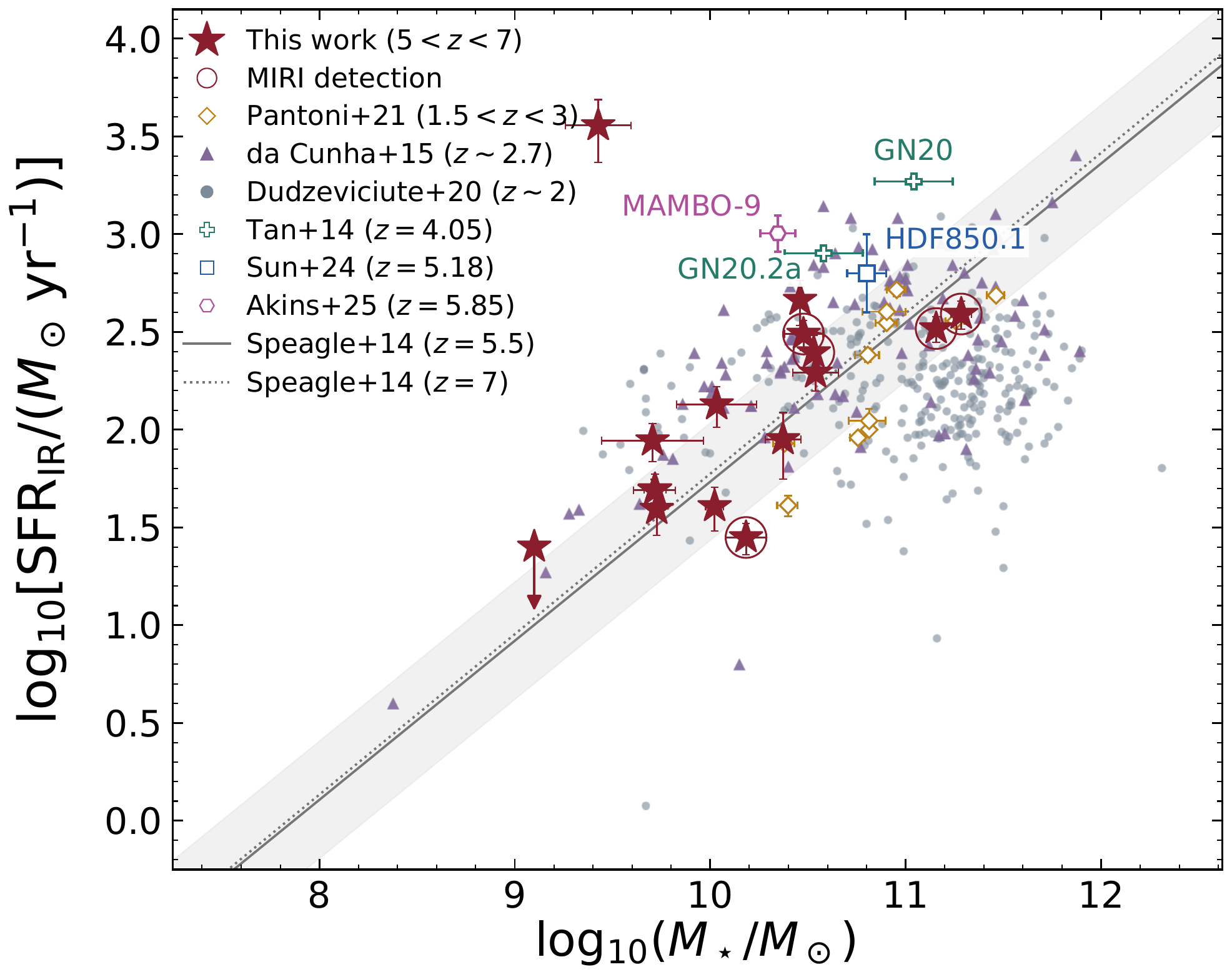}
    \includegraphics[width=0.49\textwidth]{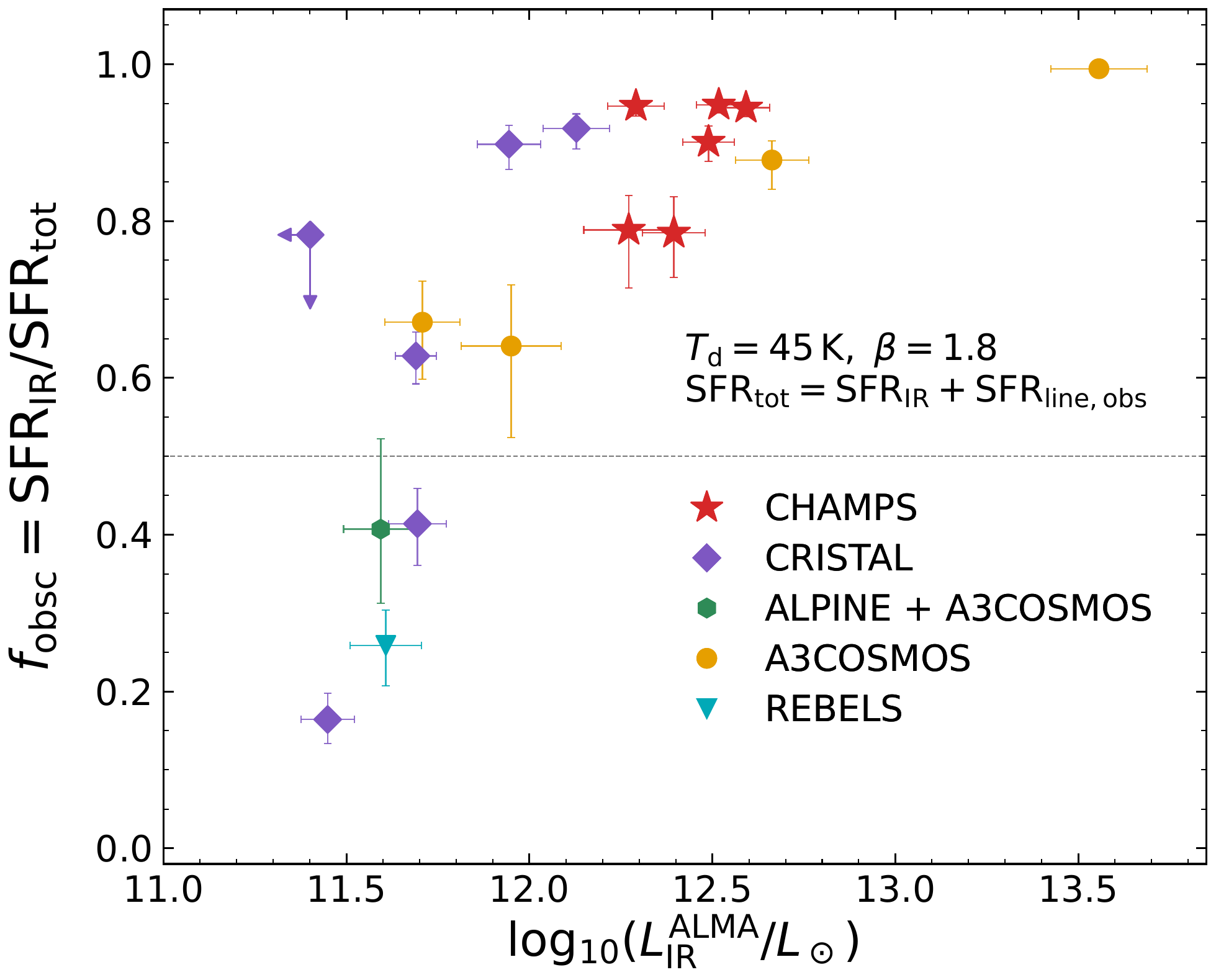}
    \caption{\small $Left.$ ALMA-inferred infrared SFR versus stellar mass for the 18 C3D DSFGs, compared with SMG/DSFG samples from \citet{pantoni21,dacunha15,dudzeviciute20} and $z>4$ systems from \citet{tan14,sun24,akins26}. Solid and dotted curves show the total-SFR main-sequence relation of \citet{speagle14} extrapolated to $z=5.5$ and $z=7$, respectively. Our plotted SFRs include only the obscured contribution. $Right.$ Estimated obscured star-formation fraction, $f_{\rm obsc}={\rm SFR}_{\rm IR}/({\rm SFR}_{\rm IR}+{\rm SFR}_{\rm obs})$, versus ALMA-inferred $L_{\rm IR}$, where ${\rm SFR}_{\rm obs}$ is the dust-uncorrected line-based SFR. The ALMA-based estimates assume $T_{\rm d}=45$ K and $\beta=1.8$. }
\label{fig:dsfg_properties}
\end{figure*}

We characterize the 18 DSFGs using their stellar masses, infrared SFRs, and obscured star-formation fractions. Stellar masses and host-galaxy infrared luminosities are estimated by fitting the UV-to-FIR SEDs with \textsc{CIGALE} \citep{boquien19} at fixed spectroscopic redshifts. The fiducial model adopts a delayed SFH, Bruzual \& Charlot stellar populations \citep{bc03}, a Chabrier IMF \citep{chabrier03}, a fixed metallicity of $Z=0.008$, nebular emission, a modified Calzetti attenuation law \citep{calzetti00}, the dust templates of \citet{draine14}, and the AGN torus library of \citet{fritz06}. The fitted photometry includes the C3D JWST photometry described in Section~\ref{sec:photometry}, additional multi-band photometry from COSMOS2025 \citep{shuntov25}, and ALMA continuum constraints. We adopt the Bayesian means, with $L_{\rm IR}$ integrated over rest-frame 8--1000~$\mu$m and excluding the AGN contribution. The sample spans $\log(M_\star/M_\odot)\simeq9.1$--11.3 and $\log(L_{\rm IR}/L_\odot)\simeq10.7$--12.8 (Table~\ref{tab:dsfg18}). Figure~\ref{fig:dsfg_properties} compares the ALMA-inferred ${\rm SFR}_{\rm IR}$ and stellar masses with the GOODS-S DSFGs of \citet{pantoni21}, ALESS SMGs \citep{dacunha15}, AS2UDS SMGs \citep{dudzeviciute20}, and $z\sim$ 5 systems GN20 and its companions \citep{tan14}, HDF850.1 \citep{sun24}, and MAMBO-9 \citep{akins26}. These samples provide context across different redshifts and selection methods rather than a matched control population. We also show the total-SFR main-sequence relation of \citet{speagle14}, extrapolated to $z=5.5$ and $z=7$. Because ${\rm SFR}_{\rm IR}$ excludes unobscured star formation, a position below these curves does not necessarily imply a total SFR below the main sequence. We quantify obscuration using $f_{\rm obsc}={\rm SFR}_{\rm IR}/{\rm SFR}_{\rm tot}$ (Figure \ref{fig:dsfg_properties} right panel), with ${\rm SFR}_{\rm tot}={\rm SFR}_{\rm IR}+{\rm SFR}_{\rm UV,obs}$ \citep{algera23}. The observed host-galaxy UV contribution is not corrected for internal attenuation, avoiding double counting with the IR contribution. Values of $f_{\rm obsc}>0.5$ indicate predominantly obscured star formation, but this threshold is not imposed on the sample. Together, these diagnostics distinguish stellar mass, obscured activity, and the relative obscured contribution within our heterogeneous sample before comparing its environments.

\subsection{Hot-dust and obscured-AGN candidates among the five MIRI-detected DSFGs}\label{sec:dsfg_miri}

The five MIRI-detected DSFGs are \texttt{ID 46224, 14437, 88087, 131471,} and \texttt{593386}. All five sources are detected at $>3\sigma$ in F2100W. Four are also detected at $>3\sigma$ in F1000W, while \texttt{ID 46224} lacks F1000W coverage. Because the C3D MIRI imaging was obtained in parallel, the MIRI footprint and depth are not uniform across the full DSFG sample. We therefore define MIRI-bright sources only among objects covered by the relevant MIRI imaging and detected at $>3\sigma$, and interpret MIRI-detected versus non-detected comparisons as applying only to the covered subset.

For the five MIRI-detected DSFGs SED fittting with CIGALE, we tested six variants with delayed or delayed+burst SFHs, Chabrier or Salpeter IMFs, and with or without the AGN component. These tests give typical systematic uncertainties of $\sim0.1$--0.2 dex in $\log M_\star$ and SFR. The derived quantities are summarized in Table~\ref{tab:phot_sfr_pah_summary} and Appendix Figure \ref{fig:sed_miri}.
 
At $z\simeq5.2$--6.0, F2100W probes rest-frame $\lambda_{\rm rest}\simeq3.0$--$3.4\,\mu{\rm m}$, where the 3.3~$\mu$m PAH feature may contribute \citep{kim12,shipley16,lyu24}. We do not use this feature as a robust SFR measurement, because the current MIRI data lack a redward band to define the continuum baseline and the 3.3~$\mu$m PAH feature is an uncertain SFR indicator and exhibits substantial scatter \citep{kim12,gregg24}. Instead, we use it only as a consistency test. We define ${\rm SFR}_{\rm PAH,req}$ as the equivalent SFR required if the full F2100W excess were powered by 3.3~$\mu$m PAH emission, and
\[
f_{\rm PAH,max}=
\min\left(1,\frac{{\rm SFR}_{\rm IR}}{{\rm SFR}_{\rm PAH,req}}\right),
\]
where ${\rm SFR}_{\rm IR}$ is inferred from the ALMA dust-continuum measurement using an optically thin modified-blackbody model with the fiducial parameters $T_{\rm d}=45$ K and $\beta=1.8$. The model includes CMB heating and the CMB background following \citet{cunha13}. Thus, $f_{\rm PAH,max}$ is the maximum fraction of the F2100W excess that can be attributed to PAH emission without exceeding the ALMA-inferred obscured SFR.

For \texttt{ID 46224, 14437, 88087,} and \texttt{131471}, a PAH-only explanation would require ${\rm SFR}_{\rm PAH,req}\simeq2.1\times10^3$, $2.4\times10^3$, $3.3\times10^3$, and $9.3\times10^3~M_\odot\,{\rm yr^{-1}}$, respectively. The corresponding allowed PAH fractions are only $f_{\rm PAH,max}\simeq0.008$, 0.04, 0.04, and 0.016. PAH emission is therefore unlikely to dominate the F2100W excess in these four sources, favoring an additional hot-dust continuum and making them hot-dust or obscured-AGN candidates \citep{donley12,imanishi08,yang23,garciabernete22}. For \texttt{ID 593386}, $f_{\rm PAH,max}=0.28\pm0.21$, so PAH or warm dust may contribute more substantially; its broad H$\alpha$ component nevertheless supports a broad-line AGN candidate.

Two MIRI-detected DSFGs, \texttt{ID 14437} and \texttt{ID 131471}, are detected in the VLA 3~GHz catalog, with $S_{\rm 3GHz}=18.5\pm2.5\,\mu{\rm Jy}$ and $31.8\pm2.8\,\mu{\rm Jy}$, respectively \citep{smolcic17}. Using the ALMA-based $L_{\rm IR}$ and assuming $S_\nu\propto\nu^{-0.7}$ to convert the 3~GHz fluxes to rest-frame 1.4~GHz, we find $q_{\rm TIR}=1.18\pm0.11$ and $1.16\pm0.08$ for \texttt{ID 14437} and \texttt{ID 131471}, respectively. These values are $\sim0.8$--0.9 dex below the expected main-sequence star-forming value at $z\simeq5.6$--5.8 \citep{delhaize17}, indicating a radio excess. Together with their low $f_{\rm PAH,max}$ values and F2100W excesses, this supports their classification as hot-dust or obscured-AGN candidates, although compact obscured star formation and uncertainties in $L_{\rm IR}$, the radio spectral index, and the high-redshift IR--radio relation remain possible systematics \citep{delvecchio17}. \texttt{ID 593386} has a larger allowed PAH fraction, $f_{\rm PAH,max}=0.28\pm0.21$, so PAH or warm dust may contribute more strongly to its F2100W flux; its broad H$\alpha$ component nevertheless supports a less-obscured broad-line AGN candidate.

In Table~\ref{tab:phot_sfr_pah_summary}, we use ``Type-I candidate'' for sources with broad H$\alpha$ evidence and MIRI hot-dust excess. We use ``Type-1.x AGN candidate'' for sources with weaker or intermediate broad-line evidence plus additional AGN indicators, such as radio emission or a low allowed PAH fraction. Sources with strong F2100W excess inconsistent with PAH emission alone, but without a secure broad-line component, are classified as obscured-AGN candidates. Secure AGN confirmation would require additional diagnostics, such as deep X-ray detections, high-ionization emission lines, robust broad-line decomposition, full mid-infrared SED/template fitting, or a radio excess relative to the far-infrared--radio correlation.

\begin{figure*}
\centering
    \includegraphics[width=0.49\textwidth]{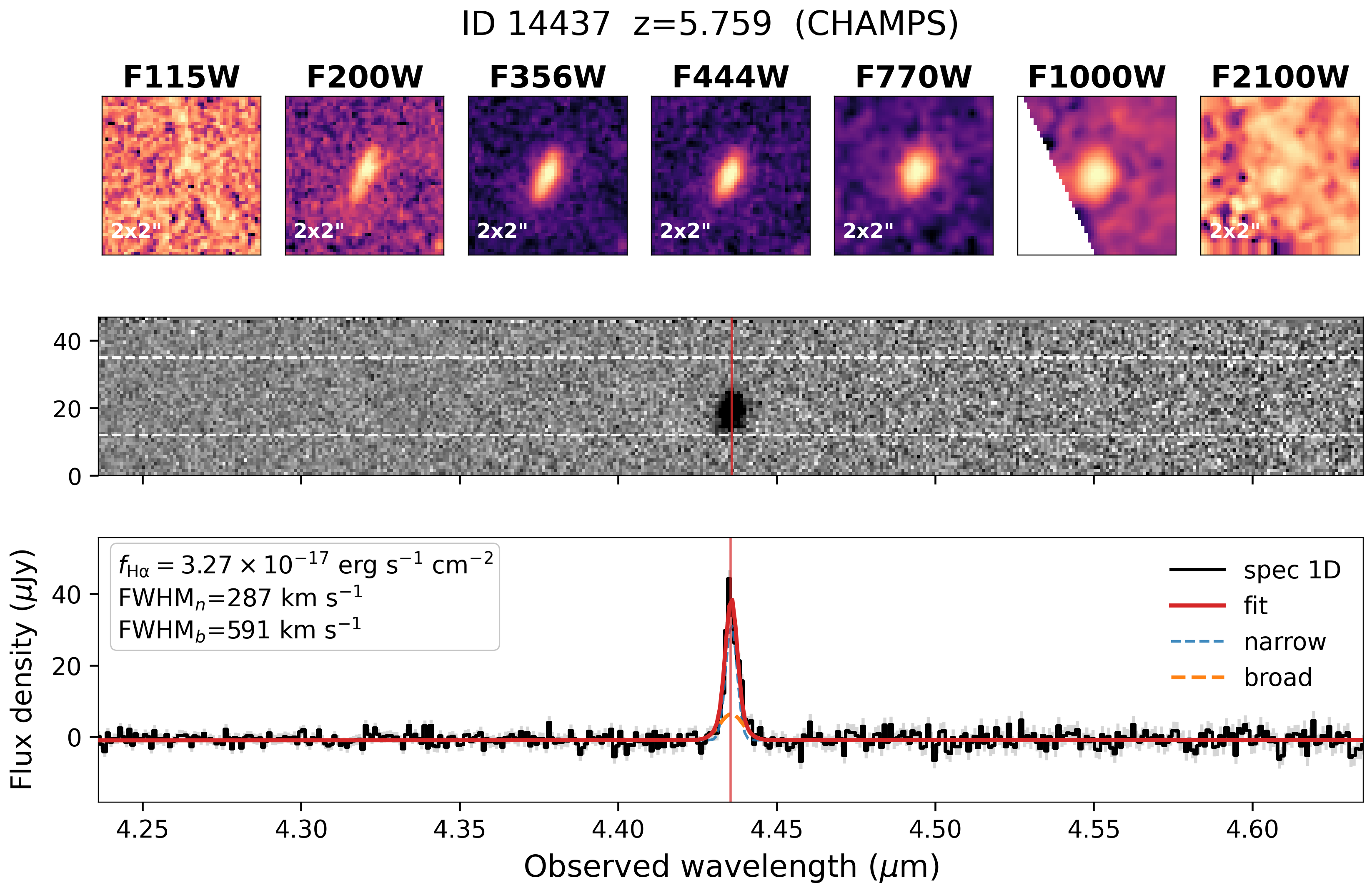}
    \hfill
    \includegraphics[width=0.49\textwidth]{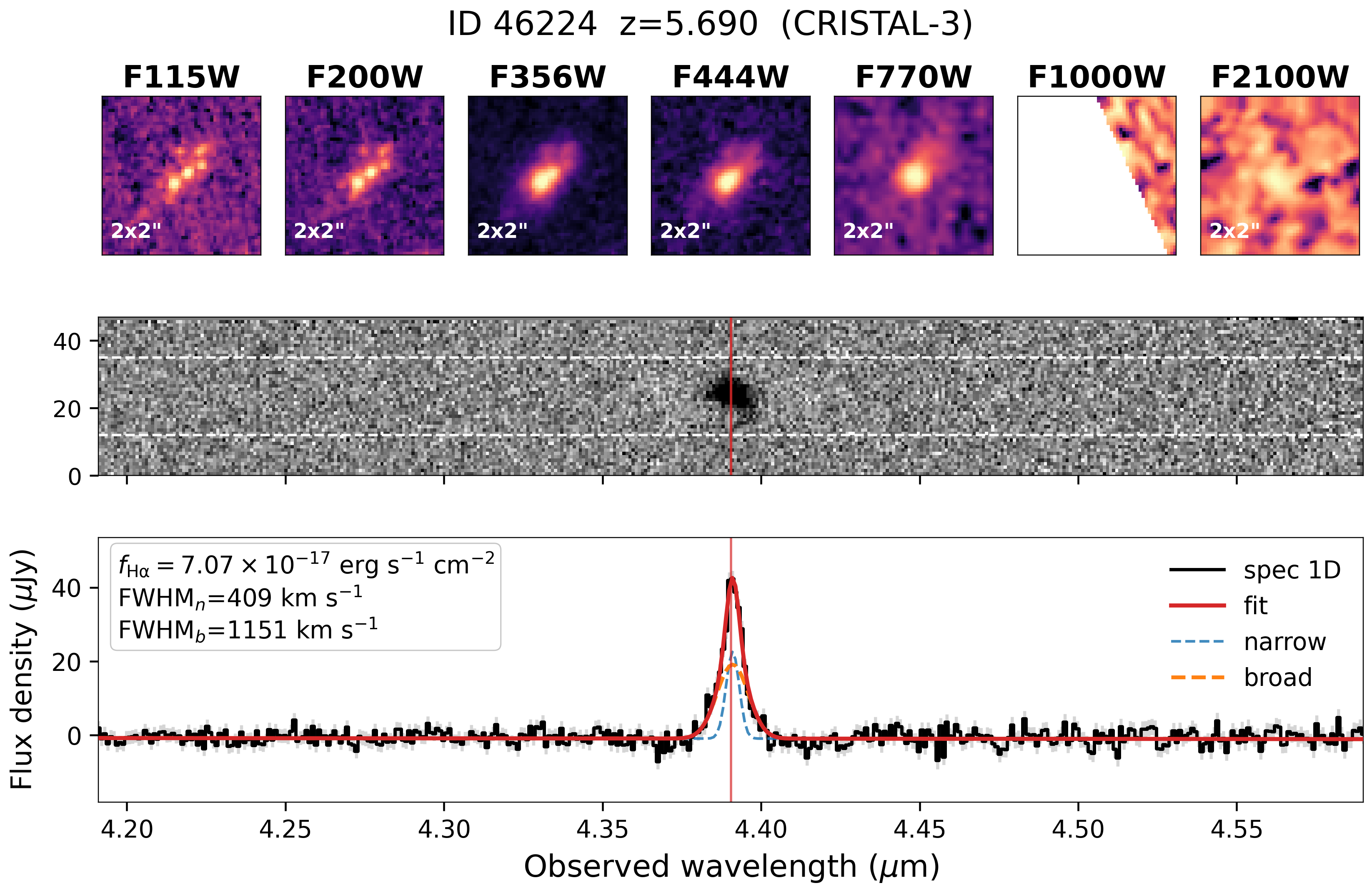}
    \includegraphics[width=0.49\textwidth]{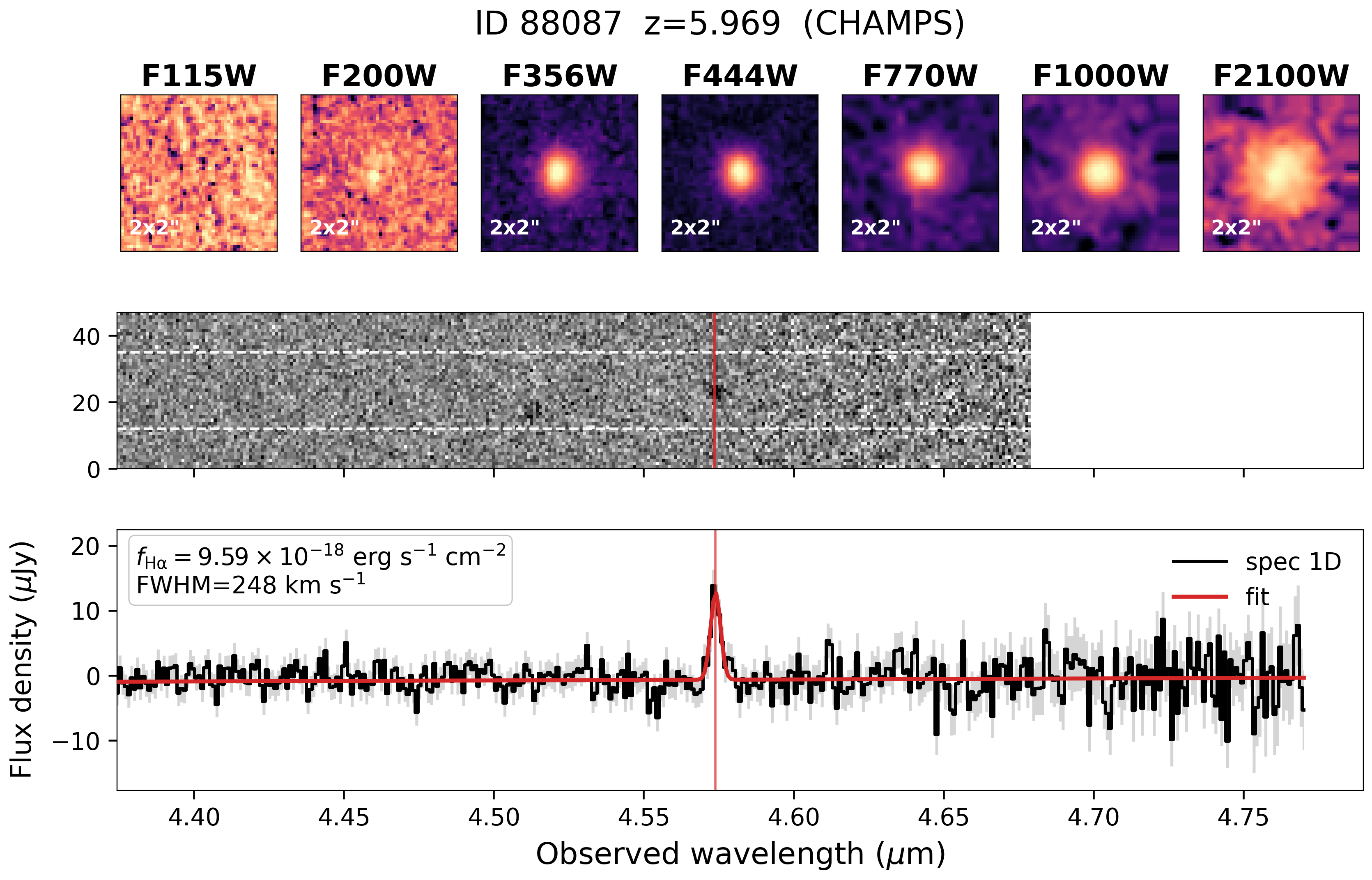}
    \hfill
    \includegraphics[width=0.49\textwidth]{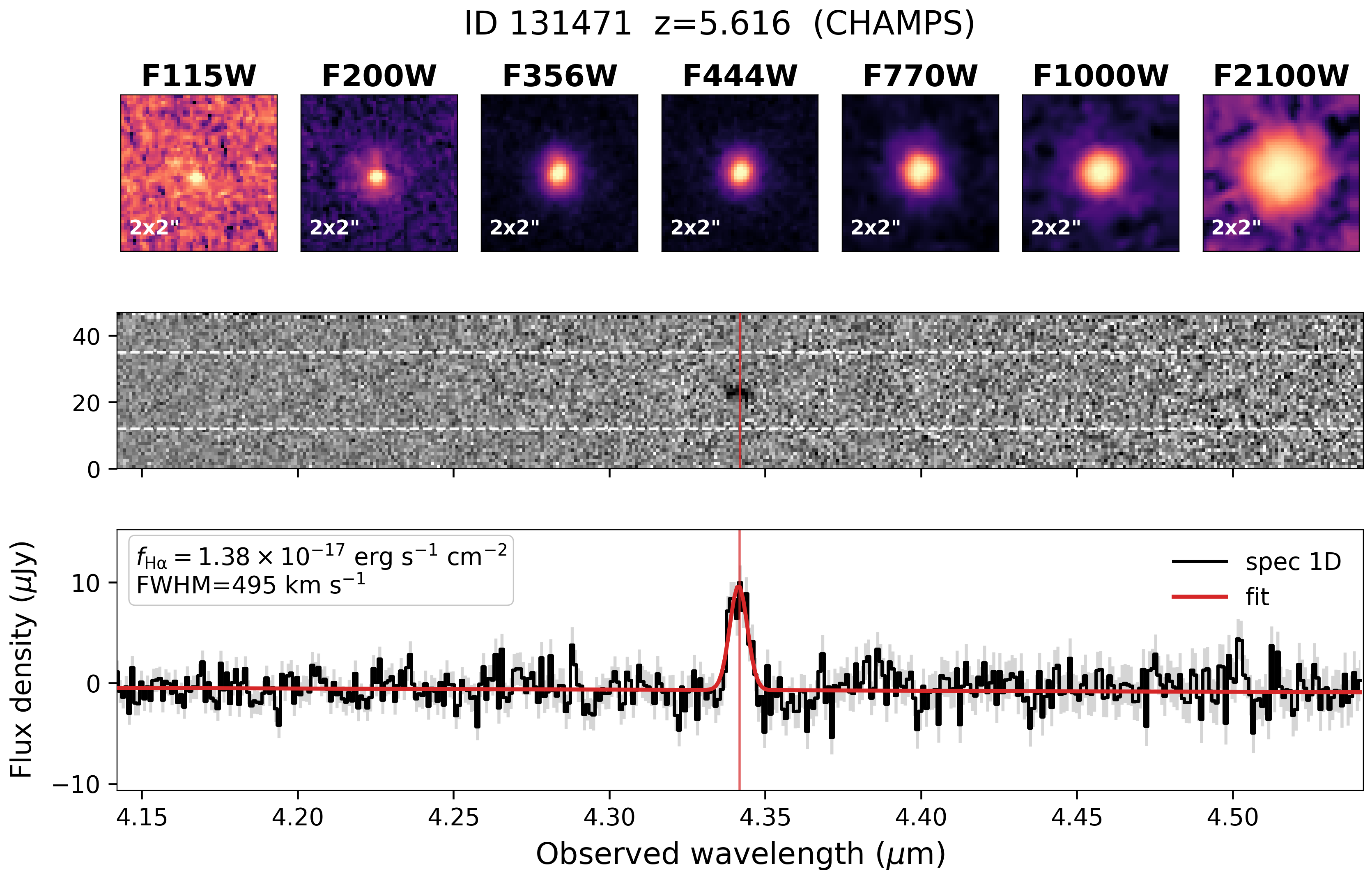}
    \includegraphics[width=0.49\textwidth]{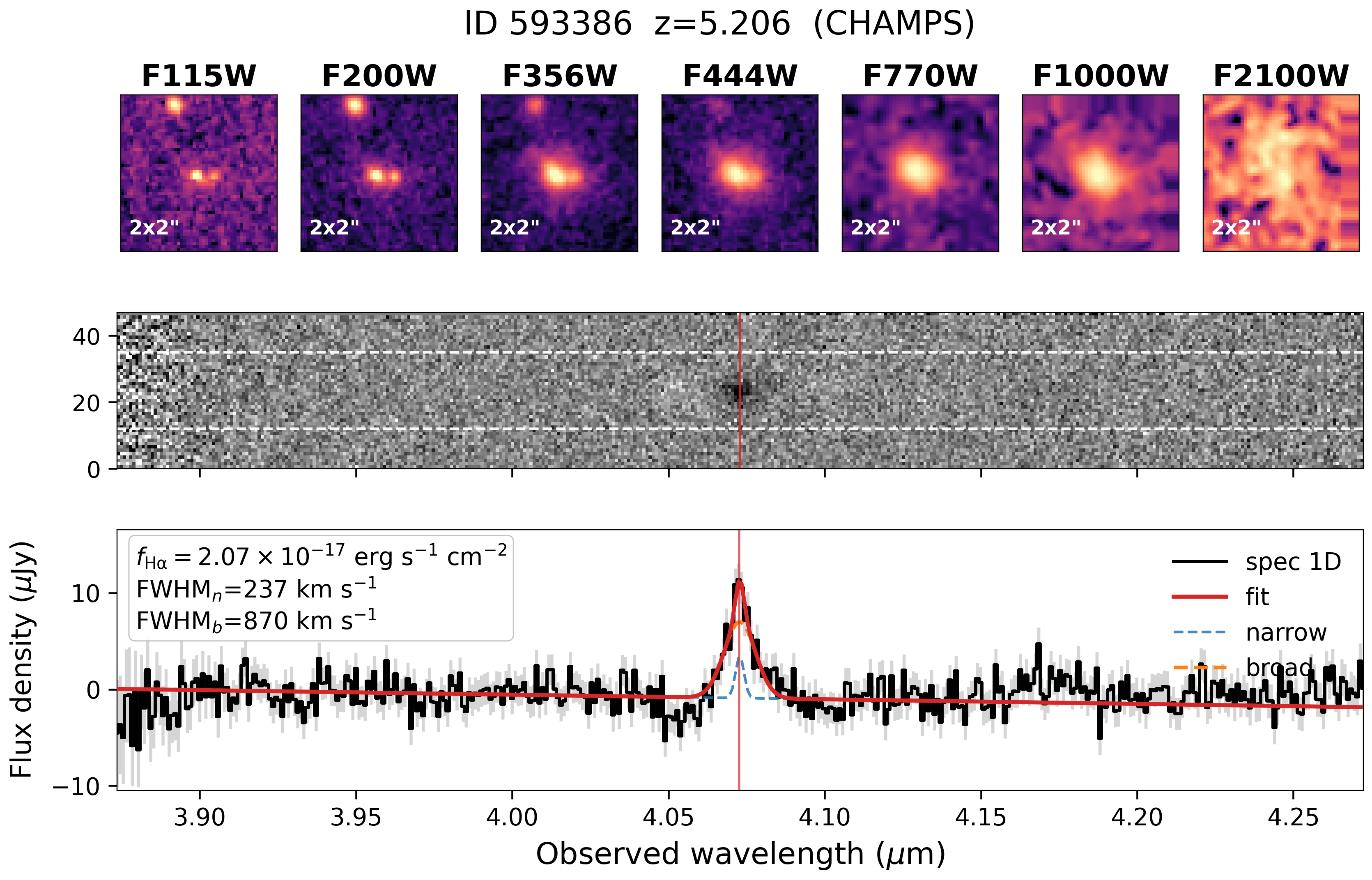}

    \caption{
    Multi-band JWST cutouts and H$\alpha$ spectra for the five DSFGs with C3D MIRI F2100W detections. Four also have robust F1000W detections.
    }
    \label{fig:five_miri_dsfg_cutouts}
\end{figure*}

\begin{table*}
\centering
\footnotesize
\caption{Photometry and derived properties for the five MIRI-detected DSFGs. SED-based quantities are from the fiducial \textsc{CIGALE} fits described in Section~\ref{sec:dsfg_miri}. Variant fits suggest typical systematic uncertainties of $\sim0.1$--0.2 dex in $\log M_\star$ and SFR; the quoted table uncertainties are formal fitting uncertainties only. ${\rm SFR}_{\rm PAH,req}$ is not a measured SFR, but the equivalent SFR required if the full F2100W excess were powered by 3.3~$\mu$m PAH emission. We define $f_{\rm PAH,max}=\min(1,{\rm SFR}_{\rm IR}/{\rm SFR}_{\rm PAH,req})$.}
\label{tab:phot_sfr_pah_summary}
\setlength{\tabcolsep}{2.8pt}
\renewcommand{\arraystretch}{1.08}

\begin{tabular}{l|cccc|cc|c|c|l}
\hline
Source ID & $f_{770}$ & $f_{1000}$ & $f_{2100}$ & $S_{\nu,{\rm 3GHz}}$ & SFR$_{\rm SED}$ & SFR$_{\rm IR}$ & $\log M_\star$ & $f_{\rm PAH,max}$ & Comment \\
 & ($\mu$Jy) & ($\mu$Jy) & ($\mu$Jy) & ($\mu$Jy) & ($M_\odot\,{\rm yr}^{-1}$) & ($M_\odot\,{\rm yr}^{-1}$) & ($M_\odot$) &  &  \\
\hline
46224  & $1.63 \pm 0.06$ & --- & $8.86 \pm 2.05$ & --- & $67 \pm 19$ & $17 \pm 3$   & $10.19 \pm 0.10$ & $0.008 \pm 0.003$ & Type-I candidate \\
14437  & $2.73 \pm 0.01$ & $1.70 \pm 0.39$ & $8.56 \pm 1.96$ & $18.5 \pm 2.5$ & $53 \pm 35$ & $96 \pm 21$  & $10.53 \pm 0.02$ & $0.04 \pm 0.02$ & Type-1.x AGN candidate \\
88087  & $5.09 \pm 0.02$ & $4.05 \pm 0.30$ & $14.30 \pm 1.02$ & --- & $202 \pm 10$ & $130 \pm 20$ & $11.16 \pm 0.02$ & $0.04 \pm 0.01$ & Obscured AGN candidate \\
131471 & $9.80 \pm 0.01$ & $9.73 \pm 0.38$ & $40.53 \pm 1.18$ & $31.8 \pm 2.8$ & $531 \pm 32$ & $148 \pm 24$ & $11.29 \pm 0.05$ & $0.016 \pm 0.003$ & Obscured AGN candidate \\
593386 & $2.72 \pm 0.03$ & $2.79 \pm 0.26$ & $4.37 \pm 1.14$ & --- & $59 \pm 7$ & $111 \pm 20$ & $10.48 \pm 0.10$ & $0.28 \pm 0.21$ & Type-I; PAH may contribute \\
\hline
\end{tabular}
\vspace{0.5em}
\parbox{\textwidth}{\footnotesize\raggedright
$^{a}$ For 593386, only the narrow H$\alpha$ component shown in Figure~\ref{fig:five_miri_dsfg_cutouts} is used to estimate the SFR. The broad H$\alpha$ component is attributed to the AGN broad-line region. The fluxes of the broad and narrow components are presented in Table~\ref{tab:dsfg18}. Formal uncertainties on ${\rm SFR}_{\rm H\alpha}$ are not listed because the H$\alpha$ flux uncertainties are not included in this table.
}
\vspace{0.5em}
\end{table*}


\begin{figure}
 \includegraphics[width=0.48\textwidth]{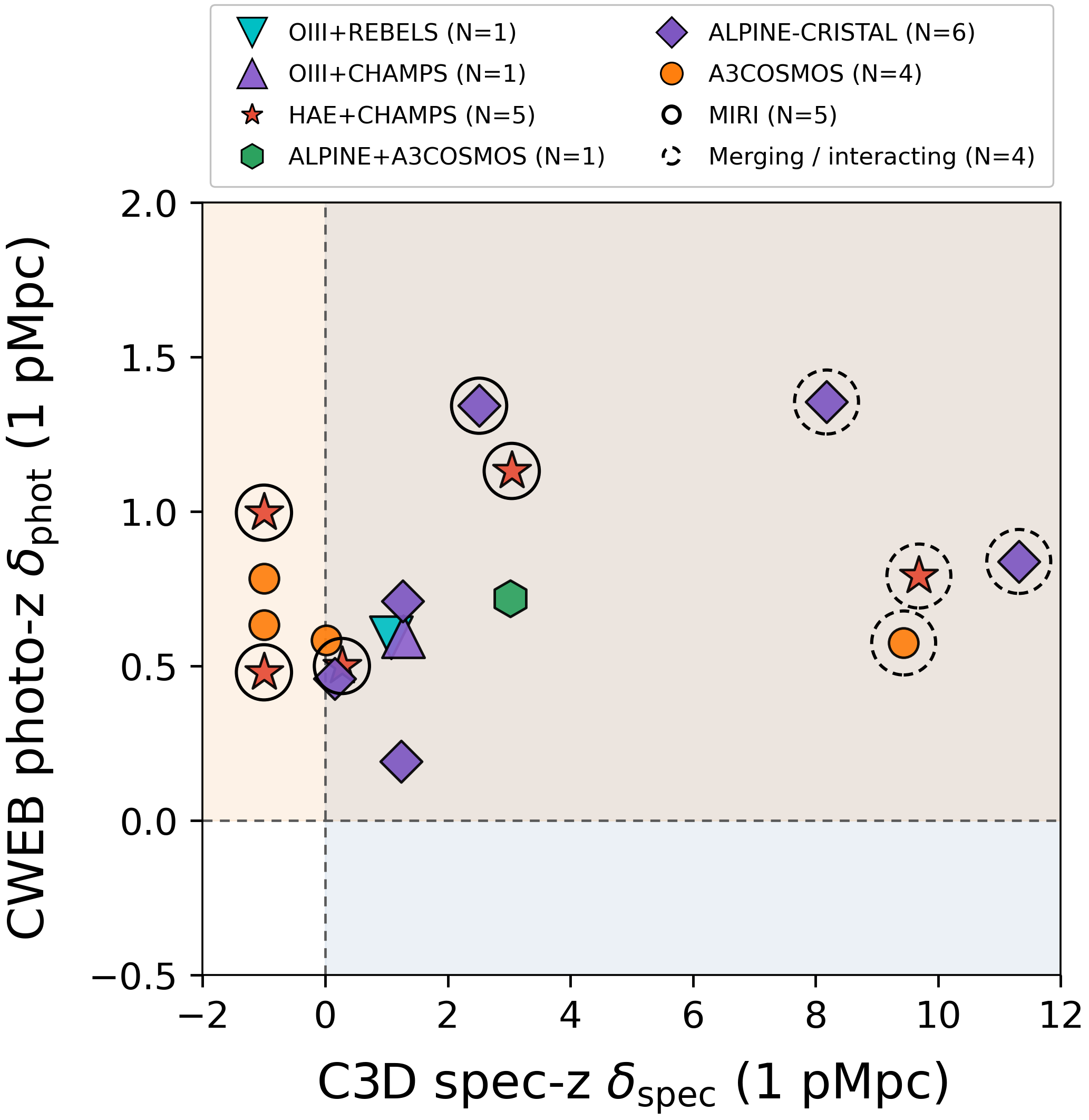}
  \caption{\small{Comparison between the C3D spectroscopic line-emitter overdensity, $\delta_{\rm spec}$, and the CWEB photo-$z$ overdensity, $\delta_{\rm CWEB}$, measured within 1 proper Mpc for the 18 DSFGs or DSFG candidates in this work. Open black circles mark MIRI-detected sources, and dashed circles mark merger/interacting candidates. The shaded blue and orange regions indicate overdense structures identified from the spectroscopic and photo-$z$ samples.
}}\label{fig:dsfg_miri_delta}
\end{figure}

\begin{figure*}
\centering
 \includegraphics[width=0.85\textwidth]{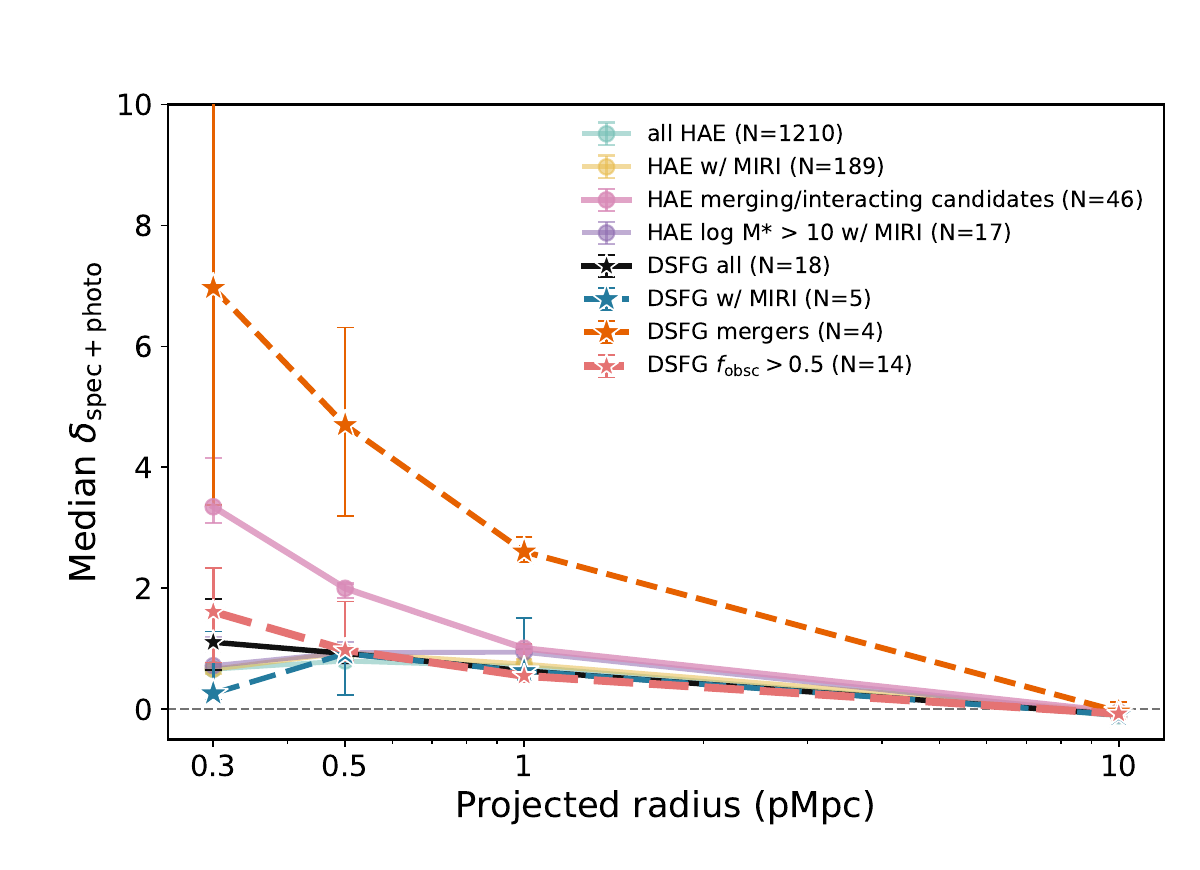}
  \caption{\small{Median aperture overdensity, $\delta_{\rm spec+phot}$, as a function of projected radius for different HAE and DSFG subsamples. This metric combines the spectroscopic line-emitter overdensity and the CWEB photo-$z$ overdensity using their separately estimated field expectations. Spectroscopic matches are removed from the CWEB counts to avoid double counting. The light-red dashed curve shows the $N=14$ DSFG subsample with $f_{\rm obsc}>0.5$. Error bars show bootstrap uncertainties on the median at each radius. The merger/interacting candidates show the strongest small-scale overdensity.}}\label{fig:dsfg_delta}
\end{figure*}

\subsection{Emission line galaxy overdensity around different DSFG}\label{sec:spec_delta}

For the spectroscopic overdensity analysis, we use the observed, completeness-uncorrected H$\alpha$ LF derived in Section~\ref{sec:ha_lf} and the \oiii~LF from \citet{meyer25b} to estimate the expected blank-field number of emission-line galaxies. This choice matches the observed neighbor counts, which are not completeness-corrected.

We measure the spectroscopic environment around each DSFG using cylindrical apertures centered on the DSFG redshift, $z_{\rm g}$. Neighboring H$\alpha$ emitters are counted within projected radii $R=\{0.3,0.5,1.0,10.0\}$ proper Mpc and a line-of-sight window $|\Delta v|<1000~{\rm km~s^{-1}}$, excluding the central target. The radii are chosen to sample compact companion or node scales ($R=0.3$--0.5 pMpc), the immediate protocluster-scale environment ($R=1$ pMpc), and the surrounding large-scale overdense complex ($R=10$ pMpc). Such scales are motivated by simulations showing that high-redshift protoclusters are extended, unvirialized structures distributed across multiple halos, filaments, and substructures \citep{chiang13,shattow13,muldrew15,contini16,muldrew18,overzier16}. The $|\Delta v|<1000~{\rm km~s^{-1}}$ window is used as a redshift-coherent membership cut rather than a virial velocity estimate; similar cuts have been used in protocluster-identification studies to balance member recovery against projection effects \citep{shattow13}.

For each target, the expected number of blank-field H$\alpha$ emitters is
\[
N_{\rm exp}^{\rm H\alpha}(R,z_{\rm g}) =
V_{\rm cyl}(R,\Delta v,z_{\rm g})
\int_{L_{\rm lim}(z_{\rm g})}^{\infty} \Phi(L,z_{\rm g})\,dL ,
\]
where $V_{\rm cyl}$ is the comoving cylindrical volume and $\Phi(L,z)$ is the observed H$\alpha$ LF. The spectroscopic H$\alpha$ overdensity is then
\[
\delta_{\rm H\alpha}(R) =
\frac{N_{\rm obs}^{\rm H\alpha}(R)-N_{\rm exp}^{\rm H\alpha}(R)}
     {N_{\rm exp}^{\rm H\alpha}(R)} .
\]
For the two \oiii-selected DSFGs, we compute the analogous quantity using the \oiii~LF. This spectroscopic overdensity traces compact, redshift-coherent structure around each DSFG while reducing foreground and background projection.

\subsection{Photometric-redshift selected galaxy overdensity around DSFGs}\label{sec:phot_delta}

For the broader galaxy environment, we use the CWEB photometric-redshift catalog and measure a probabilistic projected overdensity rather than imposing a hard photometric-redshift cut. Around each target and projected radius $R$, we select CWEB galaxies in the aperture and assign each galaxy a membership weight based on its photometric-redshift probability distribution, $p_i(z)$. The effective number of photo-$z$ selected galaxies is
\[
N_{\rm eff}^{\rm CWEB}(R) =
\sum_{i \in R}
P_i(z_{\rm g}) ,
\]
where
\[
P_i(z_{\rm g}) =
\int_{z_{\rm g}-\Delta z_v}^{z_{\rm g}+\Delta z_v} p_i(z)\,dz,
\]
and $\Delta z_v=(1+z_{\rm g})(1000~{\rm km~s^{-1}})/c$. This is the PDF-weighted analogue of projected-density methods based on photometric-redshift slices: instead of assigning each galaxy a binary membership flag, we allow each source to contribute according to its redshift probability. CWEB galaxies already matched to spectroscopic line emitters are removed from the photo-$z$ counts to avoid double counting. We estimate the expected field count, $N_{\rm exp}^{\rm CWEB}(R,z_{\rm g})$, empirically from the CWEB footprint using the same aperture size, redshift weighting, and survey mask/coverage selection. The photometric galaxy overdensity is then
\[
\delta_{\rm CWEB}(R) =
\frac{N_{\rm eff}^{\rm CWEB}(R)-N_{\rm exp}^{\rm CWEB}(R)}
     {N_{\rm exp}^{\rm CWEB}(R)} .
\]

The CWEB expectation is measured within the usable C3D--CWEB overlap region, using the same aperture, mask, coverage, and photo-$z$ probability weighting. Thus, $\delta_{\rm CWEB}$ and $\delta_{\rm spec+phot}$ are relative to the average environment in the C3D analysis footprint, not to the full COSMOS-Web field or the cosmic mean. As a consistency check, we also repeat the calculation using a hard photometric-redshift slice, selecting galaxies with $|z_{\rm phot}-z_{\rm g}| \leq 0.2$ and computing the same projected overdensity from binary galaxy counts. This hard-slice test gives broadly consistent median overdensities with the PDF-weighted estimator, but shows larger fluctuations on small scales because galaxies with uncertain photo-$z$ solutions are forced into an in/out classification. We therefore adopt the PDF-weighted CWEB overdensity as our fiducial photo-$z$ environment metric, while using the hard-slice result only as a robustness check.


\section{Discussion}\label{sec:discussion}

\begin{figure*}[ht]
\centering
  \includegraphics[width=0.7\textwidth]{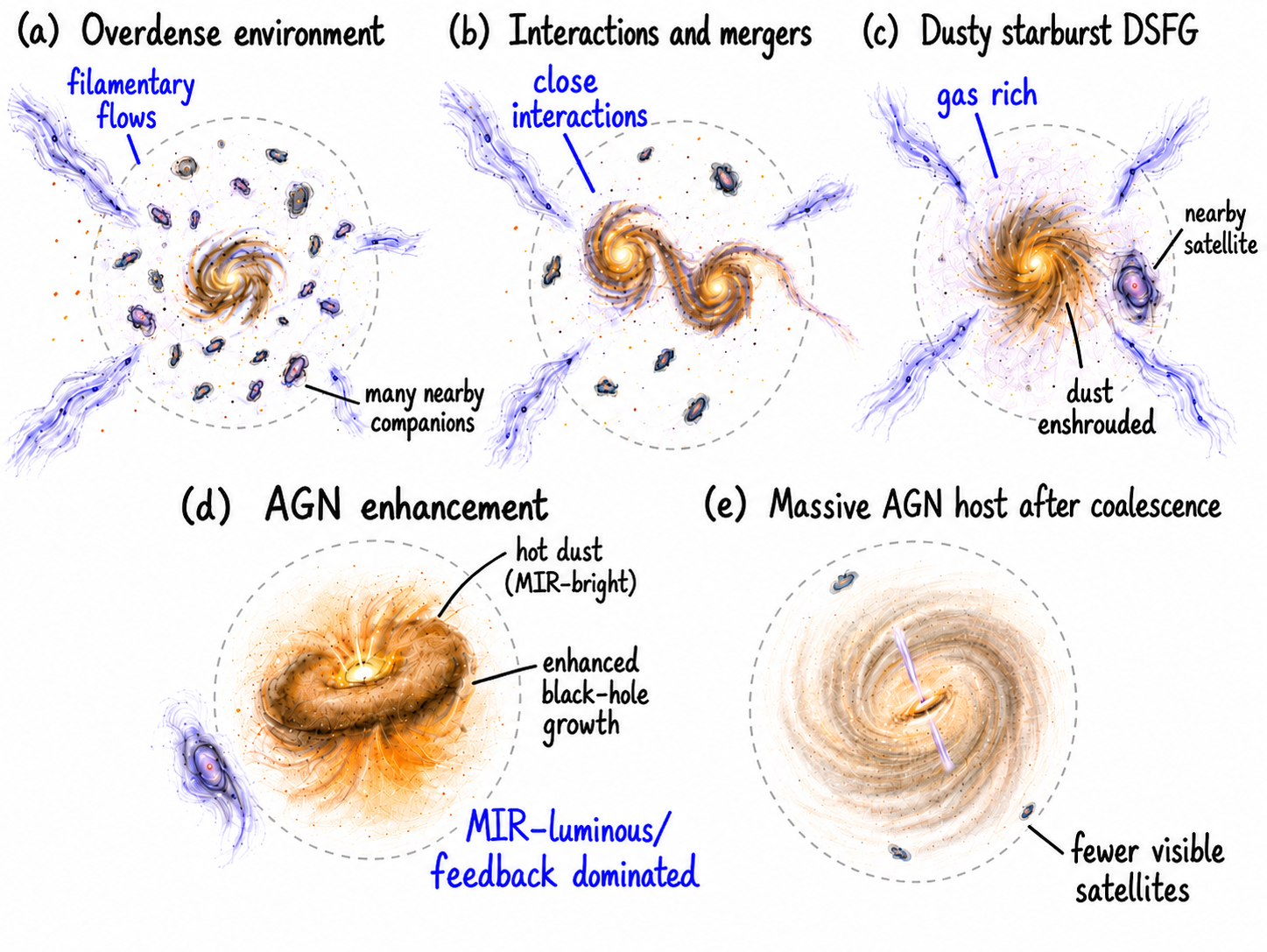}
  \caption{\small{Illustration of a possible scenario linking dense small-scale environments, dusty star formation, and hot-dust or obscured-AGN activity. The panels illustrate a possible evolutionary scenario motivated by the observed environmental trends. Companion numbers, sizes, and separations are schematic. (a) A filament-fed overdense environment containing many potential companions. (b) Galaxies, including dwarf companions, begin to interact and merge within this environment. AGN activity may already be present at this stage. (c) A dusty starburst DSFG that may have undergone a merger and may be accompanied by a possible luminous close companion. The companion is shown schematically and does not imply a confirmed gravitationally bound satellite or an enhanced satellite population for the full DSFG sample. (d) A possible MIRI-bright, hot-dust or obscured-AGN phase with enhanced black-hole growth. (e) A possible massive post-coalescence host with fewer visible satellites. Our observations show that merger/interacting DSFGs have the strongest compact HAE excess, whereas MIRI-bright DSFGs are not necessarily located at the strongest HAE peaks. The proposed evolutionary connections between these phases remain tentative.}}\label{fig:cartoon}
\end{figure*}

\subsection{Galaxy overdensity around DSFGs with different features}

In Figure~\ref{fig:dsfg_miri_delta}, we compare the photo-$z$-selected galaxy overdensity with the spectroscopically confirmed line-emitter overdensity around the 18 DSFGs in this work. The two environment tracers provide complementary views. The photo-$z$-selected sample has a broader effective line-of-sight response because each galaxy is weighted by its redshift PDF, whereas the spectroscopic H$\alpha$ and [O~{\sc iii}] emitters trace the compact population of actively star-forming companions within the adopted redshift slice. Both estimators use the same velocity interval, $|\Delta v|<1000~{\rm km~s^{-1}}$, corresponding to a full line-of-sight depth of approximately $3.4$~pMpc at $z\simeq5.25$. The spectroscopic measurement is approximately a hard count within this interval because the spectroscopic redshift uncertainties are small, whereas the photo-$z$ measurement has a less sharply localized effective depth because of the photo-$z$ uncertainties. A hard-slice photo-$z$ estimate within this interval is included as a robustness check. As a population, the DSFGs show enhanced environments in the combined $\delta_{\rm spec+phot}$ metric relative to the parent narrow-line HAE sample, but the strength of the photo-$z$ and line-emitter overdensities varies substantially from source to source. The spectroscopic line-emitter maps show the clearest dependence on DSFG properties: the sources with merger or interaction features are preferentially located in compact line-emitter overdensities, whereas the massive MIRI-bright systems are generally not located at the strongest line-emitter peaks.

To place the DSFGs in context given the still limited sample size, we compare them with the parent narrow-line HAE sample. We divide both the DSFGs and HAEs into subsamples based on merger or pair morphology in the F444W imaging, MIRI/F1000W or F2100W detection, stellar mass with $\log(M_\star/M_\odot)>10$, and the combination of high stellar mass and MIRI detection. For the 18 DSFG sample, we use the same close-pair information together with visual F444W morphology. We classify four DSFGs as merger/interacting systems: 291002, 43542, 79153 and 94974. For the parent HAE catalog, we identify merger/interacting HAEs in two steps. First, we pre-select candidate pairs with $0.15^{\prime\prime}\leq r_{\rm proj}<5^{\prime\prime}$ and $|\Delta v|<500~{\rm km~s^{-1}}$. The lower limit excludes unresolved components at the F444W imaging resolution, while the upper limit sets the close-companion search radius. This yields 35 pairs involving 63 unique HAEs. We then visually inspect the F444W images and 2D/1D spectra. A central multi-component source is retained only if it has a resolved secondary component with a peak-flux ratio of at least 0.25 and an offset of at least $0.05^{\prime\prime}$ from the image center. A double-peaked H$\alpha$ profile must have a velocity separation of at least $120~{\rm km~s^{-1}}$ and a secondary-to-primary peak ratio of at least 0.25. Internal clumps or asymmetric structures are excluded unless a resolved companion or supporting spectral evidence is present. This procedure yields 46 HAE merger/interacting candidates.

The overdensity estimator used in Figure~\ref{fig:dsfg_delta} combines spectroscopically confirmed line emitters with photo-$z$-selected galaxies in the corresponding redshift slices. It therefore traces both the compact excess of line emitters and the more complete galaxy population
sampled by the photometric-redshift catalog. Here $\delta_{\rm spec+phot}$ is not the sum of $\delta_{\rm spec}$ and $\delta_{\rm CWEB}$. It is computed by combining the observed spectroscopic and CWEB photo-$z$ counts, after removing spectroscopic matches from the CWEB counts, and comparing the sum to the sum of their separately estimated field expectations. The spectroscopic expectation comes from the observed, completeness-uncorrected H$\alpha$ or [O~{\sc iii}] luminosity function, while the CWEB expectation is measured empirically using the same aperture, mask, and photo-$z$ probability weighting. We use $\delta_{\rm spec+phot}$ as a relative environmental ranking statistic, while the spectroscopic-only overdensity remains the more conservative tracer of confirmed redshift-coherent structure. The clearest result is that the strongest small-scale environmental signal is associated with merger or interaction morphology. Both merging HAEs and merging DSFGs show enhanced overdensities relative to their parent populations, most clearly at $R_{\rm proj}\lesssim0.5$~pMpc. At $R_{\rm proj}=0.3$~pMpc, merging HAEs have a median overdensity of $\delta_{\rm spec+phot}=3.35^{+0.81}_{-0.26}$, compared to $0.66^{+0.04}_{-0.03}$ for the full HAE sample. Merging DSFGs show an even stronger excess, with $\delta_{\rm spec+phot}=6.97^{+4.28}_{-3.59}$, compared to $1.11^{+0.71}_{-0.45}$ for the full DSFG sample. At $R_{\rm proj}=0.5$~pMpc, the corresponding values are $\delta_{\rm spec+phot}=2.00^{+0.08}_{-0.15}$ for merging HAEs and $4.70^{+1.61}_{-1.51}$ for merging DSFGs. By contrast, the five MIRI-detected DSFGs discussed in Section~\ref{sec:dsfg_miri} are not more overdense than the full DSFG sample. At $R_{\rm proj}=0.3$~pMpc, the MIRI-detected DSFGs have a median overdensity of $\delta_{\rm spec+phot}=0.26^{+1.02}_{-0.10}$, lower than both the full DSFG median of $1.11^{+0.71}_{-0.45}$ and the HAE-with-MIRI median of $0.66^{+0.09}_{-0.07}$. At $R_{\rm proj}=0.5$~pMpc, the MIRI-detected DSFGs have $\delta_{\rm spec+phot}=0.92^{+0.02}_{-0.68}$, essentially identical to the full DSFG median of $0.92^{+0.06}_{-0.16}$. These trends suggest that the environmental enhancement of DSFGs varies with morphology and MIRI properties. The merger/interacting and MIRI-detected subsamples contain four and five sources, respectively, so larger samples are needed to constrain the amplitude and scale dependence of these differences more precisely.

The error bars in Figure~\ref{fig:dsfg_delta} show the $16$th--$84$th percentile bootstrap confidence interval on the median. For each subsample and aperture radius, we resample the sources with replacement and recompute the median overdensity. The plotted uncertainties are the central $68\%$ interval of the resulting bootstrap distribution and therefore represent the statistical uncertainty on the median, not the full source-to-source scatter.

This distinction suggests that DSFG environments cannot be described by a single overdensity value. Figure~\ref{fig:dsfg_delta} shows a modest population-level enhancement in $\delta_{\rm spec+phot}$ for the full DSFG sample relative to the parent HAE population at small projected radii, while the strongest compact HAE excess is concentrated in the merger/interacting subset. The remaining DSFGs do not show the same compact HAE enhancement, but may trace more extended or less centrally concentrated overdense environments. A DSFG can be embedded in a large-scale galaxy overdensity while lacking a strong excess of visible H$\alpha$ companions. However, this interpretation is not established for the sources with low HAE counts, and a genuinely lower local density remains equally consistent with the data. Such a case may indicate that the surrounding structure is more evolved, more dust-obscured, lower in line equivalent width, or below the \ha~detection threshold. The median $5\sigma$ H$\alpha$ line-flux limit is $4.42\times10^{-18}~{\rm erg~s^{-1}~cm^{-2}}$, corresponding to $\log L_{\rm H\alpha}\simeq42.1$--$42.4$ across the HAE redshift range. At $z\simeq5.2$, integrating our \ha~luminosity function predicts $\sim0.4$ emitters above the flux limit within $R=0.5$~pMpc, compared with $\sim2.6$ emitters when the LF is extrapolated down to $L_{\rm H\alpha}=10^{41}~{\rm erg~s^{-1}}$. Thus, roughly two additional faint emitters per aperture could lie below the detection limit on average. At $R=1$~pMpc, the corresponding expected counts are $\sim1.6$ and $\sim10.6$, respectively. These estimates are illustrative because the LF is constrained mainly above $\log L_{\rm H\alpha}\simeq42.5$, and they do not account for dust-obscured companions with suppressed H$\alpha$ emission. Conversely, DSFGs with merger features appear to identify the compact nodes where line-emitting companions are still abundant and interactions are ongoing.

\subsection{A possible phase-dependent link between dust-obscured star formation, AGN activity, and structure growth}

The results above show two robust observational trends: merger/interacting DSFGs are associated with the strongest compact line-emitter overdensities, while the MIRI-bright DSFGs are not necessarily located at the strongest HAE peaks. These population-level trends do not by themselves establish
a common evolutionary sequence, especially given the heterogeneous
selection of the DSFG sample. A possible interpretation is that these populations trace different phases or locations within early structure growth, as illustrated schematically in Figure~\ref{fig:cartoon}. In this picture, overdense regions increase the probability of close encounters and gas-rich interactions. These interactions can drive gas inflows, enhance dust-obscured star formation, and contribute to rapid stellar-mass growth. The same gas inflows may also feed black-hole growth, but the peak of the visible companion excess and the peak of the MIRI-bright hot-dust or obscured-AGN phase need not occur at the same time.

Hydrodynamical simulations of gas-rich mergers show that tidal torques can drive gas toward the central kiloparsec, triggering compact dusty starbursts and rapid black-hole accretion before feedback, gas consumption, or geometric clearing alters the observed phase \citep{dimatteo05,hopkins06,hopkins08,narayanan10,hayward13}. In such models, close companions are most readily identified before or during coalescence, when multiple galaxies remain spatially resolved and actively star-forming. The most obscured black-hole growth or hot-dust phase may occur later, after the nuclei have merged, when companions are blended with the central system or when remaining satellites are too dusty or line-faint to be selected as HAEs. Spatially resolved JWST and ALMA observations of the late-stage merger IRAS~20551--4250 similarly reveal disturbed warm and cold molecular gas, while showing that the ionized outflow is insufficient to remove the molecular reservoir or quench ongoing star formation \citep{kakkad26}. This system illustrates that merger-driven inflows, obscured activity, and feedback may overlap rather than follow a simple sequence.

This scenario provides one possible explanation for why merger-selected DSFGs show the strongest compact HAE excess, whereas MIRI-bright DSFGs do not show an enhanced number of visible H$\alpha$-emitting companions. However, the present data do not uniquely establish a time sequence. MIRI-bright systems may instead occupy different locations within the same large-scale structure, or have companions with greater dust attenuation, lower equivalent widths, or line fluxes below the H$\alpha$ detection limit. Local LIRGs and ULIRGs provide an observational analogy: gas-rich interactions, compact dusty starbursts, and buried AGN activity are often linked, although the most infrared-luminous or obscured-AGN phases are short-lived \citep{sanders96,koss18,ricci21}. A related theoretical connection is found in the GAEA semi-analytic model, where mergers trigger black-hole accretion and quasar-driven winds dominate early quenching in massive galaxies \citep{xie24}. This model connects interactions and nuclear activity to subsequent suppression of star formation in the host, but does not directly predict the observed difference in HAE companion counts. Thus, these comparisons motivate the evolutionary interpretation illustrated in Figure~\ref{fig:cartoon}.

\subsection{DSFGs are not always located at protocluster cores}

We further present the local \ha~overdensity map for each source, together with the number of HAEs in the corresponding redshift slice used to calculate the overdensity, in Appendix Figure~\ref{fig:dsfg_ha_delta_maps}. Many DSFGs lie close to high-density regions in these maps, but several do not coincide with the local HAE density peak. For these sources, the absence of a local HAE peak may reflect either a genuinely lower local density or incomplete HAE selection. The may appear in lower-density regions that connect substructures within a larger overdense complex. For example, CRISTAL-19 and CRISTAL-21 appear to reside within a large $\sim25$~cMpc overdense structure, but the position of CRISTAL-19 does not coincide with the most compact HAE density peak (Geng et al., in prep.).

This offset is expected if high-redshift overdensities and protoclusters are extended, unvirialized structures rather than compact, virialized clusters with a single well-defined center. Simulations show that protocluster progenitors can span several to tens of comoving Mpc at high redshift, with their final members distributed across multiple halos, filaments, and substructures; the compact main halo contains only a fraction of the eventual cluster population at early times \citep{chiang13,muldrew15,contini16,muldrew18,overzier16}. Observational selection also matters: stellar-mass limits, dust obscuration, SFR thresholds, and emission-line incompleteness can shift the apparent location of the density peak. A DSFG can therefore be physically associated with a large-scale overdensity without lying at the instantaneous peak of the \ha-emitter density field.

The peak-offset measurements support this picture. The merger/interacting DSFGs with well-defined compact HAE peaks are located close to the nearest line-emitter density peak, with $D_{\rm peak}=0.13$--0.43~pMpc and 7--8 nearby emitters in the corresponding redshift slice. By contrast, the MIRI-detected DSFGs are generally offset from the strongest line-emitter peaks, with $D_{\rm peak}=0.83$--1.98~pMpc and only 2--4 nearby emitters. Thus, the merger-selected systems appear to trace compact, actively assembling nodes, whereas the MIRI-bright systems may trace more evolved, more obscured, or more offset substructures.

The absolute value of $D_{\rm peak}$ should be interpreted with caution because the peak position depends on the density-map construction, including the smoothing scale, aperture size, survey-mask treatment, and the number of line emitters in each redshift slice. In broad or multi-peaked structures, modest changes in the smoothing kernel can move the apparent density maximum between neighboring substructures. We therefore use $D_{\rm peak}$ as a relative diagnostic of whether a DSFG lies near a compact HAE node, rather than as a precise distance to a unique protocluster center. The main qualitative result is robust to this limitation: merger/interacting DSFGs are associated with compact HAE-rich regions, while several MIRI-bright or massive DSFGs are offset from the strongest instantaneous HAE peak.

This behavior is consistent with observations of extended proto-superclusters and protoclusters, where high-redshift overdensities often consist of multiple peaks connected by filaments rather than a single compact core \citep{cucciati18,ramakrishnan24}. Narrow-band \ha~surveys of protoclusters also reveal Mpc-scale filamentary structures and show that line emitters trace only the unobscured or moderately obscured star-forming population, while a substantial fraction of star formation can remain hidden in dusty systems \citep{koyama13,koyama21,polletta21}. Compact dust-rich structures further show that UV-, optical-, and submillimeter-selected galaxies can recover different member populations within the same forming system \citep{rotermund21}. Taken together, these results support a scenario in which the line-emitter core, the dusty star-forming phase, and the future cluster center need not be spatially identical at $z>4$.

\subsection{Comparison with lower-redshift massive-galaxy environments}

\begin{figure}
\centering
    \includegraphics[width=0.49\textwidth]{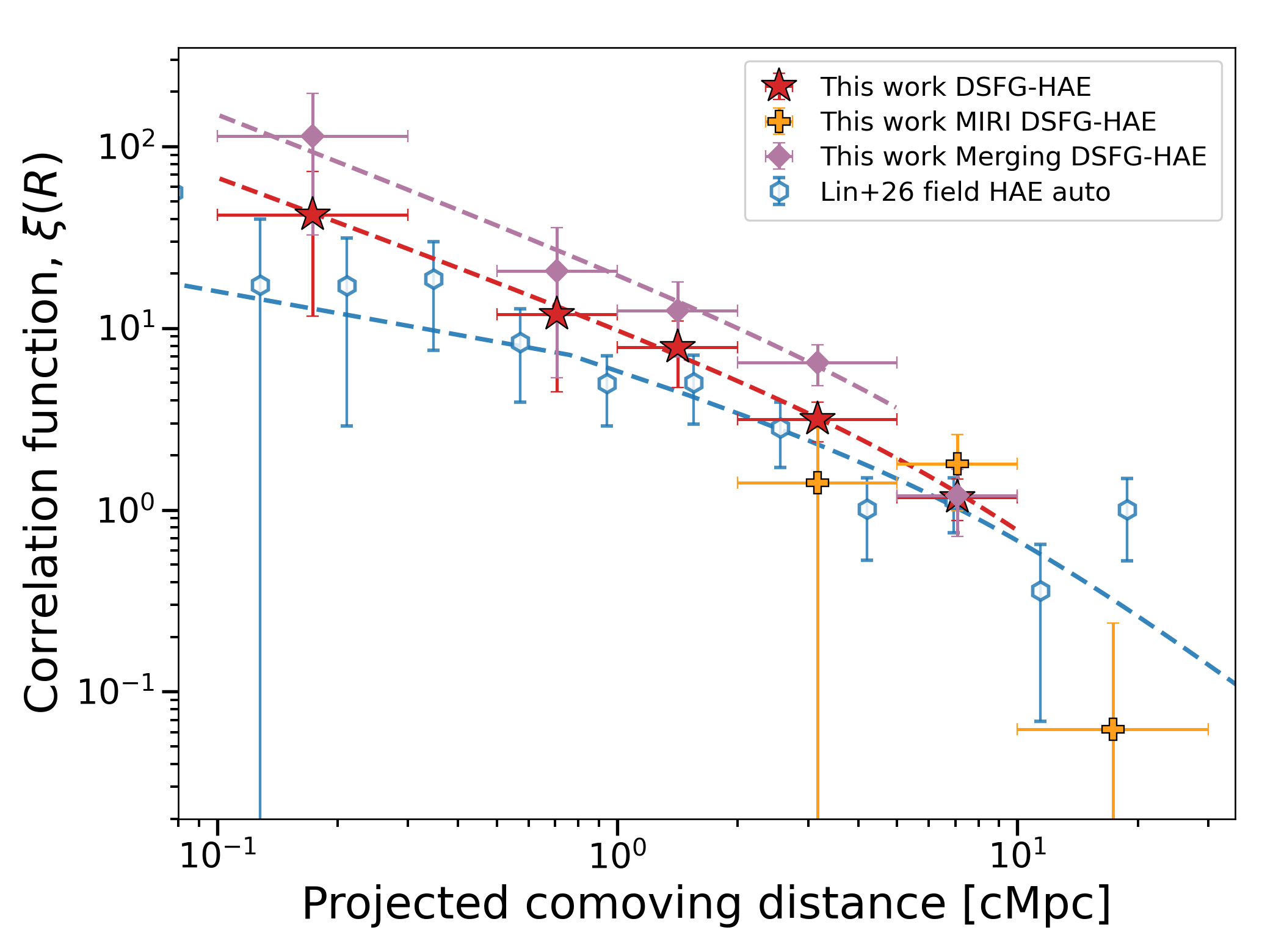}
    \caption{Projected HAE auto-correlation and DSFG--HAE cross-correlations in COSMOS-3D. Red stars, orange pentagons, and purple diamonds show the full H$\alpha$ DSFG, MIRI-bright, and merger/interacting subsets, respectively; blue hexagons show the field HAE auto-correlation from \citet{lin26}. Error bars are Poisson uncertainties, and dashed curves show fitted line-of-sight-averaged power-law models. The MIRI subset is not fitted because too few radial bins are positive. The merger subset shows the strongest small-scale HAE excess.}
    \label{fig:cc}
\end{figure}

The peak-offset and clustering measurements provide a useful link between the $z>5$ DSFG environments studied here and massive-galaxy environments at later cosmic times. At cosmic noon, massive forming structures such as the Spiderweb protocluster, Hyperion, and SSA22 show enhanced dusty star formation, AGN activity, and complex filamentary or multi-node structure rather than a single relaxed core \citep{koyama13,dannerbauer14,umehata15,umehata17,cucciati18,lemaux18,jackie26}. In these systems, dusty starbursts and AGN are distributed across dense nodes, filaments, and infalling substructures, indicating that rapid massive-galaxy growth can occur before the final cluster core is fully assembled. Our DSFGs show a similar behavior at earlier epochs: the strongest compact line-emitter excess is found around merger/interacting DSFGs near density peaks, while several massive or MIRI-bright DSFGs are offset from the current HAE peak.

Figure~\ref{fig:cc} shows the projected DSFG--HAE cross-correlation compared with the $z=5$--6 field HAE auto-correlation from the CONGRESS (GO \#3577; PIs: E.~Egami and F.~Sun) and the FRESCO (GO \#1895; PI: P.~A.~Oesch) survey \citep{lin26}, for which $R_0=6.23^{+1.68}_{-1.13}\,h^{-1}{\rm cMpc}$. We measure $R_0=6.67\pm1.45\,h^{-1}{\rm cMpc}$ for the full DSFG--HAE sample and $R_0=9.34\pm2.40\,h^{-1}{\rm cMpc}$ for the merger/interacting DSFG--HAE subset. The MIRI-detected subset is shown for comparison but is not fitted because it has too few positive radial bins. The larger $R_0$ of the merger subset indicates that interacting DSFGs drive the strongest small-scale HAE excess. This is a relative clustering result and does not by itself identify a
particular evolutionary stage. The two \oiii-selected DSFGs show a tentative excess over the \oiii~auto-correlation measurements (see Figure \ref{fig:cc_o3}) at similar redshifts from EIGER (GTO\#1243; PI: S.~J. Lilly) and ASPIRE (GO\#2078; PI: F.~Wang) surveys \citealt{eilers24,huang26}. Given that this measurement is based on only two DSFG centers and shows substantial radial scatter, we treat it only as a qualitative consistency check. We therefore do not use this two-source comparison to infer a separate halo mass or evolutionary pathway.

Because this is a cross-correlation with line emitters, and because the DSFG sample is small, we do not derive precise halo masses. Nevertheless, comparison with HAE clustering at $z\simeq4$--6 and classical SMG clustering measurements suggests characteristic halo masses of order $\log(M_{\rm h}/M_\odot)\sim12.3$--12.8 for the full DSFG--HAE sample and $\log(M_{\rm h}/M_\odot)\sim12.8$--13.3 for the merger/interacting subset \citep{hickox12,wilkinson17,stach21,lin26,shuntov26}. These values are consistent with massive galaxy or group-scale halos for the full DSFG sample, while
the merger/interacting subset is consistent with the higher halo masses inferred for strongly clustered SMG populations. This comparison does not by itself establish that these galaxies are
forming group or protocluster cores.

The corresponding physical scale is also informative. We estimate halo virial radii using $R_{\rm vir}=[3M_{\rm h}/(4\pi\Delta_{\rm vir}\rho_{\rm c}(z))]^{1/3}$, where $\rho_{\rm c}(z)=3H^2(z)/(8\pi G)$ is the critical density and $\Delta_{\rm vir}$ is the spherical-collapse overdensity relative to the critical density \citep{bryan98}. At $z=5$--6, $\Delta_{\rm vir}\simeq178$ in our adopted cosmology. Treating the illustrative halo masses above as virial masses, $\log(M_{\rm h}/M_\odot)\sim12.3$--13.3 corresponds to $R_{\rm vir}\simeq60$--150~pkpc. These are approximate halo-size estimates, not direct measurements. They are comparable to the smallest observed DSFG--density-peak offsets of $D_{\rm peak}\sim0.13$~pMpc, but much smaller than the largest offsets of $\sim2.2$~pMpc. The larger offsets therefore locate the DSFGs within extended group- or protocluster-scale environments rather than tracing the internal dimensions of individual halos. For halos with $\log(M_{\rm h}/M_\odot)\sim12.3$--13.3 at $z\sim5$--6, the expected virial radii are only $\sim50$--130~pkpc, much smaller than the observed DSFG--density-peak offsets of $D_{\rm peak}\sim0.13$--2.2~pMpc. These offsets therefore do not trace the internal sizes of individual halos or proto-BCGs. Instead, they locate each DSFG within a larger group or protocluster-scale structure. A simple halo-growth estimate based on mean mass-accretion histories suggests that such halos can evolve into group- or cluster-scale descendants by $z=0$ \citep{fakhouri10}. However, the large scatter in halo growth and the extended nature of protoclusters mean that an individual DSFG should not automatically be identified as the future BCG. Recent simulations of high-redshift massive galaxies reach a similar conclusion: massive galaxies at $z\sim5$ often become massive descendants, but do not necessarily become the most massive central galaxies at $z=0$ \citep{baxter26}.

The possible descendant also depends on the DSFG phase. However, the present measurements do not determine the evolutionary phase or descendant of any individual source. The merger/interacting DSFGs, which are close to HAE peaks and have high local emitter counts, may represent active node-building or pre-coalescence systems. This interpretation is consistent with their compact environments, but it is not uniquely selected by the data. Possible descendants include massive galaxies in future group or protocluster cores and massive satellites that later merge into a central galaxy, but the current data do not distinguish between these possibilities. They may become massive galaxies in future group or protocluster cores, or massive satellites that later merge into the central galaxy. The massive MIRI-bright DSFGs may instead represent a short-lived, more obscured transition phase, similar in a physical sense to local LIRGs and ULIRGs. Four of the five MIRI-detected DSFGs have $\log(M_\star/M_\odot)>10$, but they are generally offset from the strongest HAE peaks and have lower local line-emitter counts. This does not imply that they are isolated. Their lower local HAE counts could reflect genuinely lower local densities, more
extended substructures, or companions that are dust-obscured or below the H$\alpha$ detection limit. The MIRI-bright emission may trace a hot-dust or obscured-AGN phase, but the present data do
not establish that this phase occurs after rapid central growth or represents a specific evolutionary stage. They may reside in massive substructures where the most active companions have already merged, become dust-obscured, or fallen below the \ha~detection limit. Their MIRI-bright emission may trace a brief hot-dust or obscured-AGN candidate phase after rapid central growth, consistent with the short-lived infrared-bright stages inferred for local LIRGs/ULIRGs and obscured-AGN \citep{sanders96,koss18,ricci21}.

These massive MIRI-bright DSFGs are plausible progenitors of compact massive quiescent galaxies at $z\sim2$--3 and, eventually, massive early-type galaxies, consistent with proposed evolutionary links between high-redshift SMGs, compact dusty starbursts, compact quiescent galaxies, and local massive ellipticals \citep{toft14,barro13,ikarashi15}. However, their offsets from the present HAE peaks suggest that they should not all be interpreted as proto-BCGs. Some may instead be massive infalling satellites or secondary nodes that later merge into the central galaxy of the descendant halo. This is consistent with hierarchical models of BCG formation, in which much of the stellar mass forms early in separate progenitors and is assembled later through mergers \citep{delucia07}. In this sense, the relevant comparison is not only to the size of a final BCG, whose stellar body and intracluster-light envelope occupy tens to hundreds of kpc at low redshift, but also to the much larger protocluster assembly region from which that BCG is built.

The lower-mass DSFGs likely trace a different descendant pathway. Sources such as 465395 and 635130 lie close to line-emitter peaks but have $\log(M_\star/M_\odot)<10$, suggesting that they are unlikely to be direct progenitors of the most massive quiescent galaxies unless they experience substantial subsequent growth. These systems may instead become massive satellites, contribute to the growth of a central galaxy through later merging, or remain part of the star-forming satellite population in the forming group or cluster. This connects naturally to the lower-redshift morphology--density relation and environmental-quenching framework, in which dense environments progressively transform satellite populations and increase the passive and early-type fractions in groups and clusters \citep{dressler80,peng10,wetzel13}.

Overall, the comparison with lower-redshift massive-galaxy environments suggests that the DSFGs in this work should not be treated as a single progenitor class. The merger/interacting DSFGs may trace compact, actively assembling nodes; the massive MIRI-bright DSFGs may trace a short-lived hot-dust or obscured-AGN candidate phase on the path toward compact quiescent or massive early-type descendants; and the lower-mass DSFGs near density peaks may become future satellites or building blocks of the larger structure. Thus, the likely descendant of a $z>5$ DSFG depends not only on stellar mass, but also on its position within the density field and its current evolutionary phase.

\section{Summary}

We combine JWST/NIRCam F444W grism spectroscopy, MIRI F1000W/F2100W imaging, and ALMA-selected or ALMA-followed DSFGs in the COSMOS-3D footprint to study the environments, MIRI excesses, and possible evolutionary pathways of dusty galaxies at $z=4.9$--7.2. Our main results are as follows.

\begin{enumerate}

\item Our final sample contains 18 spectroscopically confirmed DSFGs or DSFG candidates, including 16 H$\alpha$ emitters and two [O~{\sc iii}] emitters. Five sources are detected in MIRI F2100W, and two of the MIRI-detected sources have radio counterparts.

\item The parent narrow-line HAE sample contains 1158 galaxies after broad-line removal. It has a median observed, dust-uncorrected ${\rm SFR}_{\rm H\alpha}=11.7~M_\odot~{\rm yr^{-1}}$ and a median dust-corrected ${\rm SFR}_{\rm H\alpha}=17.7~M_\odot~{\rm yr^{-1}}$. DSFGs reside in denser environments than this parent sample: within $R=0.5$ pMpc, the median $\delta_{\rm spec+phot}$ is $0.92^{+0.06}_{-0.16}$ for DSFGs and $0.80^{+0.05}_{-0.04}$ for HAEs; within $R=0.3$ pMpc, the medians are $1.11^{+0.71}_{-0.45}$ and $0.66^{+0.04}_{-0.03}$, respectively.

\item The strongest small-scale environmental signal is linked to merger or interaction morphology. The four merger/interacting DSFGs show the largest spectroscopic overdensities, with 9--12 H$\alpha$ emitters within 1 pMpc and $|\Delta v|<$ 1000 \kms~compared to an LF expectation of only $\sim1$ source, corresponding to $\delta_{\rm spec}=8$--11. Their DSFG--HAE cross-correlation reaches $R_0=9.34\pm2.40\,h^{-1}{\rm cMpc}$.

\item The MIRI-bright DSFGs do not show the same excess of visible H$\alpha$-emitting companions. They are generally offset from the strongest HAE peaks and have only 2--4 nearby HAEs in the corresponding redshift slice. Their F2100W excess is difficult to explain with the 3.3 $\mu$m PAH feature alone, making them hot-dust or obscured-AGN candidates.

\item Taken together, these results are consistent with a possible phase-dependent picture in which DSFGs trace early structure growth, but the line-emitter core, dusty star-forming phase, and future cluster center need not be spatially identical at $z>4$.

\end{enumerate}

\acknowledgements
We thank Daizhong Liu, Jorge Gonz\'alez-L\'opez, and Cheng Cheng for helpful ALMA data discussions, and Feige Wang, Xiangyu Jin, Danyang Jiang, Zijian Zhang, and Jie Chen for discussions related to the \ha~catalog. SZ and SZG acknowledge support from the National Science Foundation of China (grant no. 12673018) and the Chinese Academy of Sciences (no. E5295401). 

This work is based on observations made with the NASA/ESA/CSA James Webb space Telescope. The data were obtained from the Mikulski Archive for Space Telescopes at the Space Telescope Science Institute, which is operated by the Association of Universities for Research in Astronomy, Inc., under NASA contract NAS 5-03127 for JWST. These observations are associated with programs \#5893. Support for program \#5893 was provided by NASA through a grant from the Space Telescope Science Institute, which is operated by the Association of Universities for Research in Astronomy, Inc., under NASA contract NAS 5-03127. All of the JWST data presented in this article were obtained from the Mikulski Archive for Space Telescopes (MAST) at the Space Telescope Science Institute. The specific observations analyzed can be accessed via \dataset[doi: 10.17909/k5th-zb31]{https://doi.org/10.17909/k5th-zb31}. We acknowledge the strong support provided by the program coordinator Alison Vick and instrument reviewers Brian Brooks and Jonathan Aguilar. 

This paper makes use of the following ALMA data: ADS/JAO.ALMA\#2017.1.00428.L, ADS/JAO.ALMA\#2021.1.00280.L, ADS/JAO.ALMA\\\#2019.1.01634.L, ADS/JAO.ALMA\#2023.1.00180.L, and archival ALMA data compiled in the A3COSMOS catalog. ALMA is a partnership of ESO (representing its member states), NSF (USA), and NINS (Japan), together with NRC (Canada), NSTC and ASIAA (Taiwan), and KASI (Republic of Korea), in cooperation with the Republic of Chile. The Joint ALMA Observatory is operated by ESO, AUI/NRAO, and NAOJ. The National Radio Astronomy Observatory and Green Bank Observatory are facilities of the U.S. National Science Foundation operated under cooperative agreement by Associated Universities, Inc.

\facilities{JWST (NIRCam), JWST (MIRI), ALMA, VLA}

\appendix
\restartappendixnumbering

\section{Figures}
\label{app:dsfg_spec}

\begin{figure*}
\centering
    \includegraphics[width=0.49\textwidth]{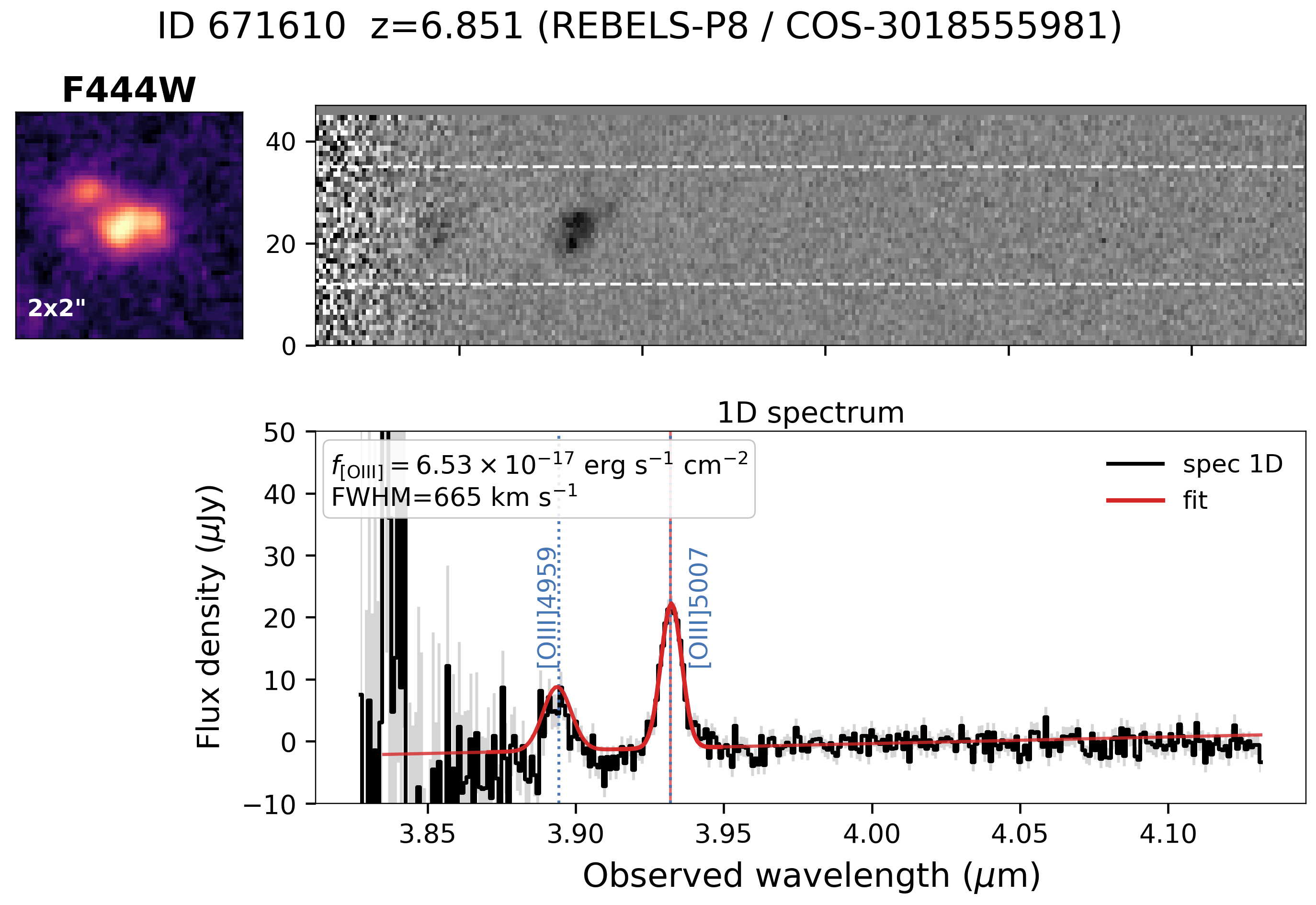}
    \includegraphics[width=0.49\textwidth]{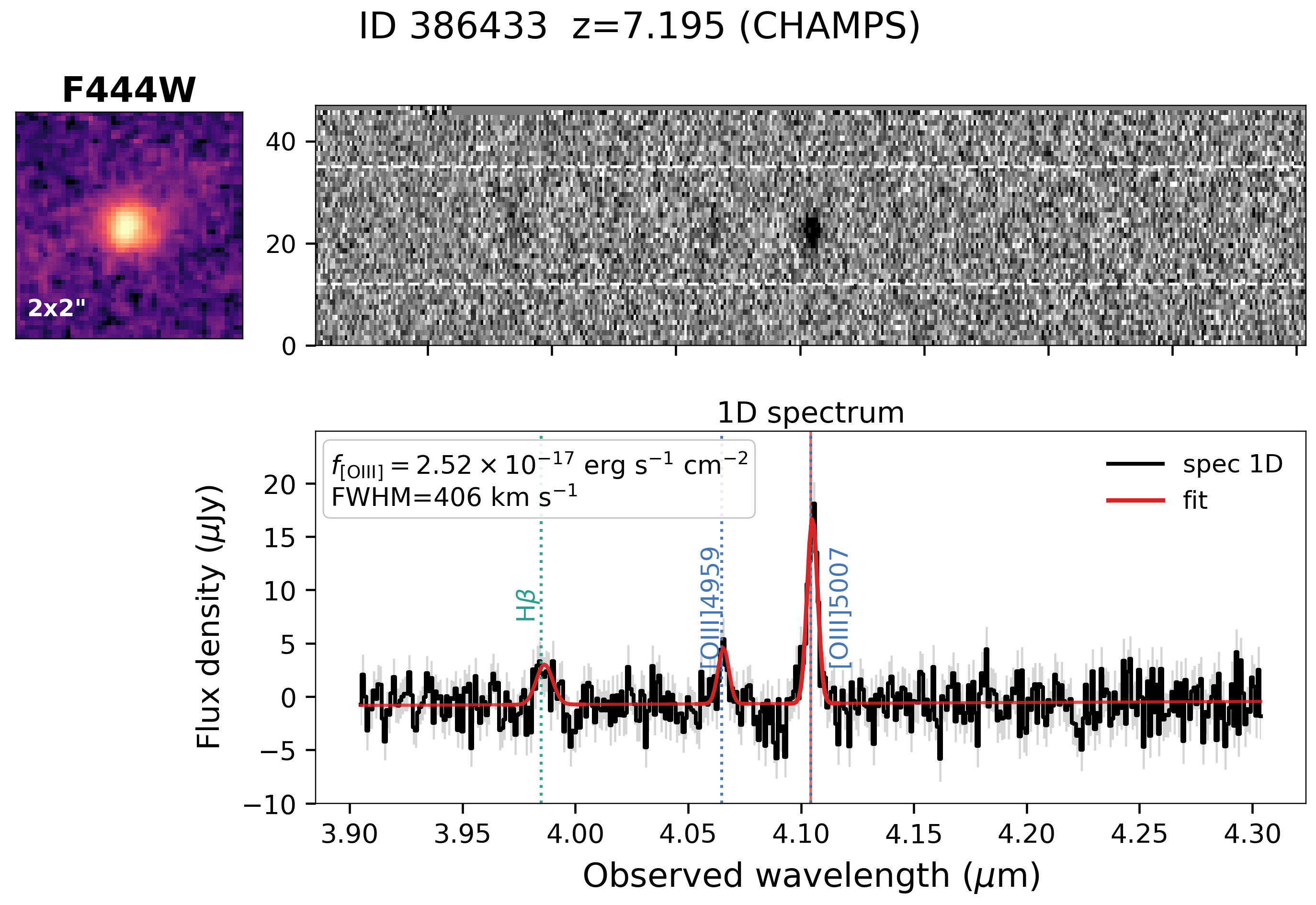}
    \includegraphics[width=0.49\textwidth]{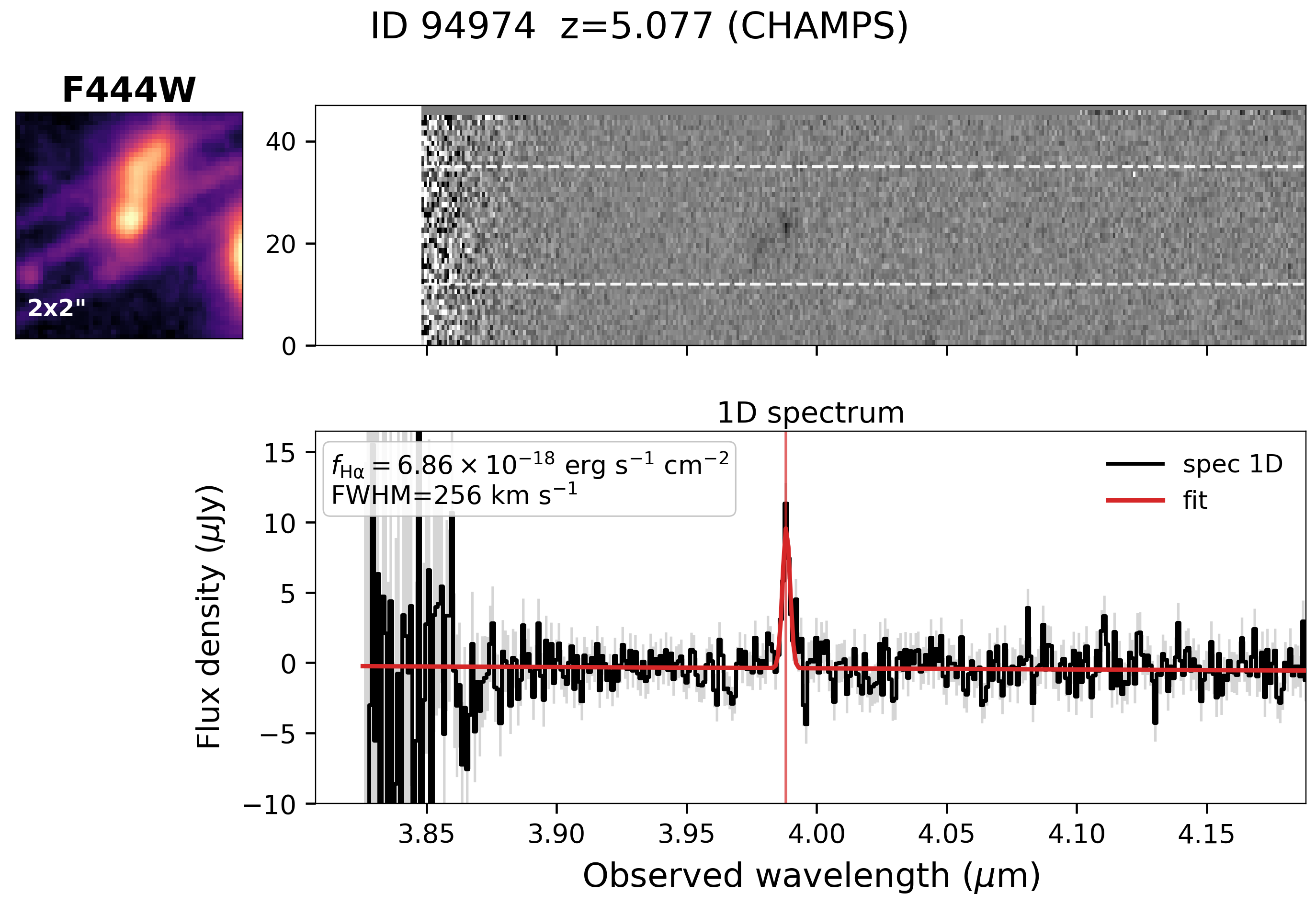}
    \includegraphics[width=0.49\textwidth]{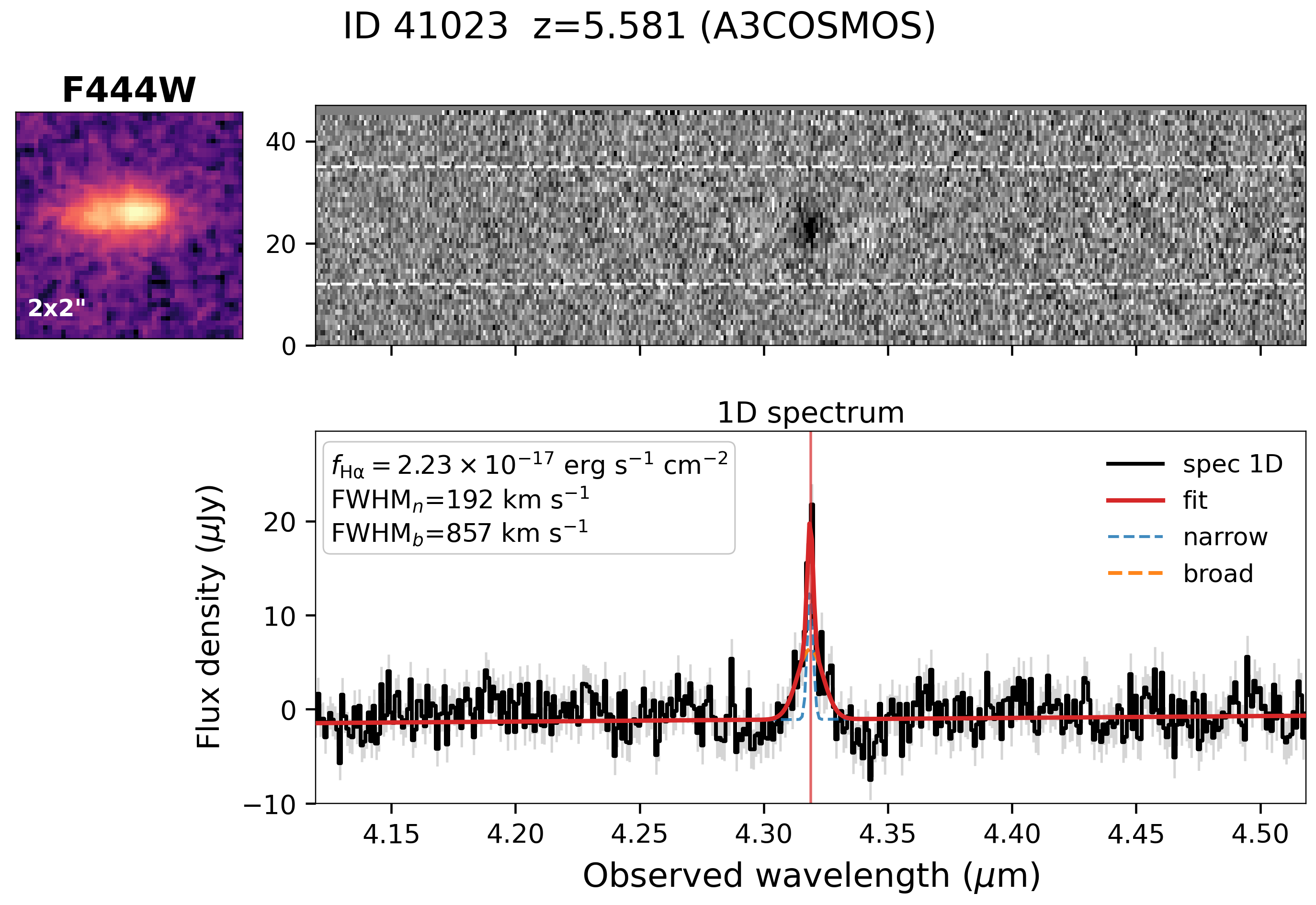}
    \includegraphics[width=0.49\textwidth]{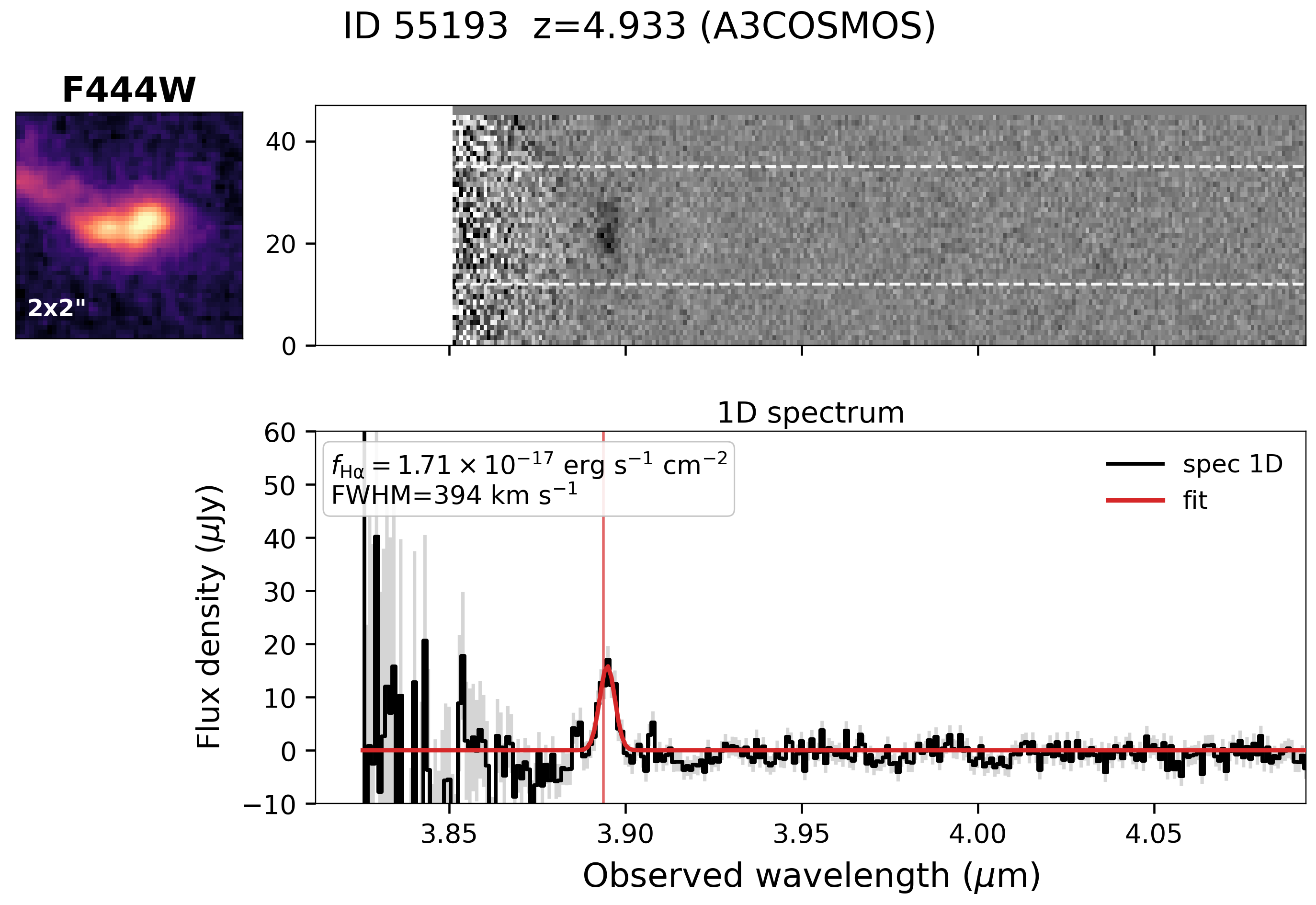}
    \includegraphics[width=0.49\textwidth]{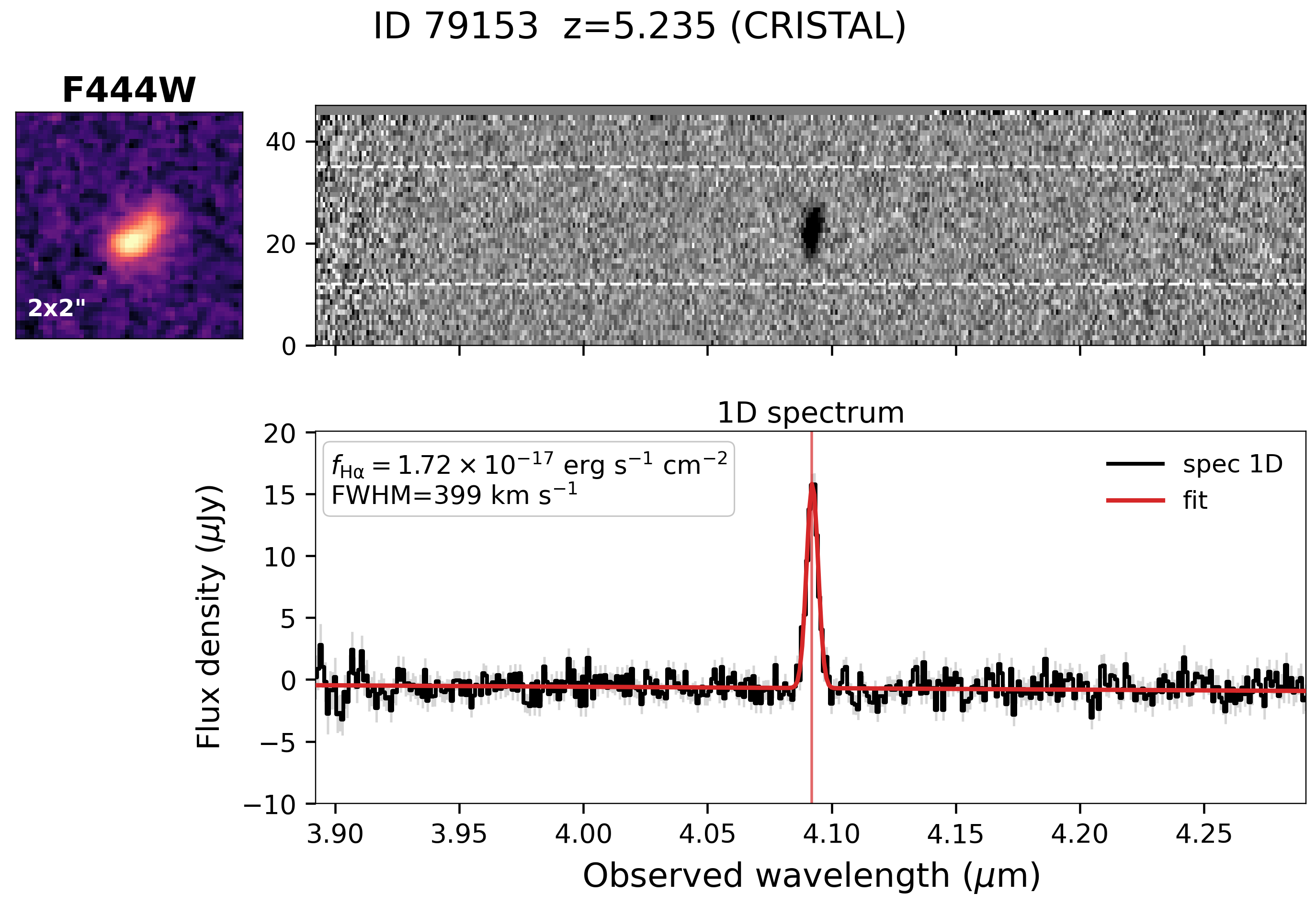}
    \includegraphics[width=0.49\textwidth]{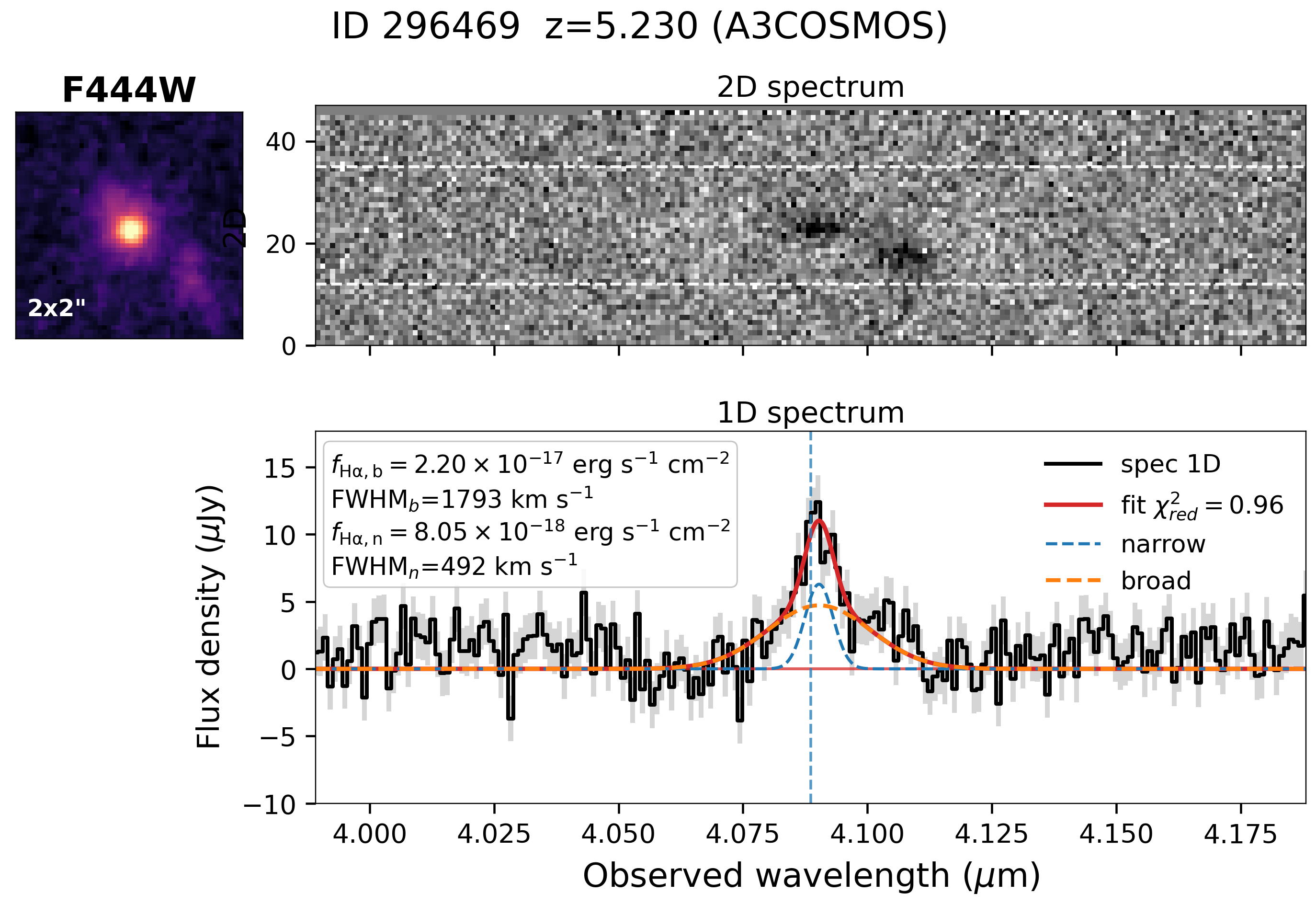}
    \includegraphics[width=0.49\textwidth]{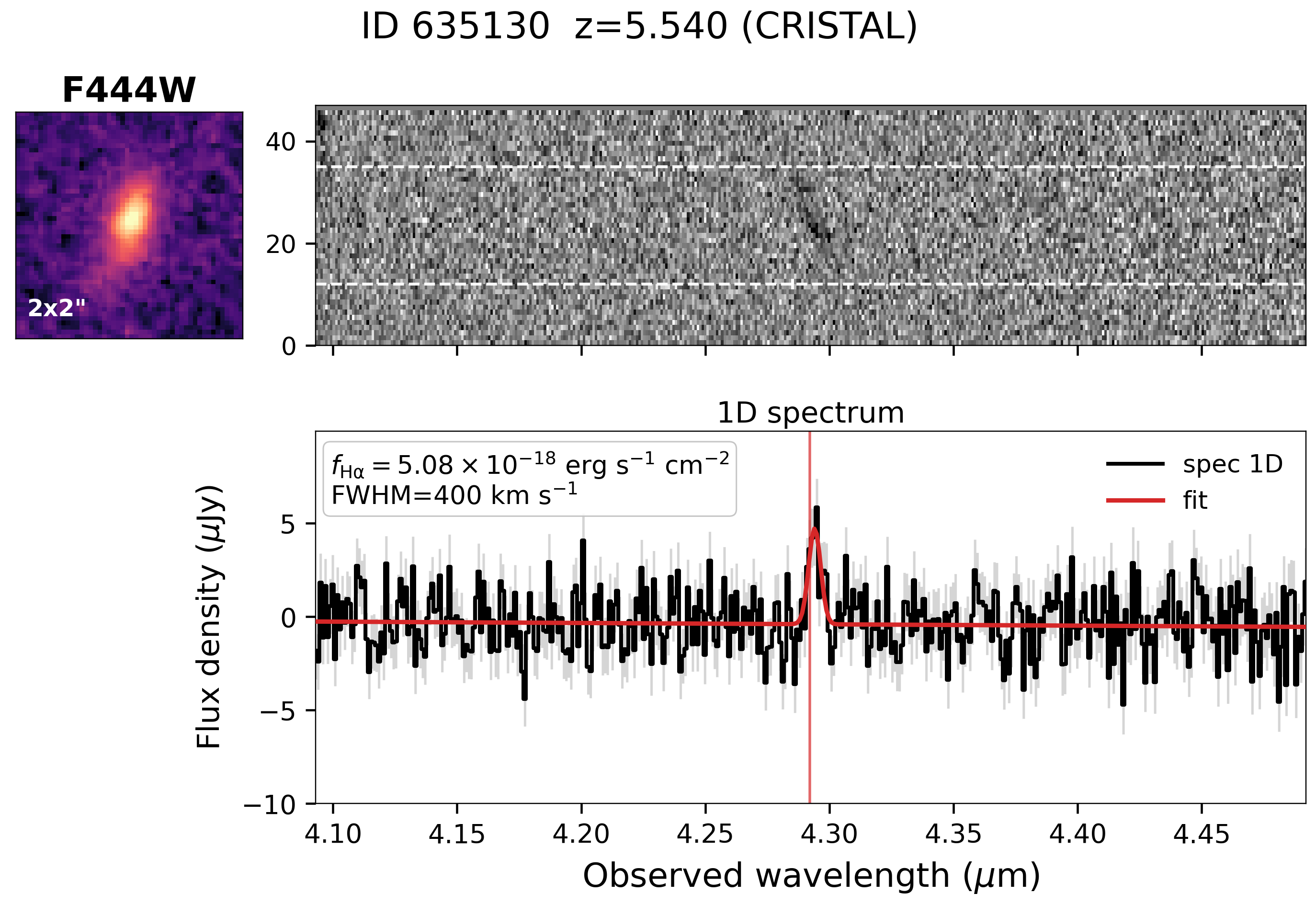}
\end{figure*}

\begin{figure*}
\centering
    \includegraphics[width=0.49\textwidth]{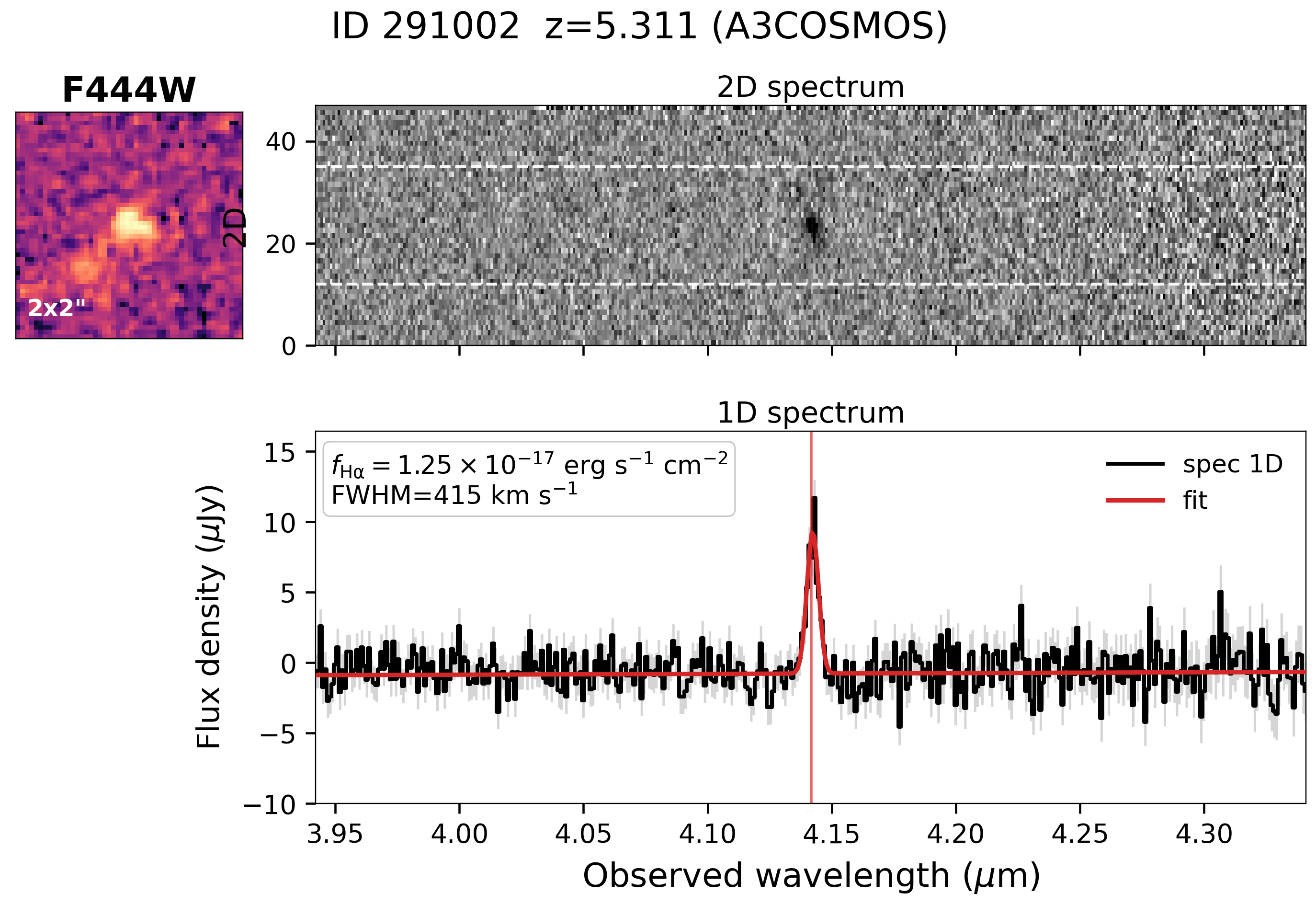}
    \includegraphics[width=0.49\textwidth]{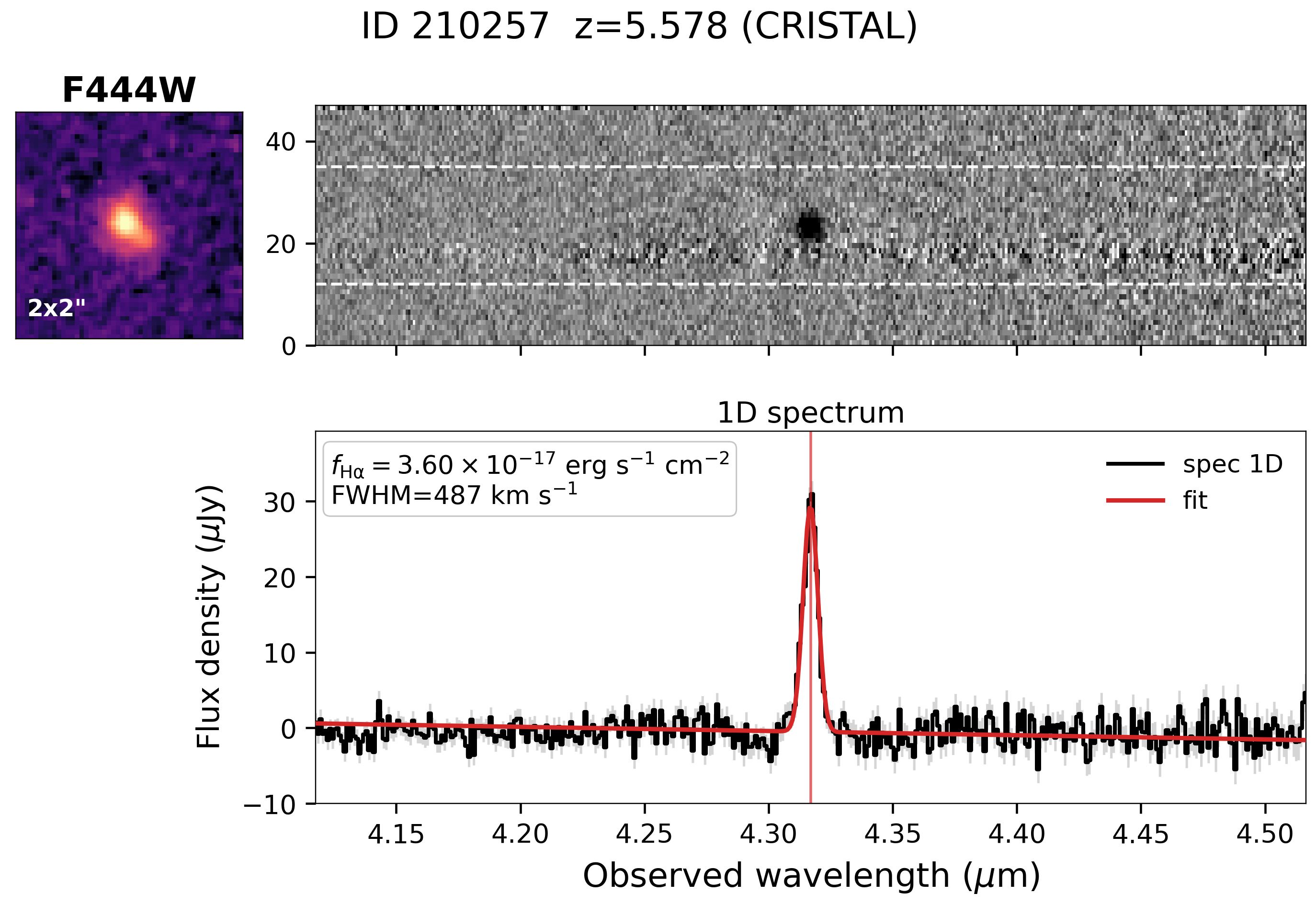}
    \includegraphics[width=0.49\textwidth]{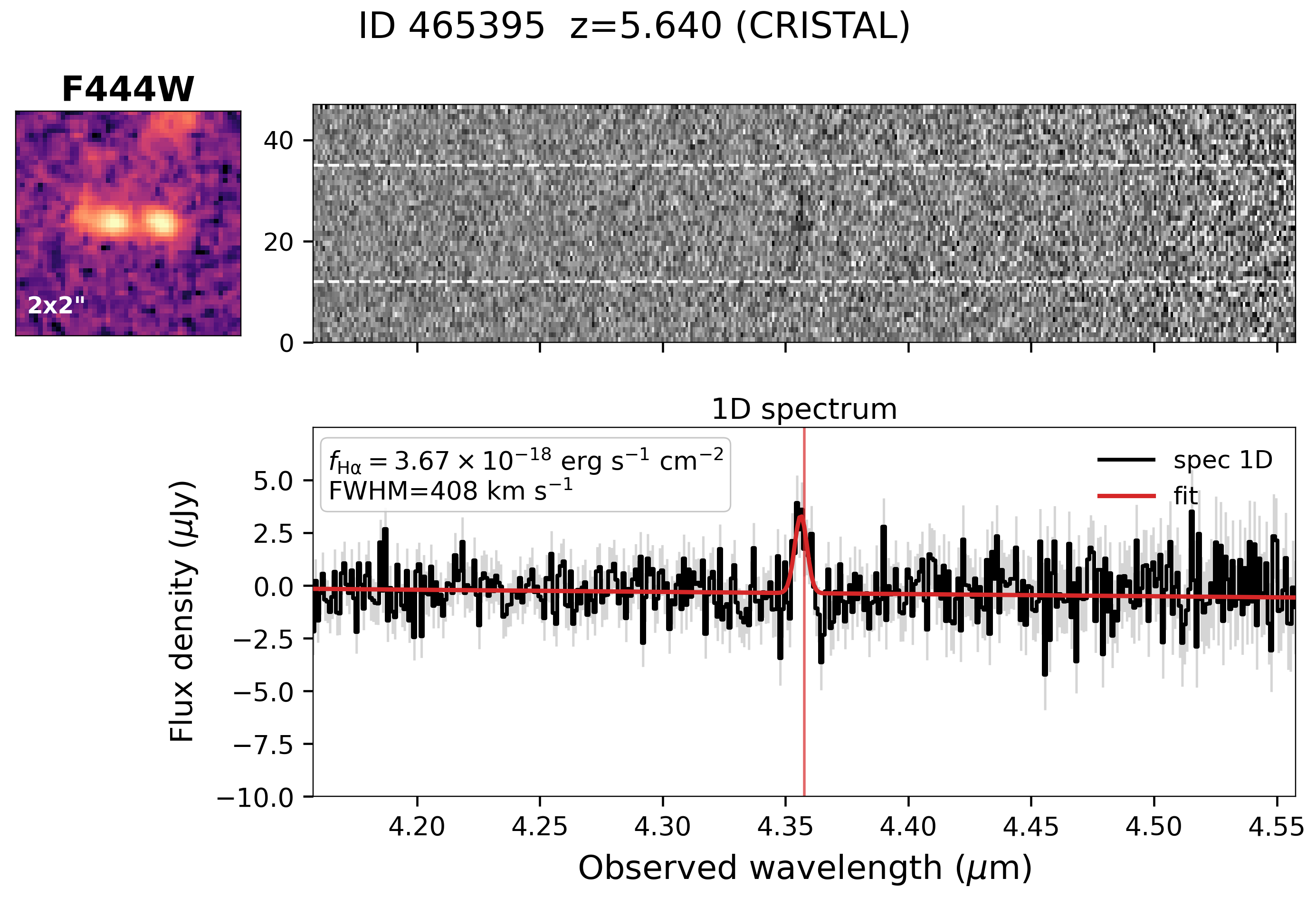}
    \includegraphics[width=0.49\textwidth]{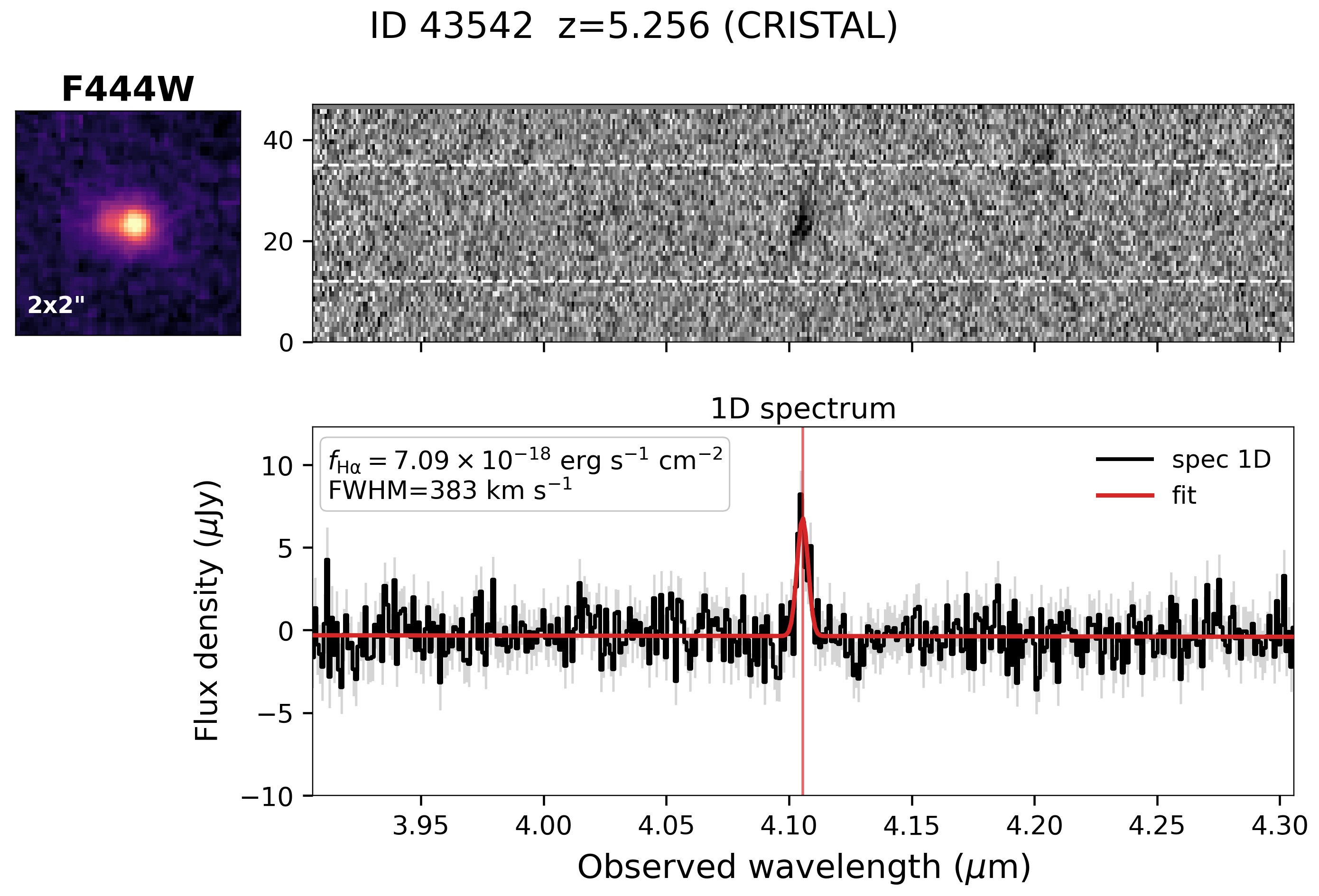}
    \includegraphics[width=0.49\textwidth]{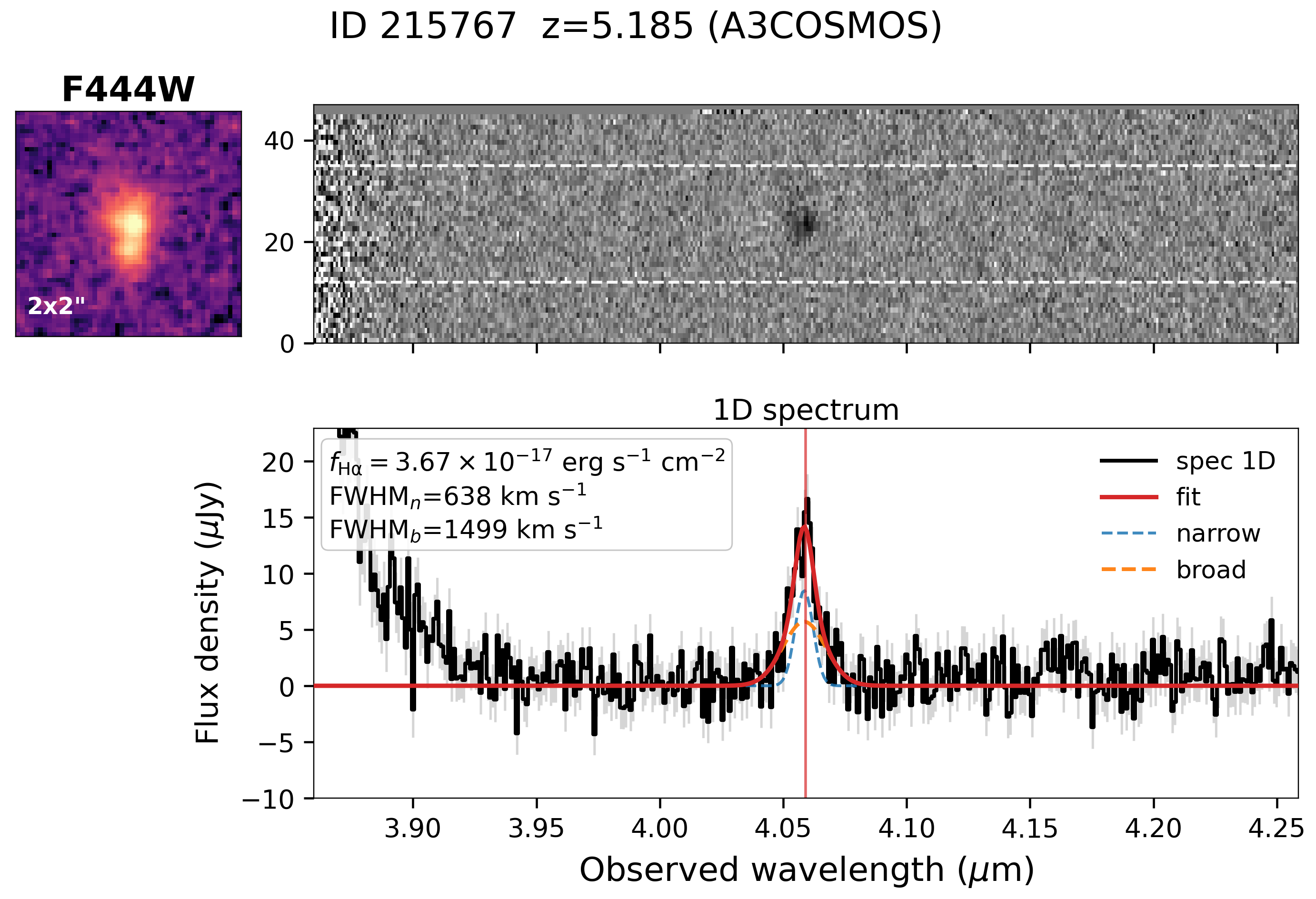}
    \caption{Multi-band JWST cutouts and H$\alpha$ spectra for the DSFGs in this work. Five already presented in Figure \ref{fig:five_miri_dsfg_cutouts}.
    }\label{fig:dsfg_cutouts}
\end{figure*}


\begin{figure*}
\centering
    \includegraphics[width=0.49\textwidth]{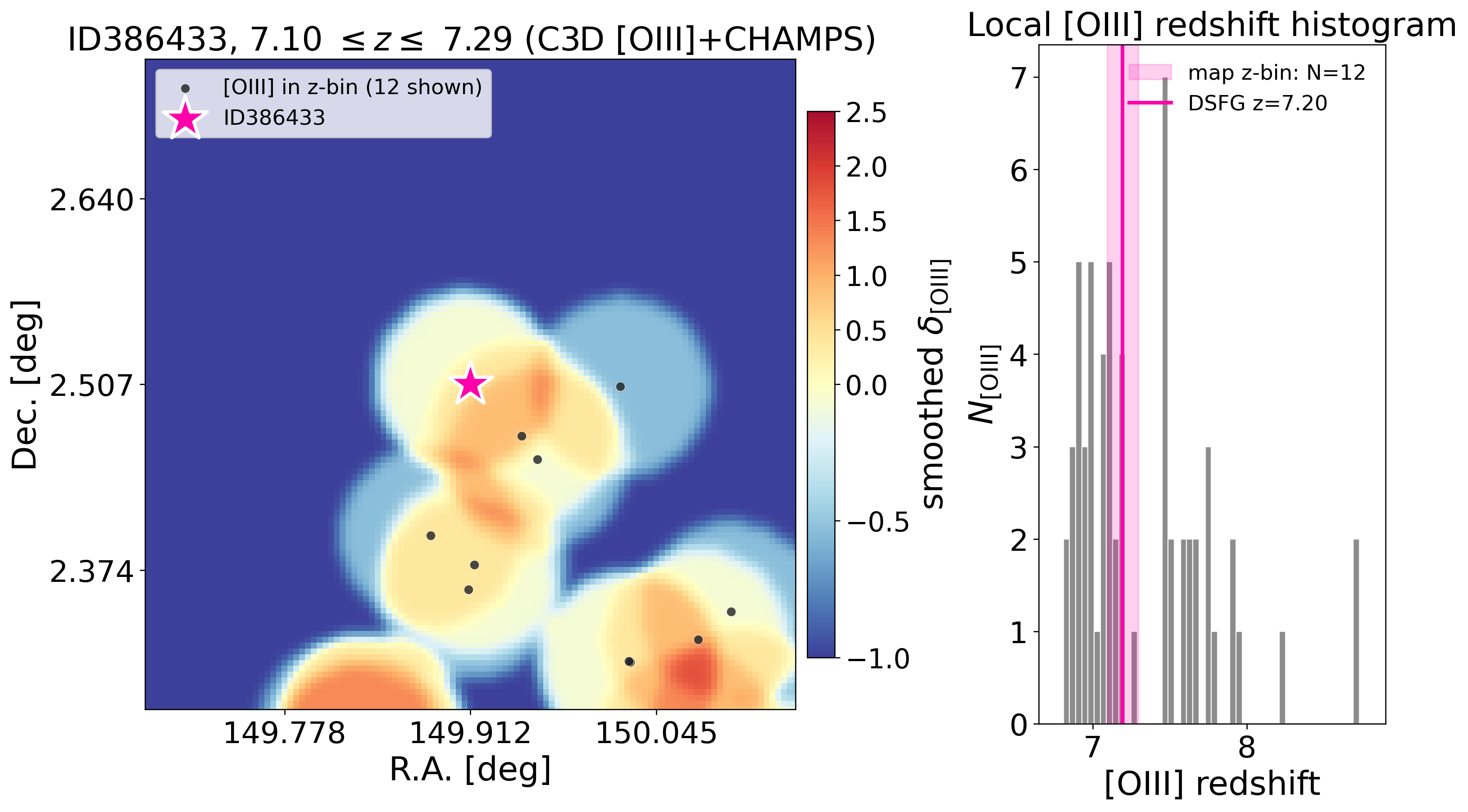}
    \hfill
    \includegraphics[width=0.49\textwidth]{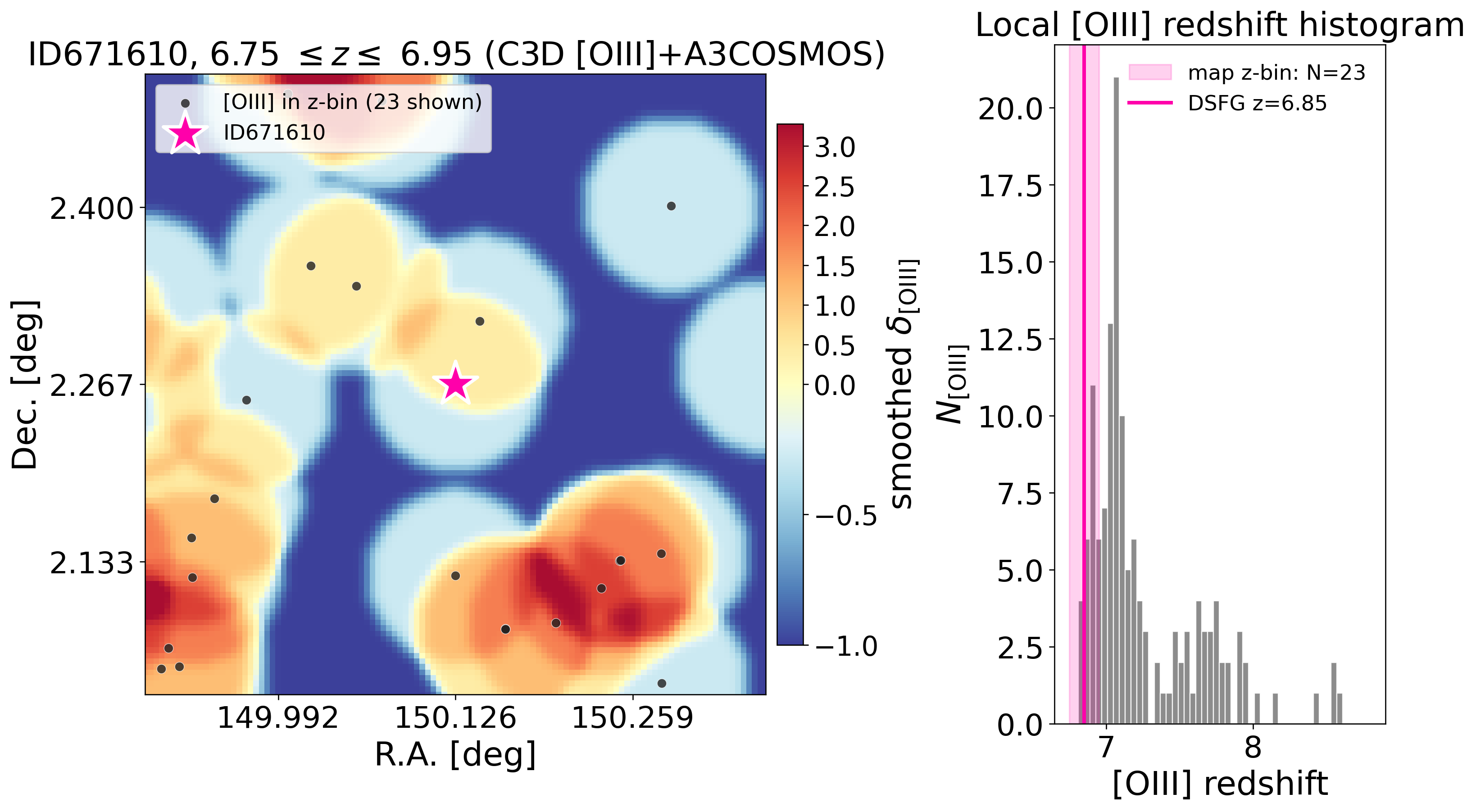}
    
    \includegraphics[width=0.49\textwidth]{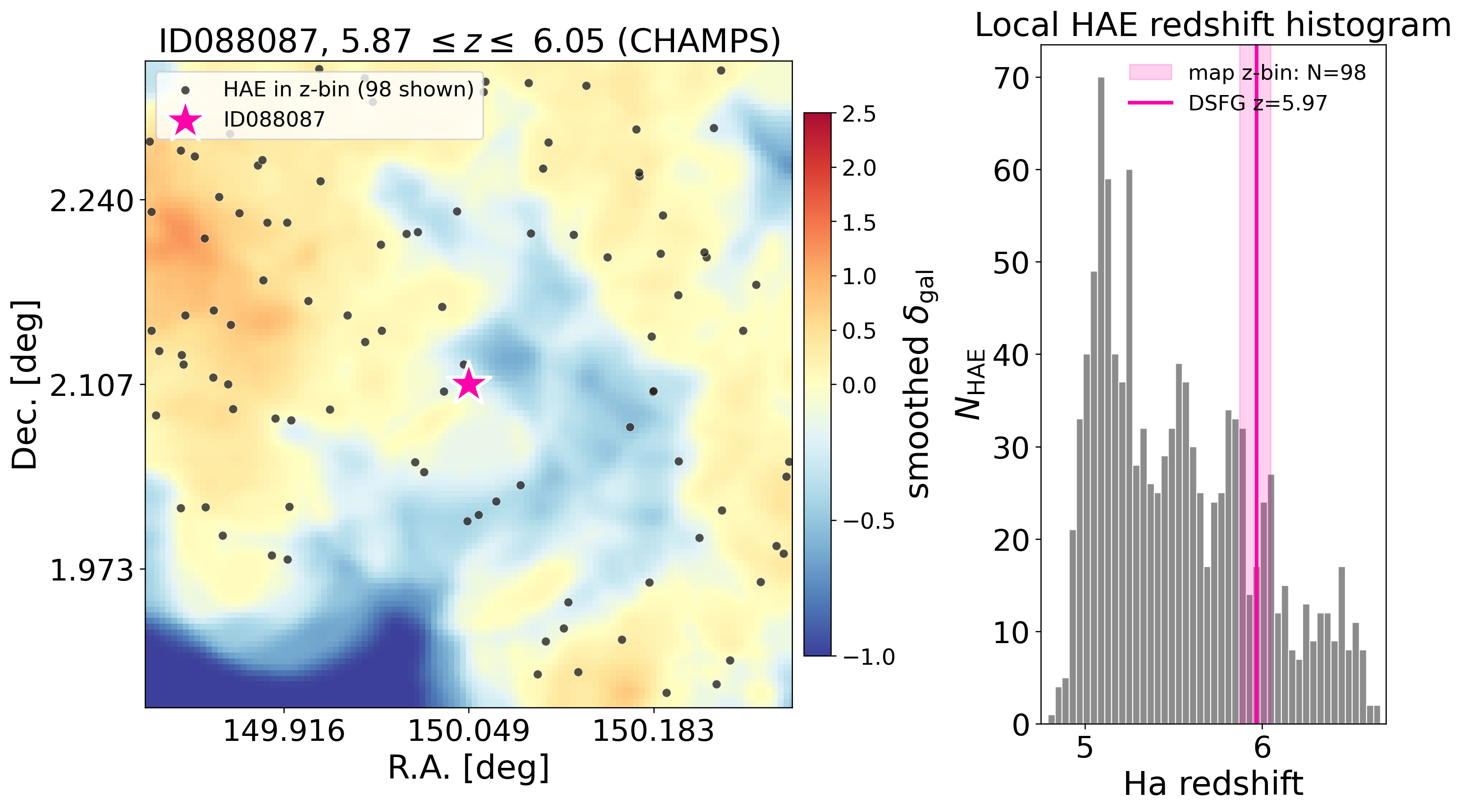}
    \hfill
    \includegraphics[width=0.49\textwidth]{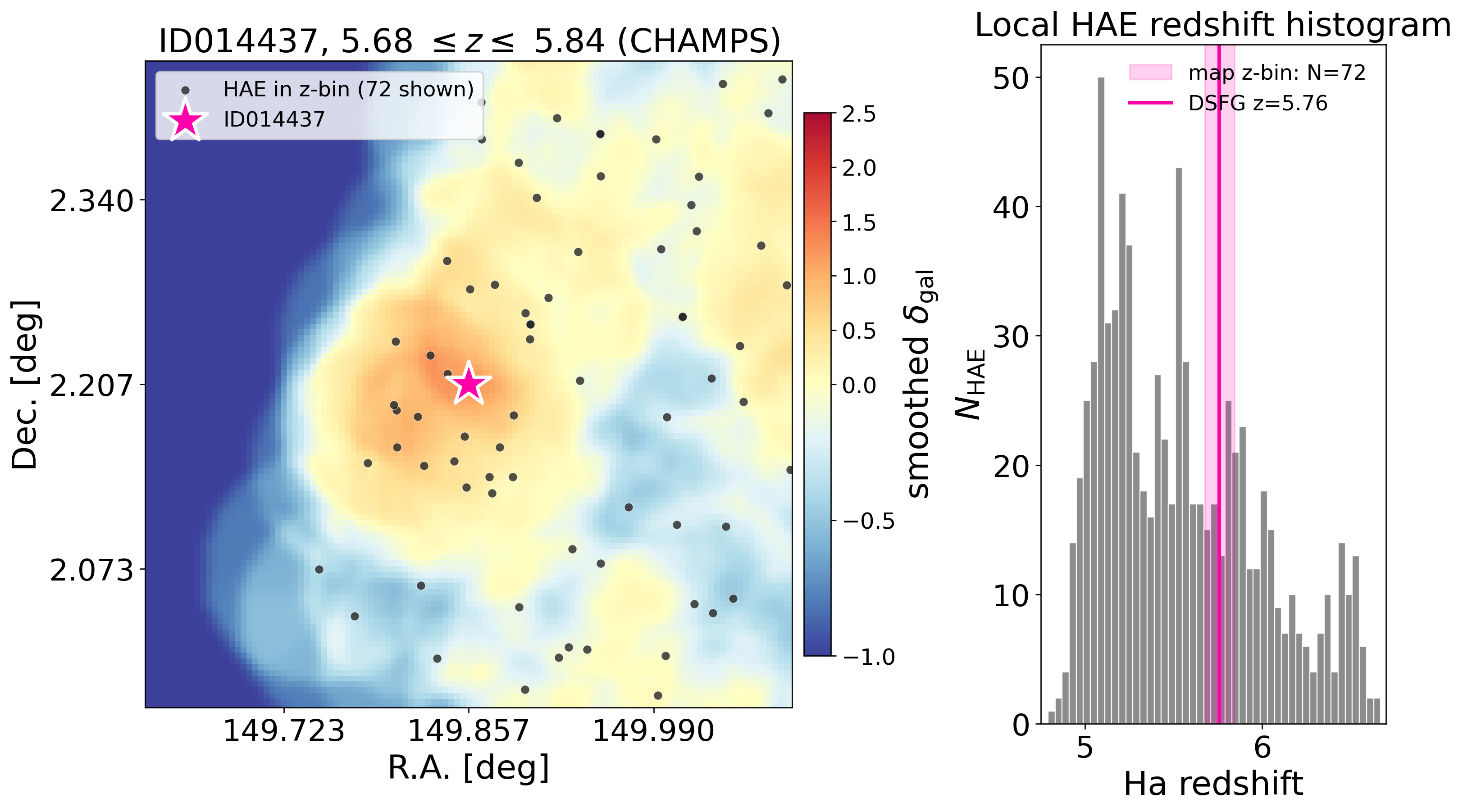}
    
    \includegraphics[width=0.49\textwidth]{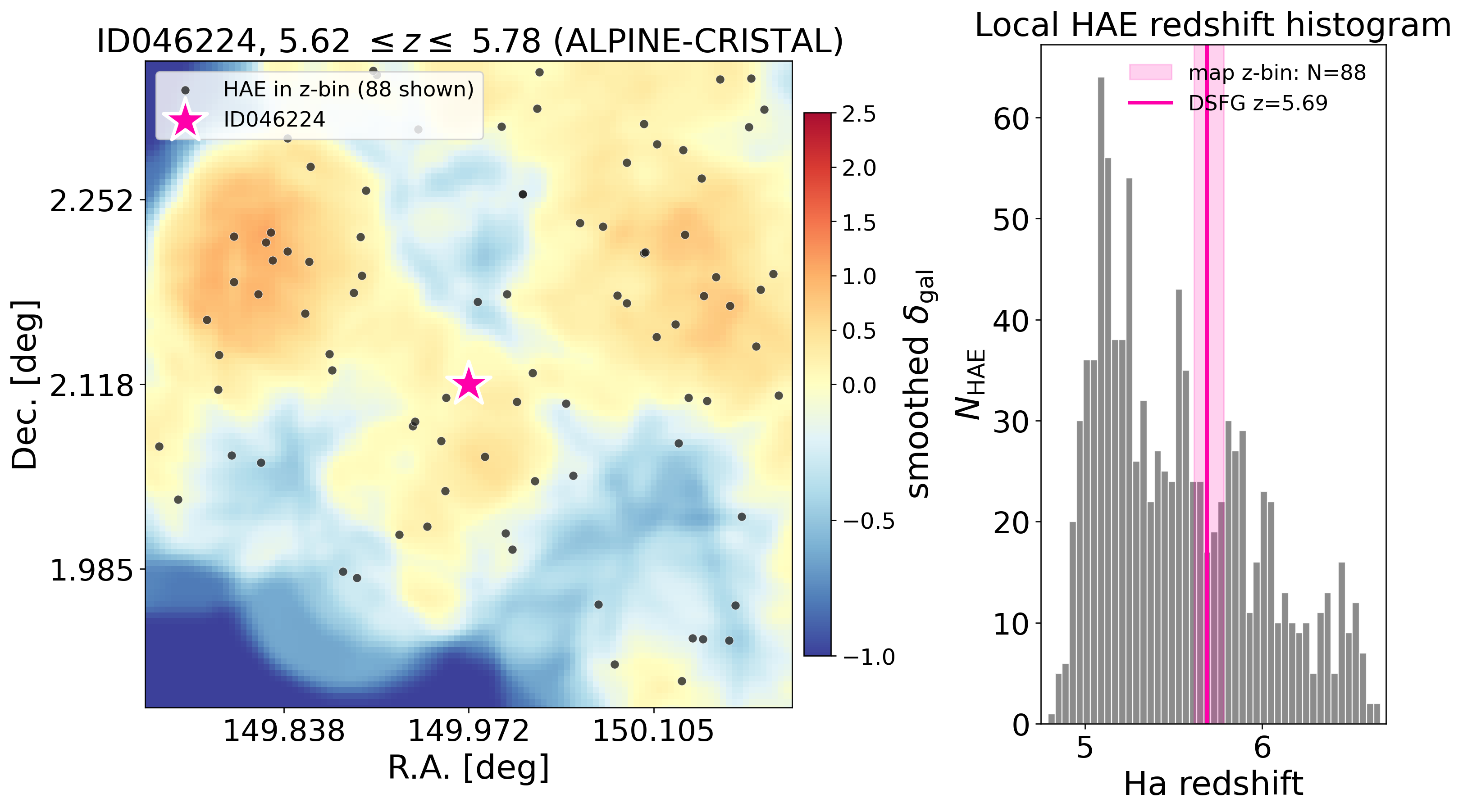}
    \hfill
    \includegraphics[width=0.49\textwidth]{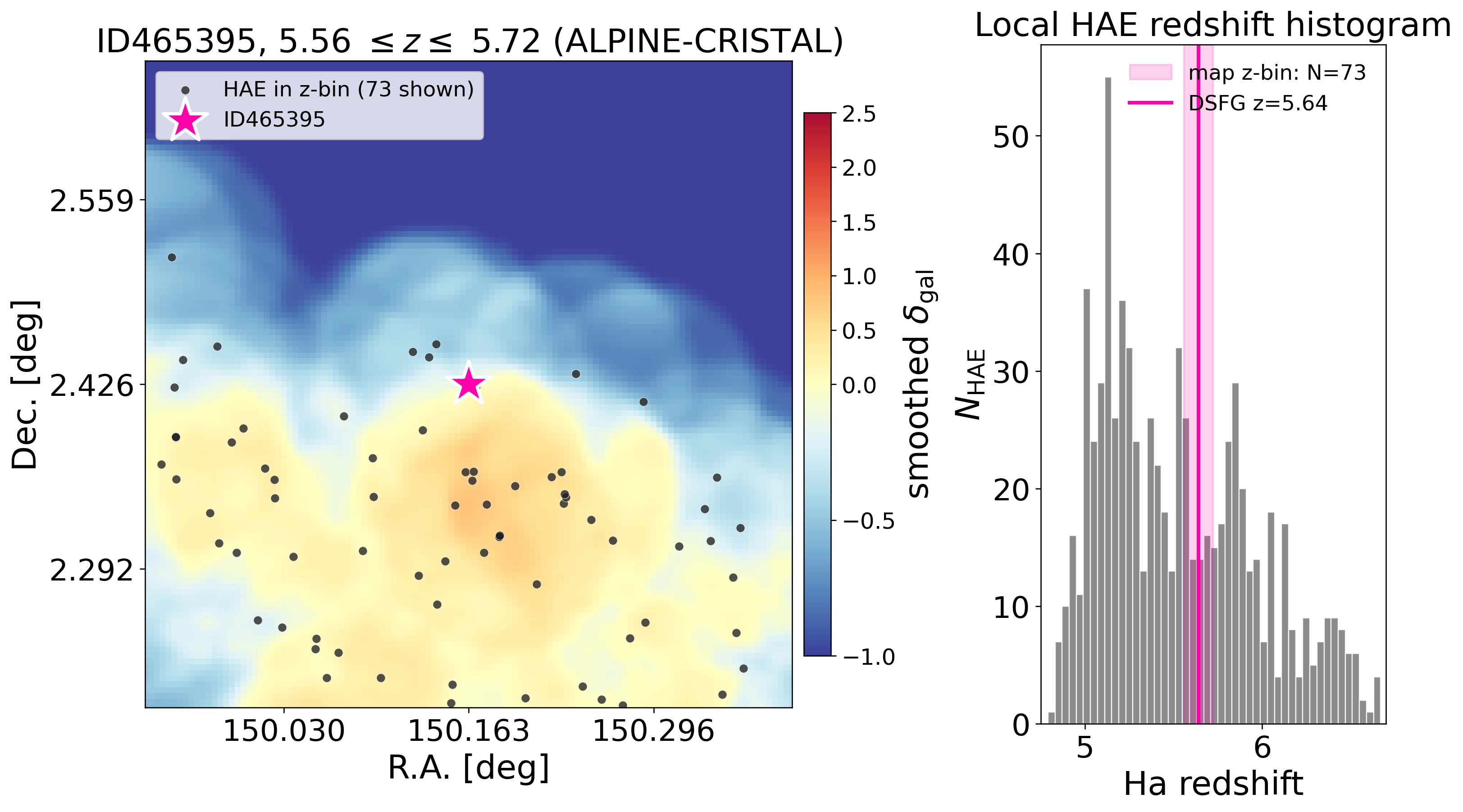}
    
    \includegraphics[width=0.49\textwidth]{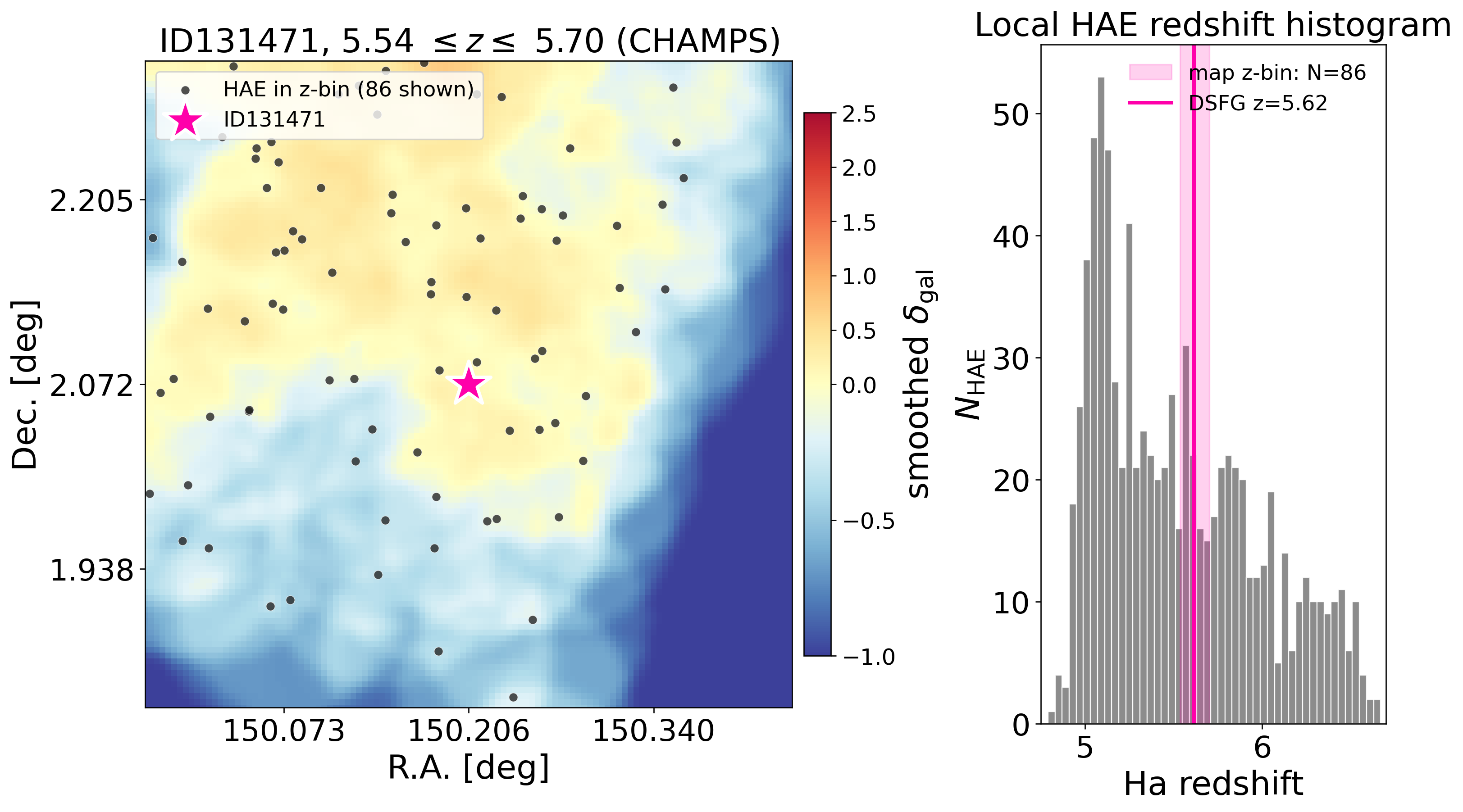}
    \hfill
    \includegraphics[width=0.49\textwidth]{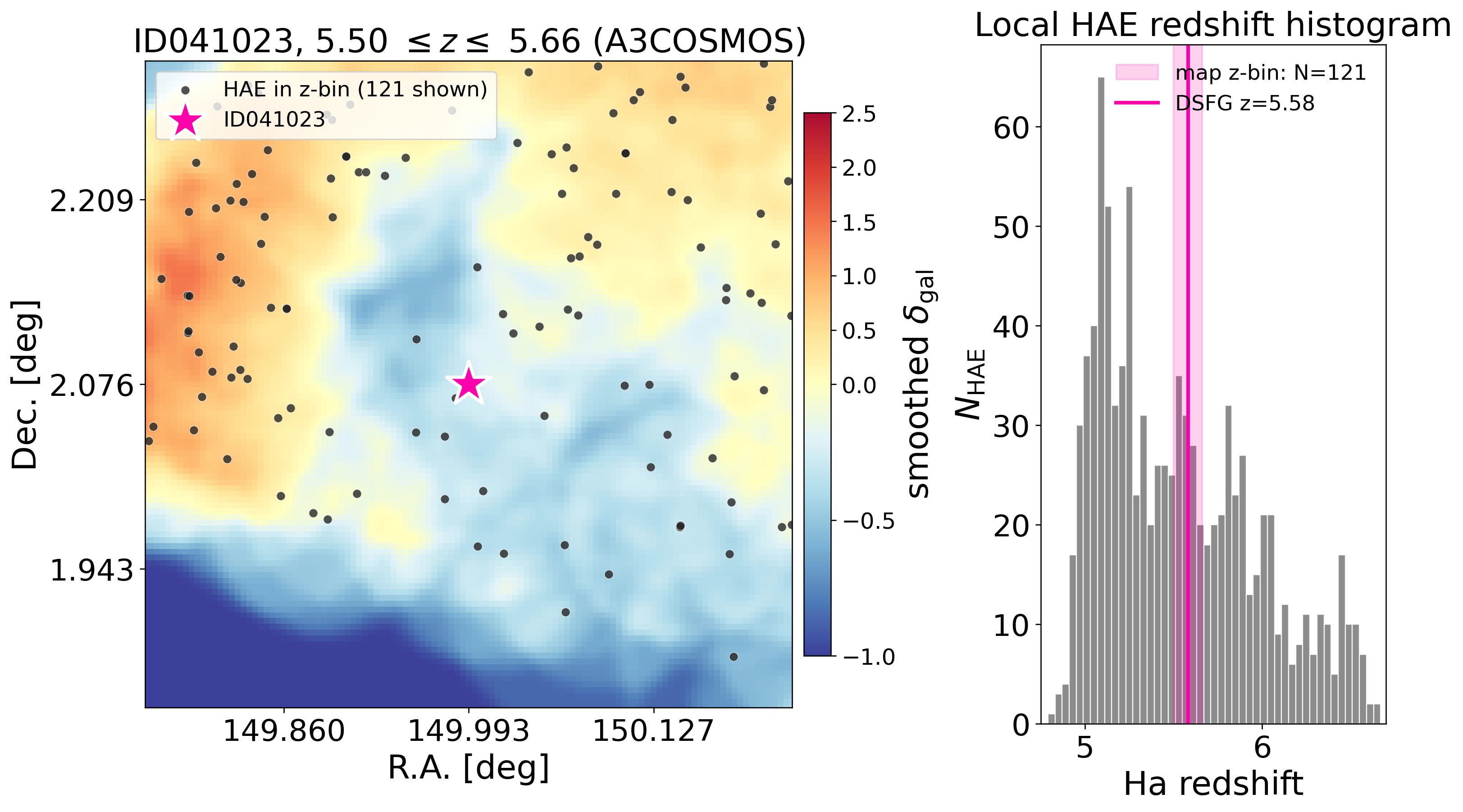}
\caption{Continued.}
\end{figure*}

\begin{figure*}
\addtocounter{figure}{-1}
\centering
\includegraphics[width=0.49\textwidth]{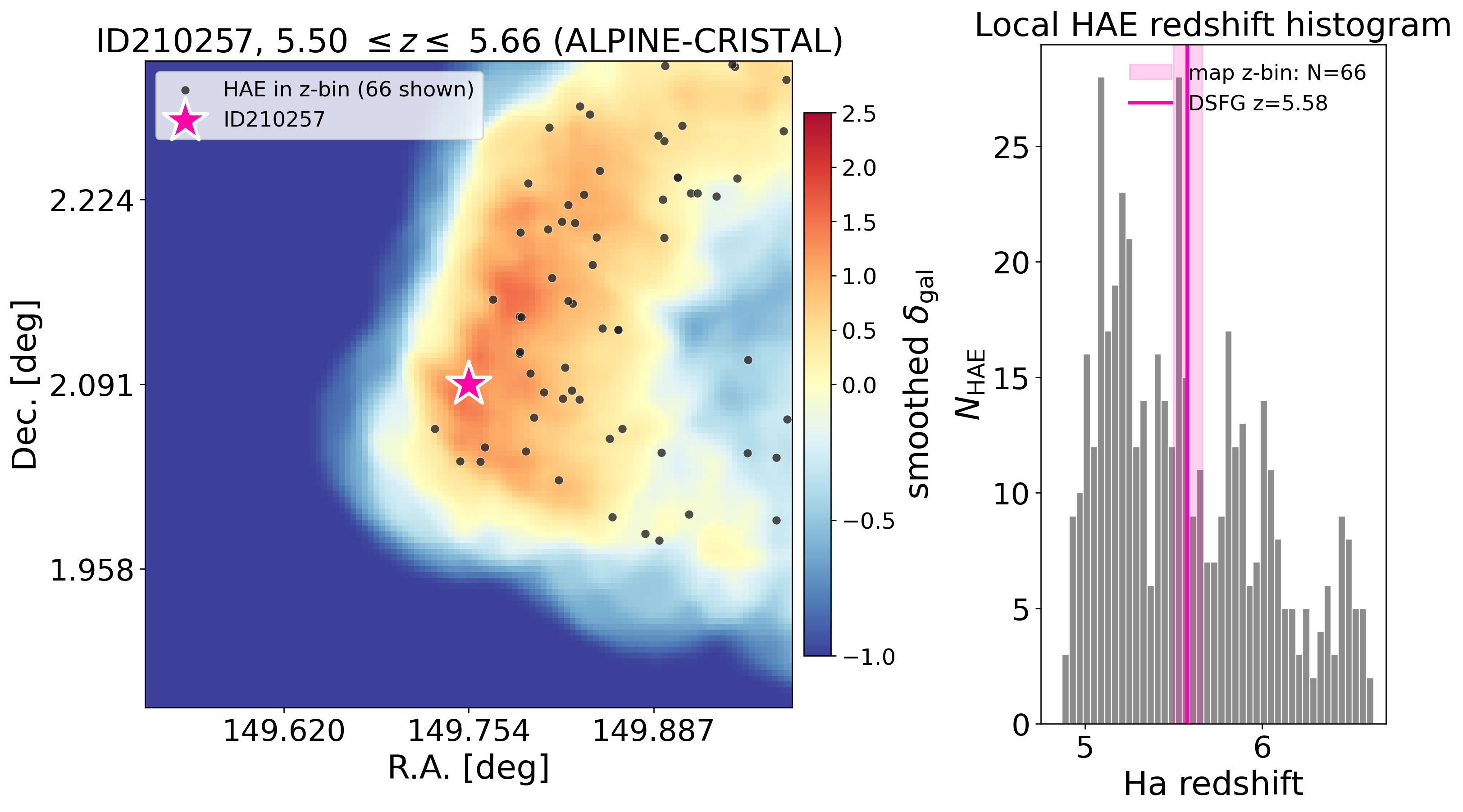}
\hfill
\includegraphics[width=0.49\textwidth]{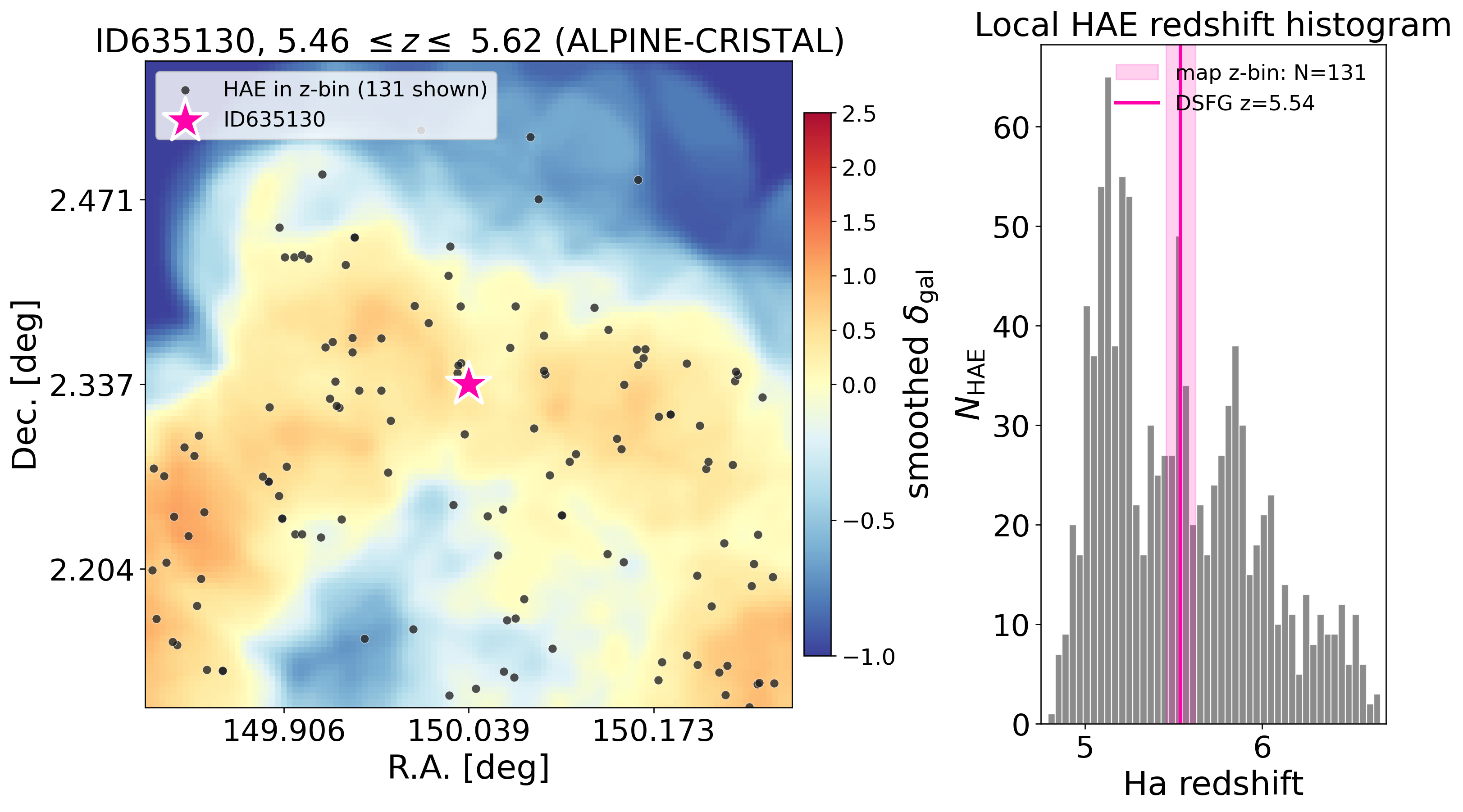}

\includegraphics[width=0.49\textwidth]{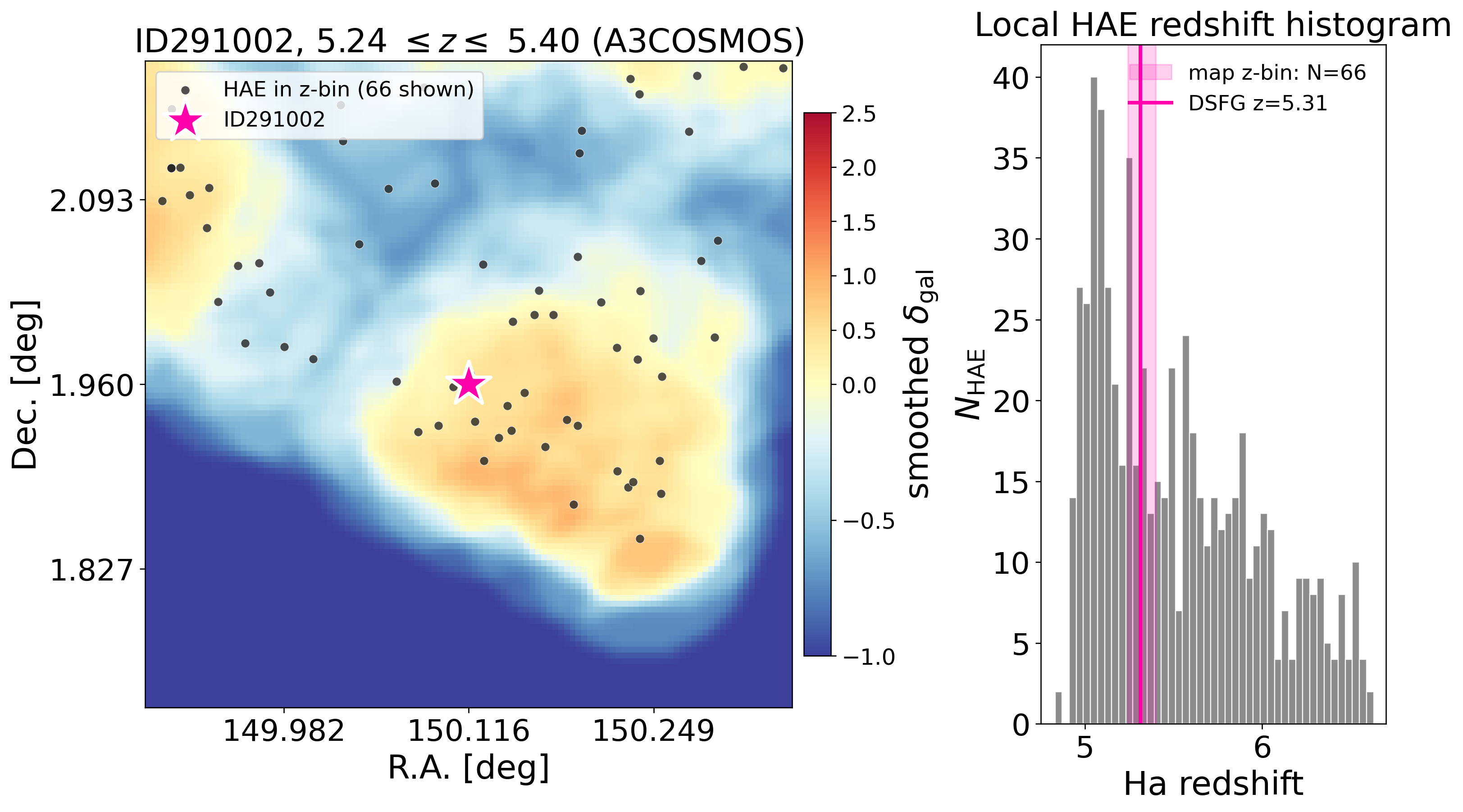}
\hfill
\includegraphics[width=0.49\textwidth]{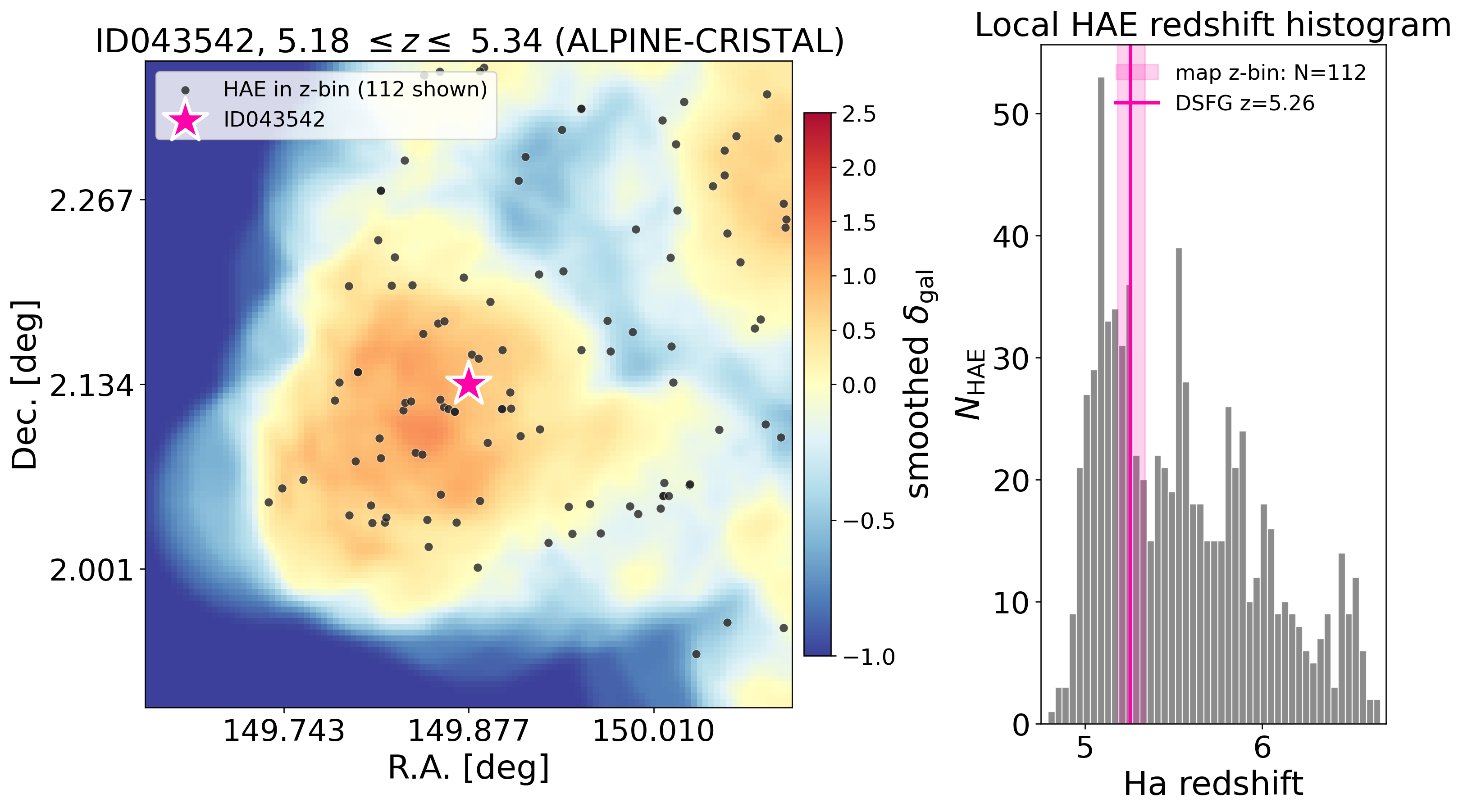}

\includegraphics[width=0.49\textwidth]{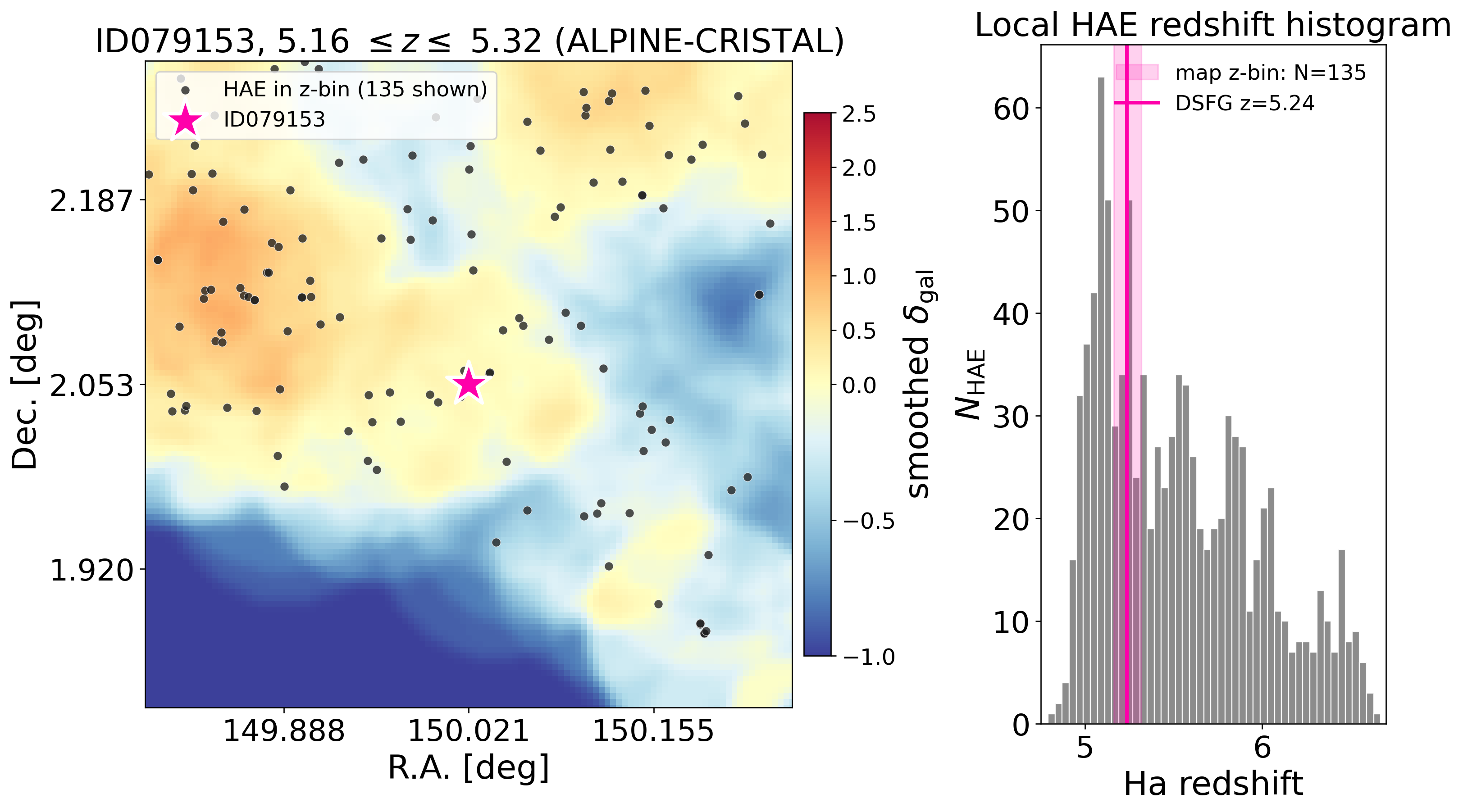}
\hfill
\includegraphics[width=0.49\textwidth]{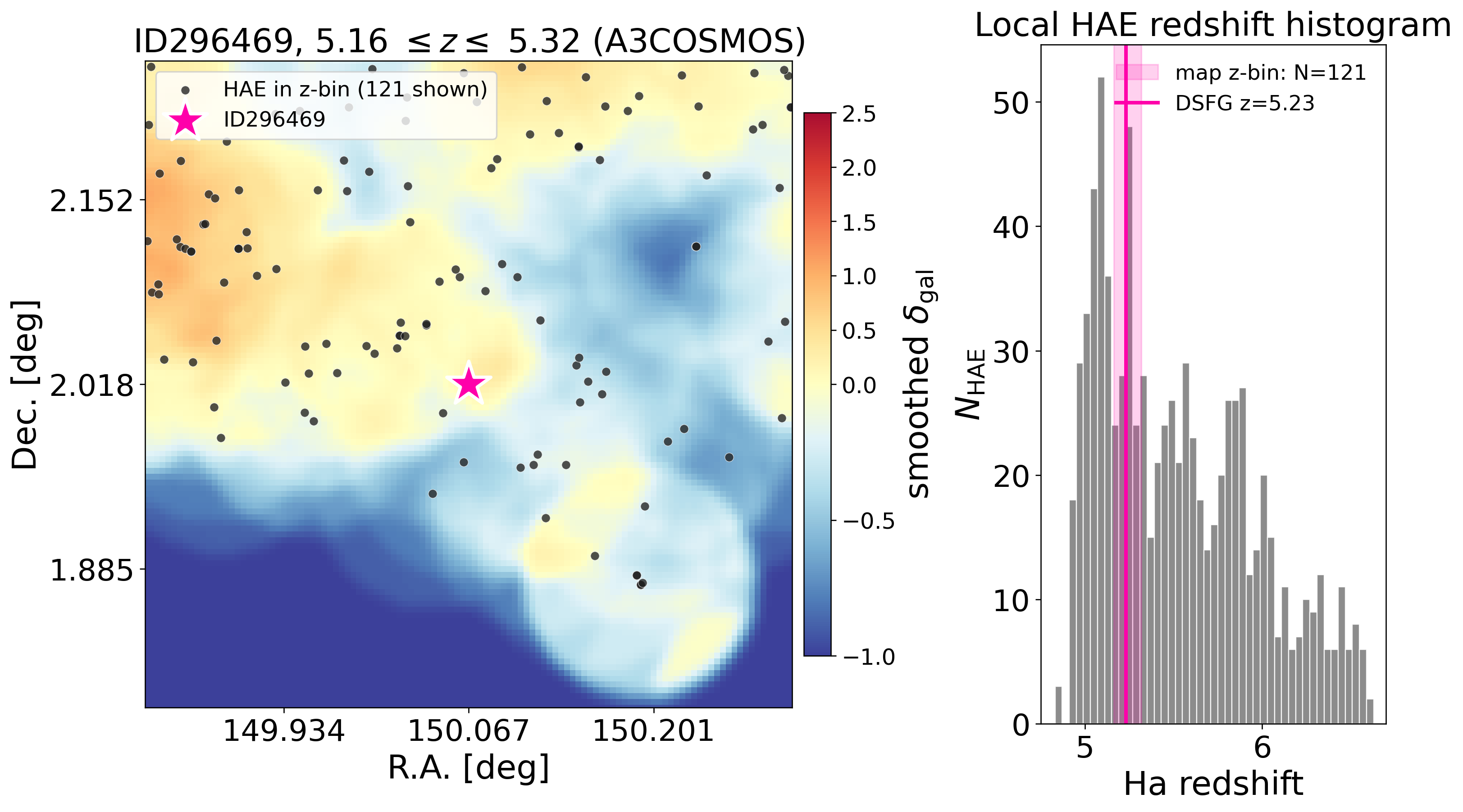}

\includegraphics[width=0.49\textwidth]{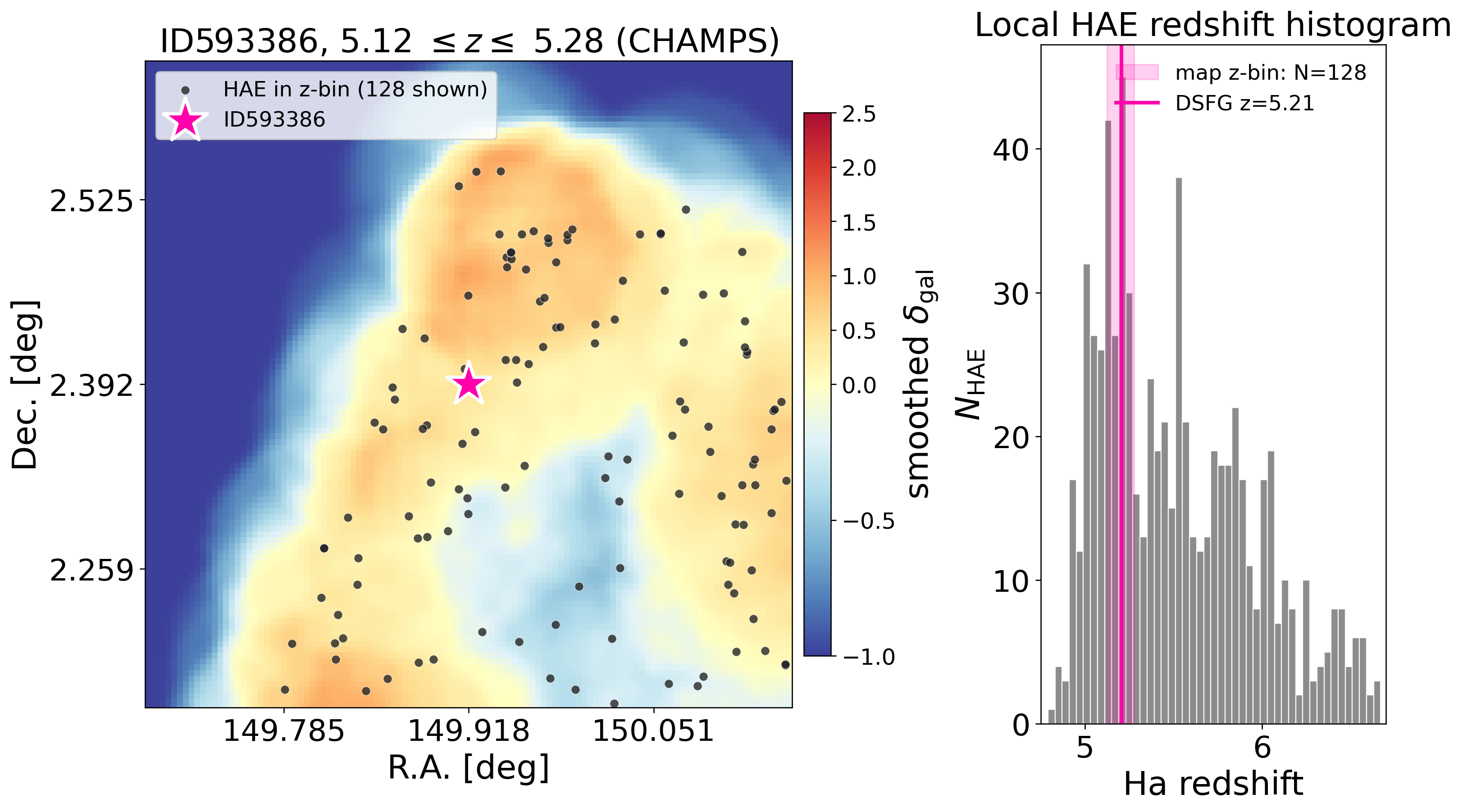}
\hfill
\includegraphics[width=0.49\textwidth]{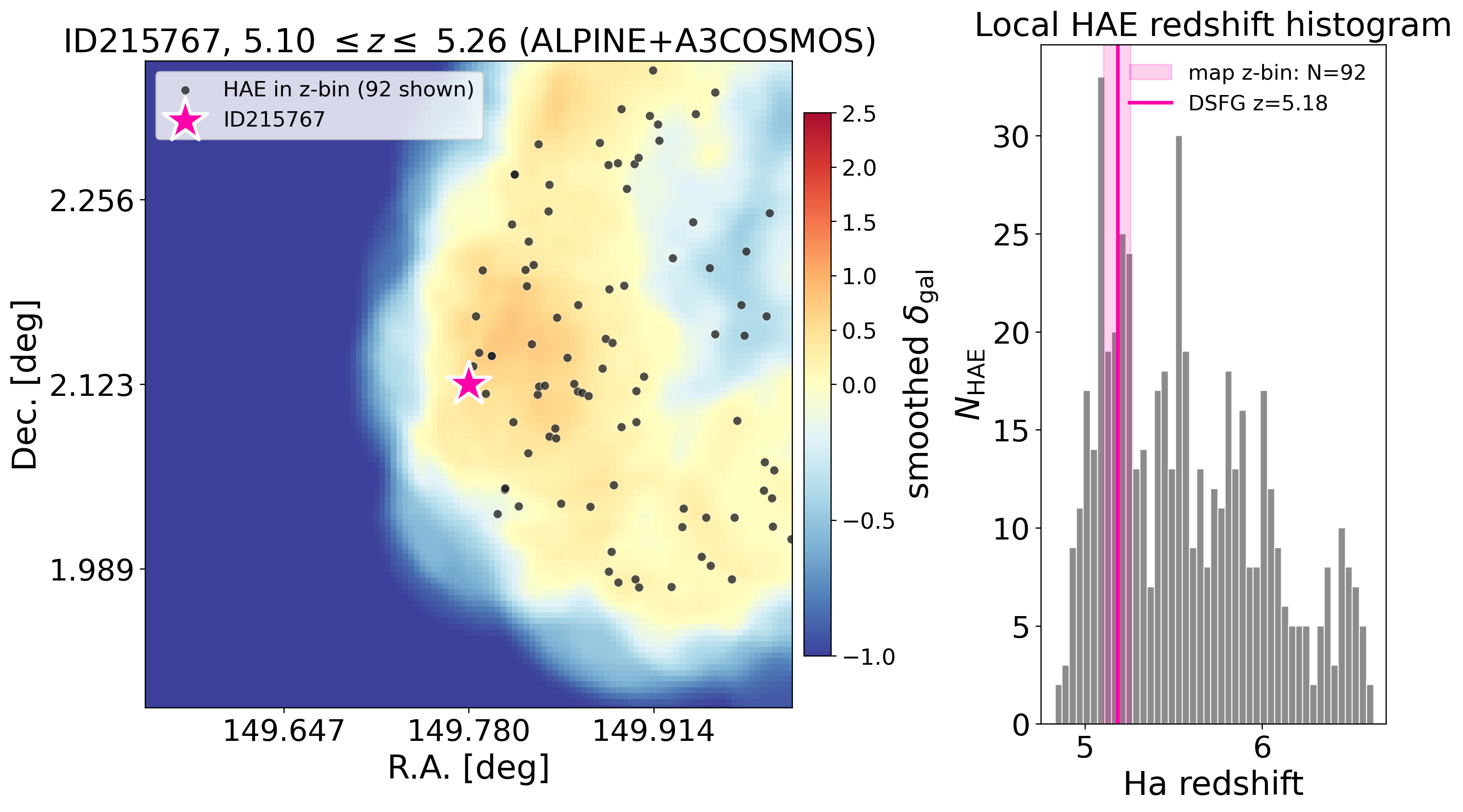}
\caption{Continued.}
\end{figure*}

\begin{figure*}
\centering
\includegraphics[width=0.49\textwidth]{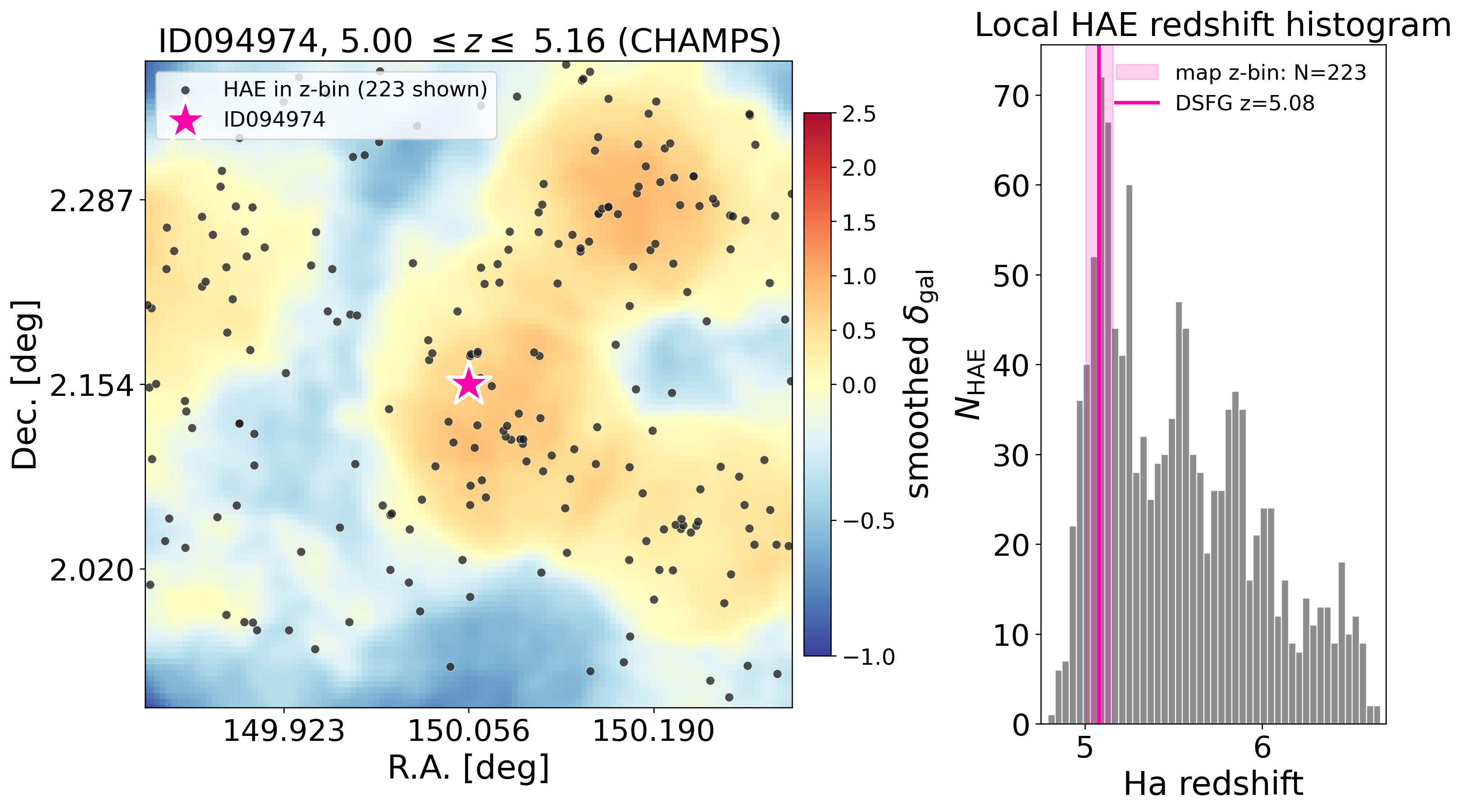}
\hfill
\includegraphics[width=0.49\textwidth]{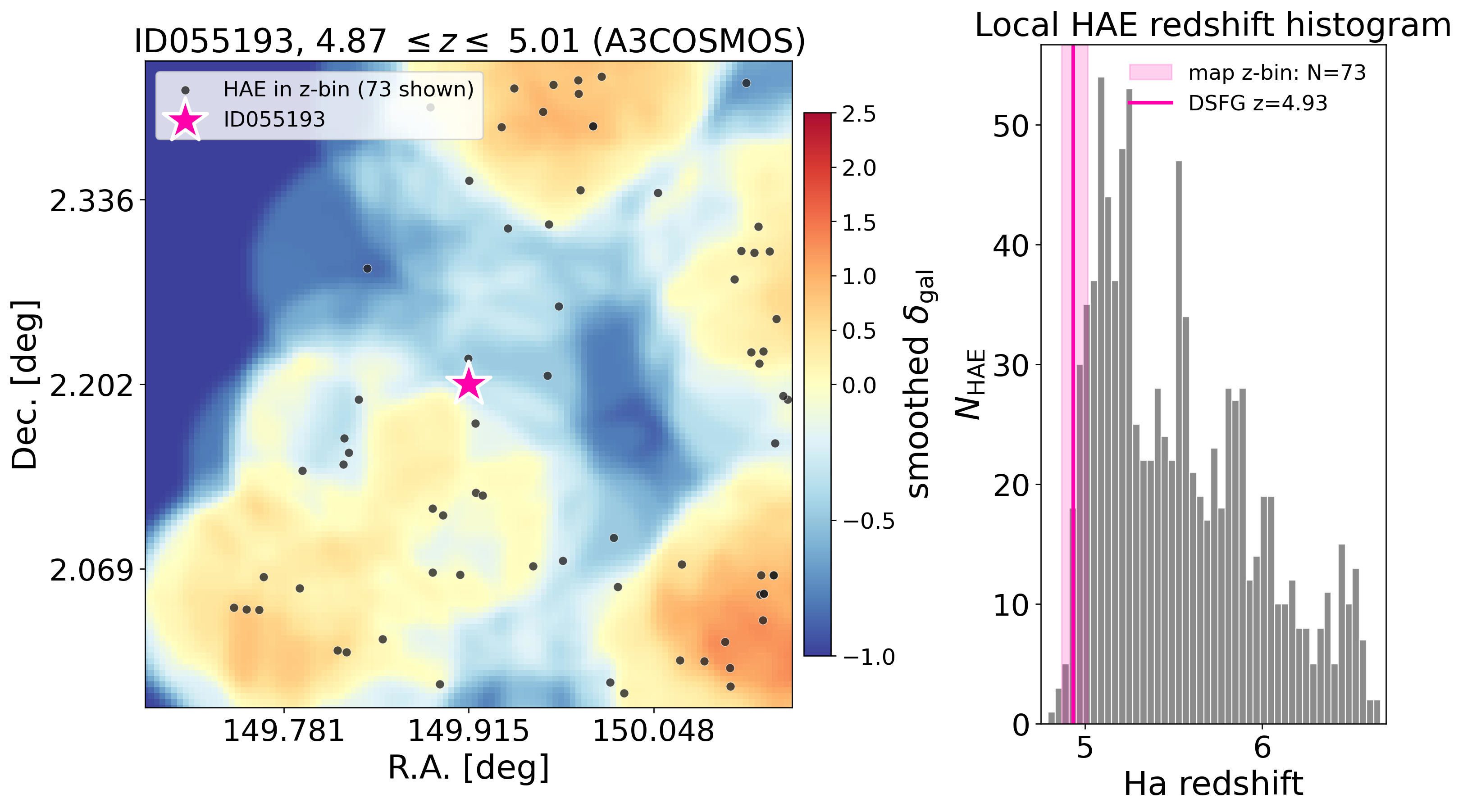}
  \caption{
Smoothed overdensity maps and local emitter redshift distributions for the 18 DSFGs. Density peaks are identified from the fiducial smoothed HAE maps in each redshift slice. Because the peak position can depend on the smoothing scale, aperture definition, and map construction, the quoted $D_{\rm peak}$ values are used as relative indicators of whether each DSFG lies near a compact HAE node, not as precise distances to a unique protocluster center.
}
    \label{fig:dsfg_ha_delta_maps}
\end{figure*}

\begin{figure*}
\centering
\includegraphics[width=0.49\textwidth]{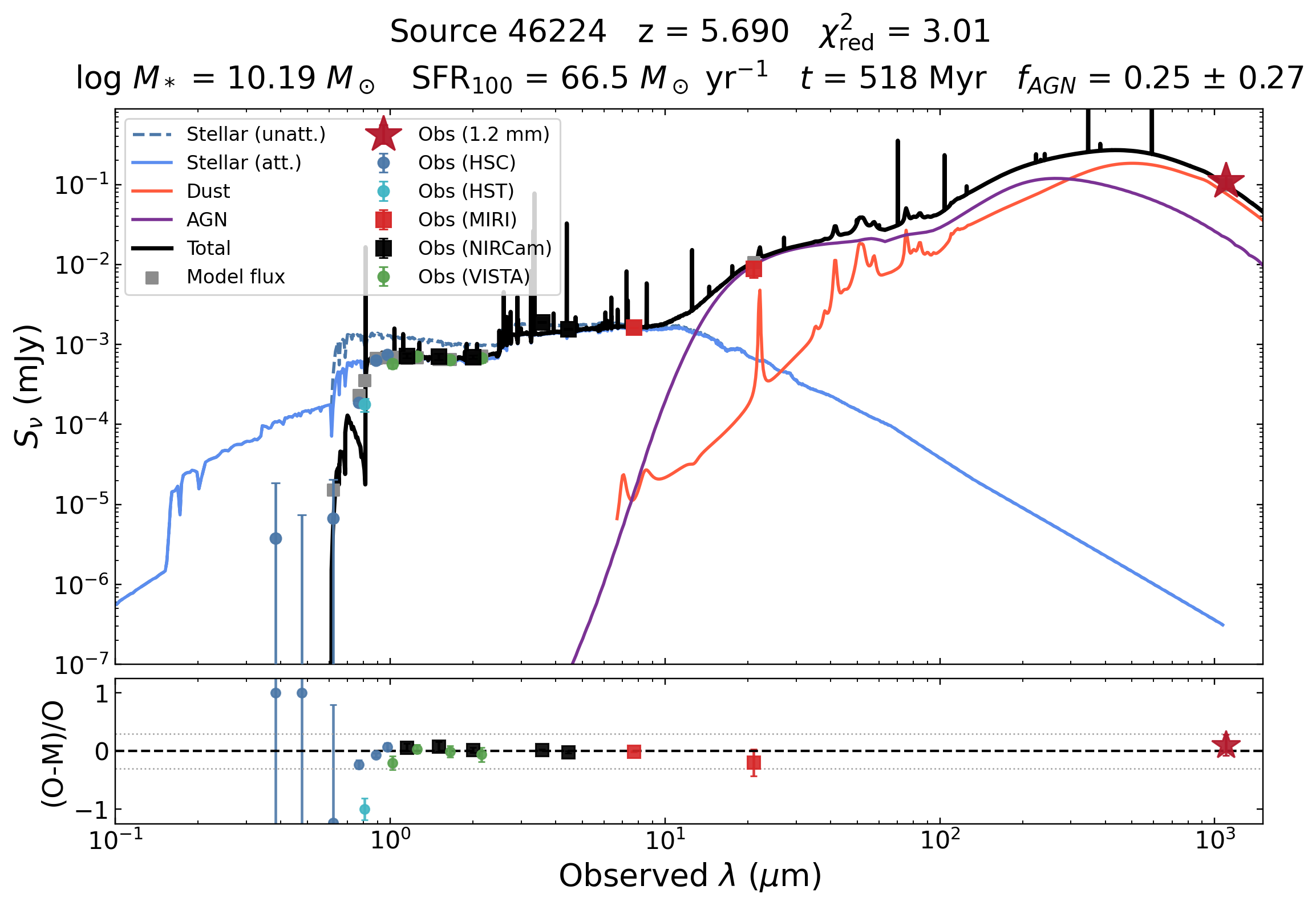}
\includegraphics[width=0.49\textwidth]{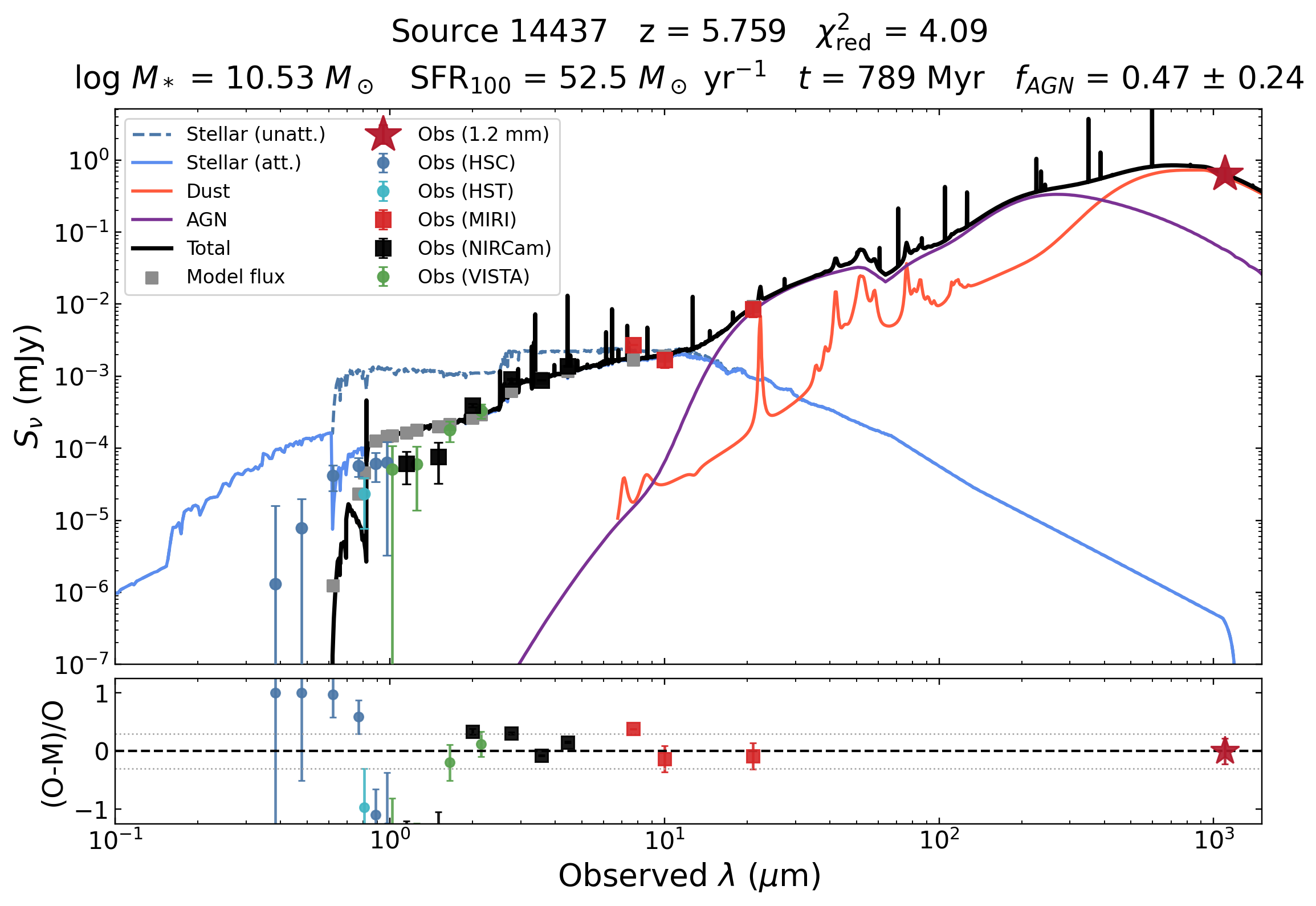}
\includegraphics[width=0.49\textwidth]{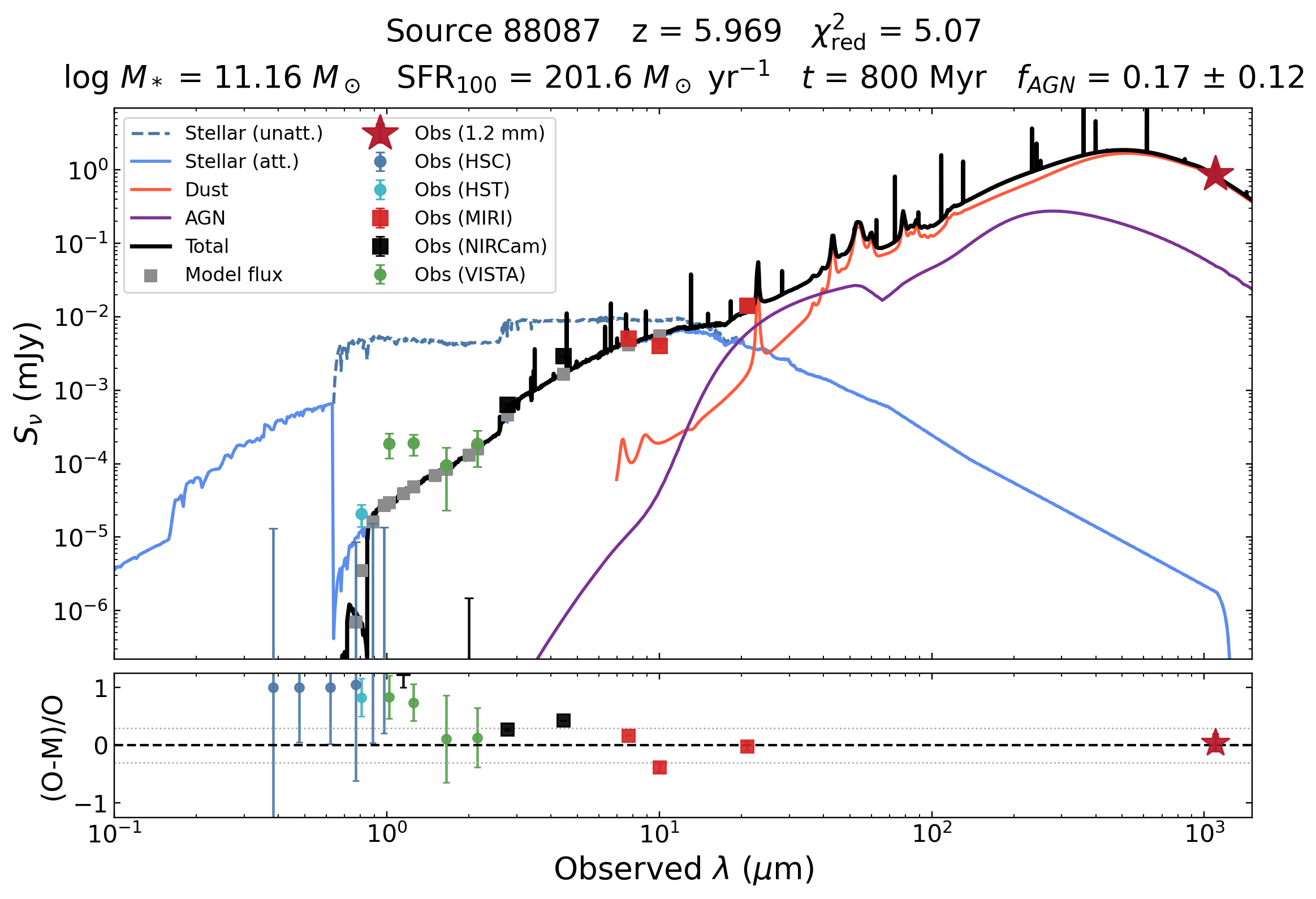}
\includegraphics[width=0.49\textwidth]{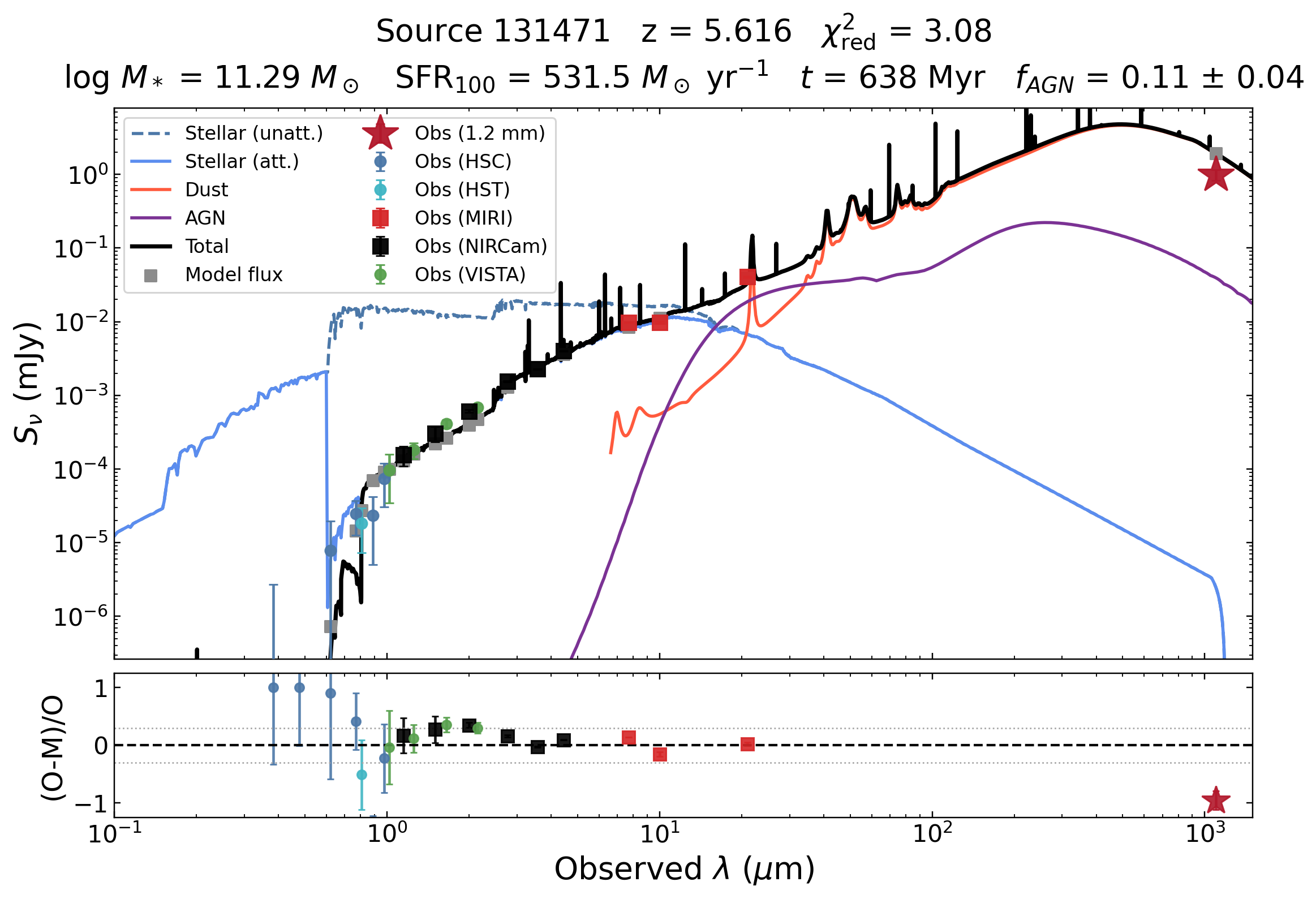}
\includegraphics[width=0.49\textwidth]{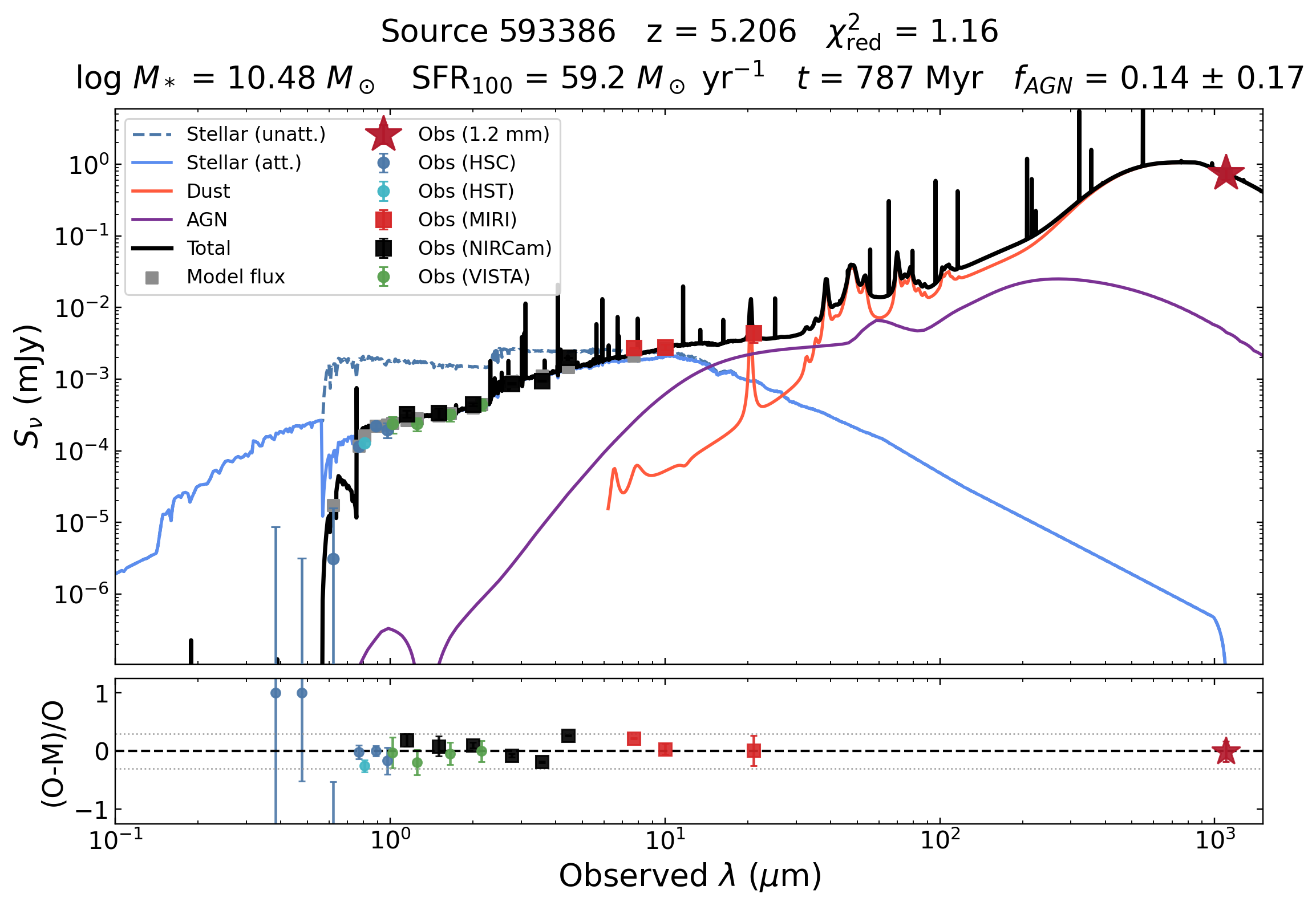}
\caption{CIGALE SED fits for the five MIRI-detected DSFGs using the Fritz AGN model \citep{fritz06} and UV-to-FIR photometry. ALMA continuum measurements are shown as red stars. Black curves indicate the total best-fit SED, while colored curves show the stellar, dust, and AGN components. Lower panels show the fractional residuals between the observed and model flux densities.}
\label{fig:miri_dsfg_sed}\label{fig:sed_miri}
\end{figure*}

\begin{figure*}
\centering
     \includegraphics[width=0.49\textwidth]{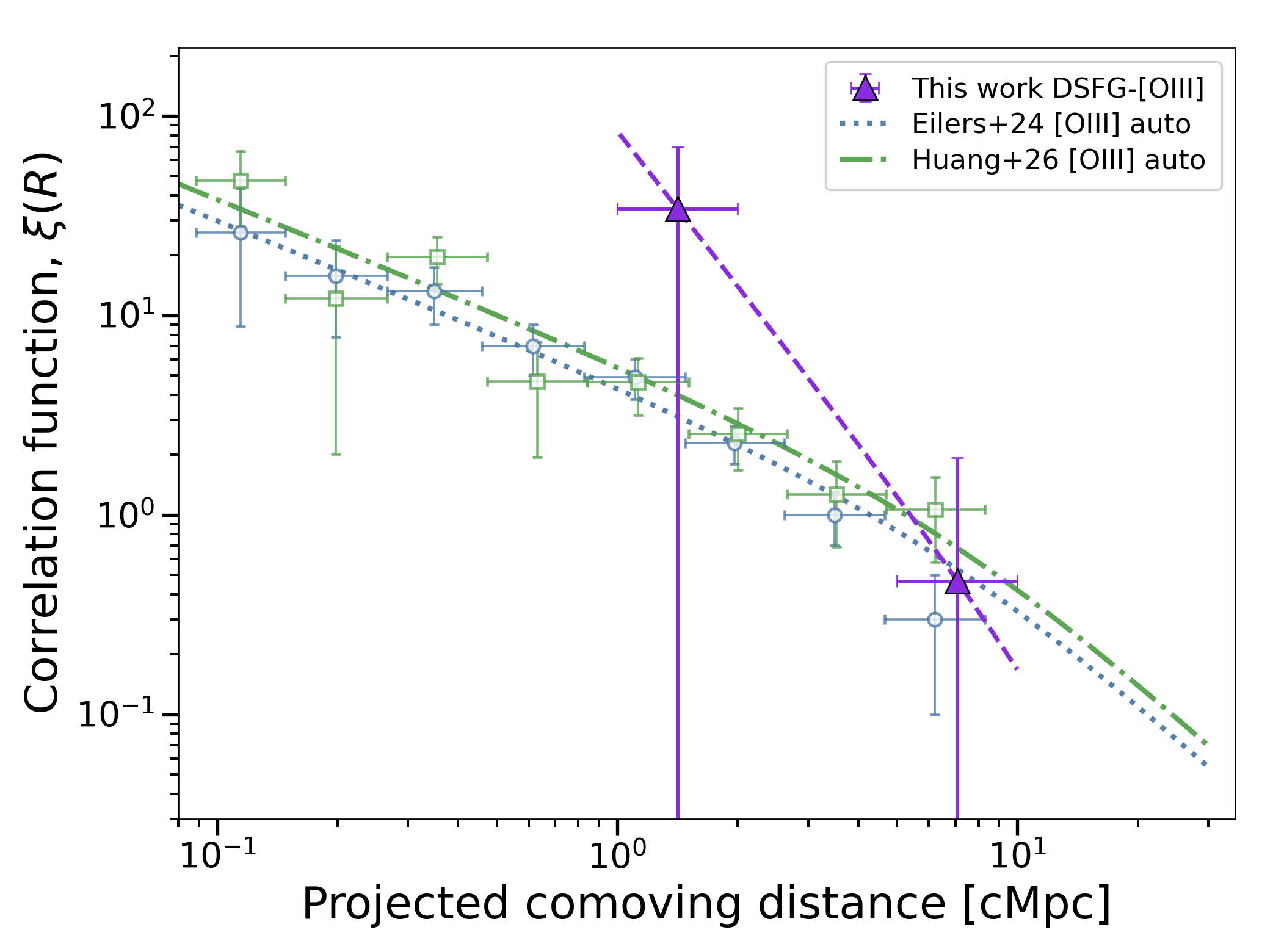}
    \caption{Projected correlation functions for the \oiii-selected DSFGs and \oiii~emitters. Purple triangles show the cross-correlation between the two \oiii-detected DSFGs and the COSMOS-3D \oiii~emitter catalog \citep{meyer25}. Error bars indicate Poisson uncertainties, and the dashed purple curve shows the best-fit line-of-sight-averaged power-law model. For comparison, we show \oiii~emitter auto-correlations from EIGER \citep{eilers24} and ASPIRE \citep{huang26}.
}\label{fig:cc_o3}
\end{figure*}

\clearpage
\bibliography{ms}{}
\bibliographystyle{aasjournal}

\end{document}